\documentclass[a4paper,12pt]{article}
\usepackage{jheppub}

\usepackage{hyperref}
\usepackage{comment}
\newcommand{\wb}{\bar w}
\newcommand{\doublecomm}[1]{\{ \{ #1 \}\}}

\newcommand{\Xd}{\buildrel{\leftrightarrow} \over {X}}

\usepackage{bm, bbm, bbold}
\usepackage{latexsym}
\usepackage{dcolumn}
\usepackage{amsmath,amsfonts,amssymb, mathrsfs, amsthm}
\usepackage[T1]{fontenc}

\usepackage{graphicx,epsfig}
\usepackage{environ}
\usepackage{mathtools}
\usepackage{braket}
\usepackage{fancyhdr}
\usepackage{hyperref}
\usepackage{graphicx,epstopdf}
\usepackage{tikz}
\usepackage{float}
\usepackage{framed}
\usepackage{soul}
\usetikzlibrary{positioning,decorations.markings, decorations.pathmorphing, calc}
\tikzset{snake it/.style={decorate, decoration=snake}}
\usetikzlibrary{angles,quotes}

\usepackage{stmaryrd}
\usepackage{slashed}
\usepackage{array,multirow}

\usepackage[framemethod=default]{mdframed}
\newmdenv[skipabove=7pt,
skipbelow=7pt,
rightline=false,
leftline=false,
topline=false,
bottomline=false,
backgroundcolor=gray!10,
linecolor=gray,
innerleftmargin=5pt,
innerrightmargin=5pt,
innertopmargin=5pt,
innerbottommargin=5pt,
leftmargin=0cm,
rightmargin=0cm,
linewidth=4pt]{eBox}

\def\a{\alpha}
\def\b{\beta}

\def\g{\gamma}

\def\e{\epsilon}

\def\l{\lambda}
\def\lb{\bar{\lambda}}

\def\p{\partial}

\newcommand{\half}{\frac{1}{2}}

\newcommand{\be}{\begin{equation}}
\newcommand{\ee}{\end{equation}}
\newcommand{\bes}{\begin{equation*}}
\newcommand{\ees}{\end{equation*}}

\NewEnviron{derivation}{
\begin{framed}
\begin{center}
{\bf-: Derivation :-}\\
\end{center}
  \BODY

\end{framed}
}

\NewEnviron{eqn}{
\begin{align}
\begin{split}
  \BODY
\end{split}
\end{align}
}

\NewEnviron{eqn*}{
\begin{align*}
\begin{split}
  \BODY
\end{split}
\end{align*}
}

\newcommand{\nno}{\nonumber}
\newcommand{\intinf}{\int_{-\infty}^{\infty}} 
\newcommand{\intsinf}{\int_{0}^{\infty}} 
\newcommand{\pd}[2]{\frac{\partial #1}{\partial #2}}

\newcommand{\tr}{\mbox{tr}}
\newcommand{\Tr}{\mbox{Tr}}
\newcommand{\red}[1]{{\color{red}#1}}
\newcommand{\blue}[1]{{\color{blue}#1}}

\usepackage{cancel}
\usepackage{tikz, pgf}
\usetikzlibrary{shapes.misc}
\usetikzlibrary{shapes}
\usetikzlibrary{fit}
\usetikzlibrary{arrows}

\usetikzlibrary{decorations.pathmorphing}	
\usetikzlibrary{decorations.markings}
\tikzset{
    sugra/.style={decorate, decoration={snake}, draw=black},
    scalarphi/.style={dashed,draw=black, postaction={decorate},
        },
    hwbou/.style={draw=blue, postaction={decorate}, ultra thick
        },
    vector/.style={draw=blue,decorate, decoration={snake}, draw},
	provector/.style={decorate, decoration={snake,amplitude=2.5pt}, draw},
	antivector/.style={decorate, decoration={snake,amplitude=-2.5pt}, draw},
   	 fermion/.style={draw=cyan, postaction={decorate},
        decoration={markings,mark=at position .55 with {\arrow[draw=black]{>}}}},
    fermionbar/.style={draw=cyan, postaction={decorate},
        decoration={markings,mark=at position .55 with {\arrow[draw=black]{<}}}},
    fermionnoarrow/.style={draw=black},
    gluon/.style={decorate, draw=red,
        decoration={coil,amplitude=4pt, segment length=5pt}},
    scalar/.style={dashed,draw=black, postaction={decorate},
        decoration={markings,mark=at position .55 with {\arrow[draw=black]{>}}}},
    scalarbar/.style={dashed,draw=black, postaction={decorate},
        decoration={markings,mark=at position .55 with {\arrow[draw=black]{<}}}},
    electron/.style={draw=black, postaction={decorate},
        decoration={markings,mark=at position .55 with {\arrow[draw=black]{>}}}},
    scalarnoarrow/.style={dashed, draw=black},
    electron/.style={draw=black, postaction={decorate},
        decoration={markings, mark=at position .55 with {\arrow[draw=black]{>}}}},
	bigvector/.style={decorate, decoration={snake, amplitude=4pt}, draw},
    photon/.style={draw=red, decorate, decoration={zigzag}, draw},
    higgs/.style={dashed, draw=black, postaction={decorate},
        },	
        goldstone/.style={draw=brown, postaction={decorate},
        },    
          ghost/.style={dashed, draw=magenta, postaction={decorate},
        decoration={markings, mark=at position .55 with {\arrow[draw=black]{>}}}
        },  
          antighost/.style={dashed, draw=magenta, postaction={decorate},
        decoration={markings, mark=at position .55 with {\arrow[draw=black]{<}}}
        },  
          mphoton/.style={decorate, decoration={snake}, draw=violet},
            realscalar/.style={draw=black}, 
           mgluon/.style={decorate, draw=blue,
        decoration={coil,amplitude=4pt, segment length=5pt}},
         weylfermion/.style={draw=orange, postaction={decorate},
        decoration={markings,mark=at position .55 with {\arrow[draw=black]{>}}}},
         weylfermionbar/.style={draw=orange, postaction={decorate},
        decoration={markings,mark=at position .55 with {\arrow[draw=black]{<}}}}, 
   	wboson/.style={draw=blue,decorate, decoration={snake,amplitude=4pt}, draw},  
    zboson/.style={draw=violet, decorate, decoration={snake}, draw},   
    lepton/.style={draw=black, postaction={decorate},
        decoration={markings,mark=at position .55 with {\arrow[draw=black]{>}}}},
    leptonbar/.style={draw=black, postaction={decorate},
        decoration={markings,mark=at position .55 with {\arrow[draw=black]{<}}}}, 
        graviton/.style={draw=blue,decorate, decoration={snake,amplitude=4pt}, draw},  
        gravitino/.style={draw=red, postaction={decorate},
        decoration={snake, markings, mark=at position .55 with {\arrow[draw=black]{>}}}},
    gravitinobar/.style={draw=red, postaction={decorate},
        decoration={snake, markings,mark=at position .55 with {\arrow[draw=black]{<}}} },    
}

\usepackage[final]{showlabels} 
\allowdisplaybreaks 

\subheader{\hfill \begin{tabular}{r} \texttt{QMUL-PH-24-24} \end{tabular}}

\begin{document}  
\title{\bf Light-cone actions and correlators of self-dual theories in AdS$_4$}

\author[a]{Chandramouli Chowdhury,}
\author[b]{George Doran,}
\author[c]{Arthur Lipstein,} 
\author[b]{Ricardo Monteiro,} 
\author[c]{Silvia Nagy,} 
\author[d, e]{and Kajal Singh}

\affiliation[a]{Mathematical Sciences and STAG Research Centre, University of Southampton, \\ Highfield, Southampton SO17 1BJ, United Kingdom}
\affiliation[b]{Centre for Theoretical Physics, Department of Physics and Astronomy, \\
Queen Mary University of London, E1 4NS, United Kingdom}
\affiliation[c]{Department of Mathematical Sciences, Durham University, \\ Stockton Road, DH1 3LE, Durham, United Kingdom}
\affiliation[d]{Department of Mathematical Sciences, University of Liverpool, \\ Liverpool, L69 7ZL,
United Kingdom}
\affiliation[e]{Chennai Mathematical Institute, H1, SIPCOT IT Park, \\ Siruseri, Kelambakkam 603103, India}

\abstract{Self-dual Yang-Mills and Einstein gravity in Euclidean AdS$_4$ are useful toy models because they can be described by simple scalar Lagrangians exhibiting a new manifestation of the colour/kinematics duality, as recently shown by two of the authors. In this paper, we clarify how the self-dual sectors fit into the full theories. In particular, we explicitly construct the
light-cone action for Yang-Mills theory and Einstein gravity in AdS$_4$ in terms of positive and negative helicity fields,
where we are able to pinpoint the self-dual sector as expected. We then show that the boundary correlators of these theories take a remarkably simple form in terms of Feynman diagrams in half of flat space, acted on by certain differential operators. We also analyse their soft limits and show that they exhibit Weinberg-like soft factors, where the soft pole which appears in scattering amplitudes is replaced by a derivative with respect to the energy.}

\maketitle

\noindent
\flushbottom
\allowdisplaybreaks

\section{Introduction}

Self-dual Yang-Mills (SDYM) and self-dual gravity (SDG) provide very fruitful toy models for the study of perturbative gauge theory and gravity, respectively. For example, their tree-level amplitudes vanish above three points, while their loop-level amplitudes are rational functions at one loop and vanish at higher loops \cite{Bern:1993qk,Bern:1996ja,Bern:1998xc,Bern:1998sv}, reflecting the underlying classical integrability \cite{Penrose:1976js,Ward:1977ta,Dunajski_1998,Prasad:1979zc,Dolan:1983bp,Boyer:1985aj,Popov:1996uu,Popov:1998pc,Park:1989vq,Husain:1993dp}. Moreover, they admit simple Lagrangians in terms of scalar fields \cite{Plebanski:1975wn,PARK1990287,Bardeen:1995gk,Chalmers:1996rq,Cangemi:1996rx} which make colour/kinematics duality \cite{Bern:2008qj,Bern:2019prr} manifest \cite{Monteiro:2011pc,Boels:2013bi}, and play a fundamental role in twistor theory \cite{lionel1996integrability,Adamo:2017qyl}. 

In recent years, the study of SDYM and SDG in non-trivial backgrounds has seen  considerable interest. This includes radiative backgrounds \cite{Adamo:2017nia,Adamo:2020yzi,Adamo:2022mev,Adamo:2023zeh}, (Anti-)de Sitter space \cite{Neiman:2023bkq,Lipstein:2023pih,Neiman:2024vit}, Eguchi-Hanson space \cite{Bittleston:2023bzp,Bogna:2024gnt}, self-dual black hole backgrounds \cite{Crawley:2023brz,Adamo:2023fbj}, and cosmological backgrounds \cite{CarrilloGonzalez:2024sto}. In \cite{Lipstein:2023pih}, SDYM and SDG in Anti-de Sitter space (AdS$_4$) were shown to admit a simple Lagrangian description in terms of scalar fields.  Since the action of SDYM is conformally invariant and AdS is conformally flat, the Lagrangian is the same as in flat space, although there are nontrivial boundary conditions. On the other hand, the Lagrangian for SDG is expressed in terms of a certain mathematical structure known as a Jacobi bracket \cite{Lichnerowicz1977LesVD,JacobiVaisman,cabauBook,Bittleston:2024rqe}, which has also appeared in \cite{Bittleston:2024rqe,CarrilloGonzalez:2024sto}, making a property known as colour-kinematics duality manifest. The Jacobi bracket reduces to the well-known Poisson bracket in the flat space limit. It had previously been observed in \cite{Monteiro:2022lwm,Adamo:2021lrv} that the colour-kinematics duality in flat space encodes the $Lw_{1+\infty}$ algebra encountered in celestial holography \cite{Strominger:2021mtt,Ball:2021tmb}. The Jacobi bracket identified in \cite{Lipstein:2023pih} corresponds to a deformation of the $Lw_{1+\infty}$ algebra. Presumably, this deformation is equivalent to the deformation subsequently found in \cite{Taylor:2023ajd,Bittleston:2024rqe}, although the connection is unclear at present.

In this paper, we will continue the investigation of SDYM and SDG in AdS$_4$, with the long-term goal of developing new theoretical tools for the study of AdS/CFT and cosmology \footnote{Note that cosmological wavefunction coefficients in dS$_4$ can be obtained from boundary correlators in AdS$_4$ by Wick rotation \cite{Maldacena:2002vr,McFadden:2009fg,McFadden:2010vh}. More recently, it was shown that in-in correlators can also be computed in terms of Witten diagrams in Euclidean AdS \cite{Sleight:2021plv,DiPietro:2021sjt}.} The immediate questions we would like to address are what form do the boundary correlators take in the self-dual sector, and how to set up a systematic expansion around it, i.e.~going beyond the self-dual sector. We address these questions by first constructing the lightcone actions for YM and Einstein gravity in AdS$_4$, and showing how the self-dual sectors embed into them. One intriguing implication of these actions, which are written in terms of positive and negative helicity fields, is that they do not contain all-plus or all-minus three-point vertices. As a result, all-plus and all-minus correlators can only arise from boundary terms. The first hints of this possibility were observed at three points in axial gauge \cite{Maldacena:2011nz}, although it is very nontrivial to see it in this gauge.

After clarifying how the scalar Lagrangians for SDYM and SDG arise from lightcone Lagrangians of the full theories, we proceed to derive their Feynman rules in AdS momentum space, and compute their tree-level boundary correlators up to five points. As a warm-up, we first consider a massless $\varphi^3$ theory in half of flat space (whose kinetic term can be mapped to a conformally coupled scalar theory in AdS$_4$ using a Weyl transformation). Remarkably, we find that the Feynman diagrams of SDYM and SDG can be reduced to $\varphi^3$ correlators, by applying certain differential operators to each interaction vertex. This is suggestive of a double copy structure, although we are still far from a systematic understanding. Encouraged by the simplicity of the Feynman diagrams, we then look at their soft limits and discover another surprise: they take a universal-looking form similar to the one recently found in full YM and gravity \cite{Chowdhury:2024wwe}, which involves an energy derivative acting on lower-point diagrams. This structure may arise from asymptotic symmetries analogous to those underlying the soft limits of scattering amplitudes \cite{Strominger:2013jfa,He:2014laa} or cosmological correlators \cite{Creminelli:2012ed,Hinterbichler:2013dpa}. 

The paper is organised as follows. In section \ref{sec:lightcone}, we derive lightcone actions for YM and gravity in AdS$_4$, and demonstrate how the SDYM and SDG scalar actions found in \cite{Lipstein:2023pih} arise. In section \ref{feynmanrules}, we derive the Feynman rules for the scalar actions, and in section \ref{sec:correlators} we compute boundary correlators and show that they take a very simple form when expressed in terms of $\varphi^3$ correlators acted on by certain differential operators. We then compute soft limits in section \ref{softlimits}, and present our conclusions and future directions in section \ref{conclusion}. There are also a number of Appendices which present additional details. In Appendix~\ref{app:boundary}, we write explicitly the boundary terms accumulated when deriving the lightcone action for positive and negative helicity fields from the standard Yang-Mills action; we do not repeat the exercise for gravity. In Appendix~\ref{appendix:othercoordchoice}, we provide the lightcone action for gravity in AdS$_4$ using an alternative coordinate system to the one in the main text, which may be more suitable for some applications. In Appendix~\ref{app:5ptSDG}, we present five- and six-point computations in SDYM and SDG. Finally, Appendix~\ref{3ptappendix} describes the spinor-helicity formalism for AdS$_4$ correlators in lightcone gauge, lifts the 3-point scalar correlator of SDYM to a spinning correlator, and shows that this gives the 3-point correlator of full YM up to a boundary term. In that Appendix, we also present spinorial expressions for other correlators SDG and SDYM correlators although the relation to spinning correlators is not investigated.

\section{Light-cone actions}
\label{sec:lightcone}

In this section, we derive the `bulk' light-cone gauge actions of Yang-Mills theory and Einstein gravity and identify the terms corresponding to the self-dual sector. By `bulk', we mean that we will not describe here boundary terms in the action (leaving that for appendix~\ref{app:boundary}, and only in the case of Yang-Mills theory). Our goal is to clarify how the self-dual sectors are part of the full theories.

Ultimately, we want to consider a perturbative expansion on the AdS${}_4$ background, given in Poincar\'e coordinates by
\begin{equation}
\label{eq:AdSmetricx30}
    ds^2_{AdS}= \frac{R^2}{z^2}\left(-du dv +(dx^1)^2 +(dx^2)^2\right)\,,
\end{equation}
where 
\begin{equation}
    z=\frac{u-v}{2}\,,
\end{equation}
and 
$R$ is related to the cosmological constant as
\begin{equation}
    \Lambda = -\frac{3}{R^2}\,.
\end{equation}
We will also use the complex coordinates
\begin{equation}
    \label{eq:DoubleNullCoordsi}
    w= x^1+ix^2\,,\quad \bar{w} = x^1-ix^2\,.
\end{equation}
Nevertheless, we will proceed on a more general background until it becomes convenient to be specific.

\subsection{Yang-Mills}
\label{sec:YMaction}

The `bulk' action for Yang-Mills theory on (A)dS$_4$ matches that on flat spacetime obtained in ref.~\cite{Chalmers:1998jb}, because of the conformal invariance of the action. Nevertheless, we review here the derivation as a warm-up to the gravity case, and also to highlight our notion of helicity.

The action is
\begin{eqn}\label{YMaction}
S_{YM}[A_\mu] = - \frac14 \int d^4 x \sqrt{|g|}\ \tr \,F_{\mu\nu} F^{\mu\nu}\,,
\end{eqn}
where we define the field strength as
\begin{eqn}
F_{\mu\nu} = \p_\mu A_\nu - \p_\nu A_\mu + ig_{YM} [A_\mu, A_\nu]\,.
\end{eqn}
We will suppress the coupling constant in the following. It can be trivially reinstated by rescaling the gauge field and the action.

For any conformally flat metric,
\begin{eqn}
ds^2 = e^{2\Omega(x)}(- du dv + d w d\bar w)\,,
\end{eqn}
the action gives
\begin{eqn}
S_{YM}[A_\mu] = \frac{1}{2}\int d^4 x \;\tr\Big[ F_{uv}^2 + F_{w\bar w}^2 + 2 \big( F_{uw} F_{v \bar w}+ F_{u \bar w} F_{vw} \big) \Big]\,.
\end{eqn}
Notice that the dependence on the conformal factor dropped out due to Weyl invariance.
Imposing now light-cone gauge,
\begin{eqn}
A_u = 0\,,
\end{eqn}
which implies $F_{u\mu} = \p_u A_\mu $\,, we obtain
\begin{eqn}
S_{YM}[A_v,A_w,A_{\bar w}] & =  \frac12 \int d^4 x\; \tr \Bigg[ (\p_u A_v)^2 + 2(\p_u A_w)(- \p_{\bar w} A_v + i [A_v, A_{\bar w}])   \\
    + 2(\p_u A_{\bar w}) & (- \p_{w} A_v + i [A_v, A_{ w}]) +  2 (\p_u A_w \p_v A_{\bar w} + \p_u A_{\bar w} \p_v A_w) + F_{w\bar w}^2  \Bigg]\,. 
    \label{eq:F2lightcone}
\end{eqn}
Up to boundary terms, we can rewrite this action as
\begin{eqn}
S_{YM}[A_v,A_w,A_{\bar w}]  &= \frac12 \int d^4 x\; \tr \Bigg[ \Big( \p_u A_v - \frac{1}{\p_u} \big( \p_u \p_{w} A_{\bar w} - i [\p_u A_w, A_{\bar w}] + (w \leftrightarrow \bar w)\big) \Big)^2  \\
&  - \Big( \p_{w} A_{\bar w} - \frac{i}{\p_u} [\p_u A_w, A_{\bar w}] + (w \leftrightarrow \bar w) \Big)^2 - 4 A_w \p_u \p_v A_{\bar w} + F_{w\bar w}^2 \Bigg]\,.
\end{eqn}
We note the appearance of the inverse derivative with respect to $u$, which is typical of light-cone gauge actions. It can be defined via the following differential equation:
\begin{equation}
\label{eq:InverseDerivativeIdentity}
   \partial_{u}\left(\frac{1}{\partial_{u}}f(x)\right)=f(x)\,.
\end{equation}
Now, since $A_v$ appears quadratically in the action, we can integrate it out exactly, and we are left with the terms in the second line. Further manipulation of those terms, up to total derivatives in the integrand, leads to
\begin{eqn}
S_{YM}[A_w,A_{\bar w}] = 2\int d^4 x\, \tr &\Big(  A_w \Box A_{\bar w} -  i A_w\, \p_u   [\frac{ \p_w}{\p_u} A_{\bar w}, A_{\bar w}] -  i A_{\bar w}\, \p_u  [\frac{\p_{\bar w}}{\p_u}  A_{w}, A_{w}] \\
&  -  [\p_u A_w, A_{\bar w}] \frac{1}{\p_u^2} [\p_u A_{\bar w}, A_w]\Big)\,.
\label{eq:YMlightcone}
\end{eqn}
In our conventions, $A_{\bar w}$ ($A_w$) represent the positive (negative) helicity fields. In Appendix~\ref{app:boundary}, we collect all the boundary terms accumulated between the expressions \eqref{eq:F2lightcone} and \eqref{eq:YMlightcone}, assuming that there is a boundary at $z=\frac{u-v}{2}=0$, which will be the case of interest in this paper.

Finally, the action for self-dual Yang-Mills theory is obtained by keeping only the terms linear in the negative helicity field, which therefore becomes a Lagrange multiplier:
\begin{eqn}
S_{SDYM}[A_w,A_{\bar w}] = 2\int d^4 x\, \tr  A_w \Big( \Box A_{\bar w} -  i \p_u \big[ \frac{\p_w}{\p_u} A_{\bar w}, A_{\bar w} \big]  \Big)\,.
\end{eqn}
Redefining
\begin{eqn}
\label{eq:defPhiPhibar}
A_{\bar w} = \p_u \Phi\,, \qquad A_{w} =- \frac{1}{\p_u} \bar\Phi\,,
\end{eqn}
we can write
\begin{eqn}\label{lag_sdym}
S_{SDYM}[\Phi,\bar\Phi] = 2\int d^4 x\, \tr \ \bar \Phi \Big( \Box \Phi -  i  \big[ \p_w \Phi, \p_u \Phi \big]  \Big)\,.
\end{eqn}
Ref.~\cite{Monteiro:2011pc} observed that the derivative in the interaction term leads to Feynman rules where the kinematic part of the vertex is the structure constant of the Lie algebra of area-preserving diffeomorphisms in the $(u,w)$ null plane. This algebra mirrors the colour algebra, making manifest the colour-kinematics duality in the self-dual sector of Yang-Mills theory.

One crucial observation is that the procedure above defines an off-shell notion of helicity that is inherited from flat space. The positive helicity part of our gauge field has components $(0,A_v,0,A_{\wb})$, while the negative helicity part has components $(0,A_v,A_w,0)$. We will discuss in Appendix~\ref{3ptappendix} the associated polarisation vectors, which are distinct from those used e.g.~in \cite{Maldacena:2011nz}.

\subsection{Einstein gravity}

We now deal with the much more elaborate case of gravity, starting from the Einstein-Hilbert action,
\begin{equation}
\label{eq:EHaction}
    S_{G}=\frac{1}{2\kappa^2}\int d^4 x \sqrt{|g|}\,(R-2\Lambda)\,,
\end{equation}
where $\kappa=\sqrt{8\pi G_N}$, $R$ is the Ricci scalar and $\Lambda$ is the cosmological constant.
We will present the light-cone gauge action expressed in terms of positive and negative helicity fields. The perturbative expansion in flat spacetime has long been worked out; see e.g.~\cite{Bengtsson:1983vn,Ananth:2006fh}. However, despite significant earlier work \cite{Akshay:2014pla} (where we find a small discrepancy), this is to our knowledge the first complete derivation on an (A)dS$_4$ background.

We will begin the discussion without specifying a background and eventually specialize to flat and AdS background. We follow the approach described in Appendix C of \cite{Ananth:2006fh} by choosing light-cone coordinates $(u,v,x^1,x^2)$ such that the light-cone gauge is defined by
\begin{equation}
\label{eq:LCgauge}
    g_{u u}=g_{u i}=0\,,
\end{equation}
for $i=1,2$. These are only three conditions, so we are left with a fourth gauge choice to be made later. The components of the metric can be written as 
\begin{equation}
\label{eq:LCMetric}
    g_{\mu \nu}= 
    \begin{pmatrix}
        0 & -e^F & 0 & 0\\
        -e^F & g_{vv} & g_{v1} & g_{v2}\\
        0 & g_{v1} & e^G \gamma_{11} & e^G \gamma_{12}\\
        0 & g_{v2} & e^G \gamma_{12} & e^G \gamma_{22}\\
    \end{pmatrix},
\end{equation}
where $F$ and $G$ are real functions, and $\gamma_{ij}$ is a real, symmetric matrix of unit determinant. The determinant of the metric is, therefore,
\begin{equation}
    g=-e^{2(G+F)}\,,
\end{equation}
and the inverse metric is
\begin{equation}
\label{eq:LCMetricInverse}
    g^{\mu \nu}= 
    \begin{pmatrix}
        g^{u u} & -e^{-F} & g^{u 1} & g^{u 2}\\
        -e^{-F} & 0 & 0 &0\\
        g^{u 1} & 0 & e^{-G} \gamma_{22} & -e^{-G} \gamma_{12}\\
        g^{u 2} & 0 & -e^{-G} \gamma_{12} & e^{-G} \gamma_{11}\\
    \end{pmatrix}.
\end{equation}
Some components of the metric can be integrated out in the action, similarly to what we saw previously in the Yang-Mills case, where we integrated out $A_v$. 

To proceed with this, it is convenient to express certain components of the metric in terms of components of the inverse metric as follows:
\begin{equation}
\label{eq:Metricvcomponents}
     g_{vv}= e^{2F}(-g^{uu}+e^G\left(g^{ui}\gamma_{ij}g^{ju})\right)\,, \qquad
     g_{vi}=e^{F+G} g^{uj}\gamma_{ji}\,.
\end{equation}
After this substitution, the action depends only on the quantities appearing in \eqref{eq:LCMetricInverse}. Now, the action can be written as 
\begin{equation}
\label{eq:EHinvmetricaction}
    S_{G}=\frac{1}{2\kappa^2}\int d^4 x \sqrt{|g|}\,(g^{uu}R_{uu}+2g^{uv}R_{uv}+2 g^{ui}R_{ui}+g^{ij}R_{ij}-2\Lambda)\,.
\end{equation}
Up to total derivatives in the integrand (whose identification requires making use of the unit-determinant property of $\gamma_{ij}$), this action depends on $g^{uu}$ only through the explicit appearance of this quantity in the expression above. This means that $g^{uu}$ is a Lagrange multiplier enforcing the constraint $R_{uu}=0$. The latter takes the explicit form 
\begin{equation}
    \partial_u F\, \partial_u G - \half (\partial_u G)^2 - \partial_u^2 G +\frac{1}{4} \partial_u \gamma^{ij} \,\partial_u \gamma_{ij}=0\,,
    \label{eq:Ruu0}
\end{equation}
where $\gamma^{ij}$ denotes the matrix inverse of $\gamma_{ij}$. 

We recall that we have one more gauge choice left, which we will choose to relate $F$ and $G$ in order to simplify the equation above. However, the most convenient gauge choice turns out to depend on the background we want to expand on -- flat space or (A)dS -- so we will leave this for a later stage. Now, we turn our attention to the two quantities $g^{ui}$: one can check that the action's dependence on these is quadratic. Therefore, these two quantities can be integrated out exactly by substituting their equations of motion, which take the form
\begin{equation}
\label{eq:giuucondition}
    \partial_u(e^{2G-F} \gamma_{ij} \partial_u(e^F g^{uj}))= N_i\,,
\end{equation}
where 
\begin{equation}
\label{eq:Ni}
    N_i=e^G \left(\half \partial_u \gamma^{jk}\partial_i \gamma_{jk}-\partial_u\partial_i(F+G)+\partial_i F \partial_u G\right)+\partial_k(e^G \gamma^{jk}\partial_u \gamma_{ij})\,.
\end{equation}
These two equations can be checked to be equivalent to the constraints $R_{ui}=0$. The equations can be solved formally by introducing the inverse derivative with respect to $u$, which leads to
\begin{equation}
\label{eq:giuiEOM}
   g^{ui}= e^{-F} \frac{1}{\partial_u}\left(\gamma^{ij}e^{F-2G}\frac{1}{\partial_u}N_j\right).
\end{equation}
Inserting this solution back into the action, performing a number of integrations by parts, and using the unimodularity of $\gamma_{ij}$ yields 
\begin{equation}
\label{eq:EHLCaction}
    \begin{aligned}
    S_G= \frac{1}{2\kappa^2}\int d^4 x &\left[  e^G \left( 2 \partial_u \partial_v F +\partial_u \partial_v G -\half \partial_v \gamma^{ij} \,\partial_u \gamma_{ij} \right)\right.\\
     &-e^F \gamma^{ij}\left( \partial_i \partial_j F +\half \partial_i F\, \partial_j F -\partial_i F \,\partial_j G -\frac{1}{4} \partial_i \gamma^{kl} \,\partial_j \gamma_{kl} +\half \partial_i \gamma^{kl}\, \partial_k \gamma_{jl} \right)\\
     &\left.-\half e^{F -2G} \gamma^{ij} \frac{1}{\partial_u}N_i\, \frac{1}{\partial_u}N_j -2\Lambda \,e^{F+G}\right].
     \end{aligned}
\end{equation}
This agrees with equation (C-10) of \cite{Ananth:2006fh}, with the exception of the factor of 2 on the first term and the inclusion of the cosmological constant. We reiterate that numerous boundary terms associated to integrations by parts have been dropped.

From here onward, we will specify the background we are expanding on, reviewing first the flat space case, and introducing then the (A)dS case.

\subsubsection*{Expansion around flat space}

Considering the metric \eqref{eq:LCMetric}, the Minkowski solution corresponds to 
\begin{equation}
\label{eq:FlatSpacePerturbations}
    F=-\ln(2)\,, \quad G=0\,, \quad g_{vv}=g_{vi}=0 \,,\quad \gamma_{ij}=\delta_{ij}\,.
\end{equation}
Hence, we should ensure that when the metric perturbations are turned off, we retrieve these values.
To expand the action \eqref{eq:EHLCaction} around flat space, we will now fix the remaining gauge freedom by setting 
\begin{equation}
\label{eq:FlatSpace4thGaugeChoice}
    F=\half \,G-\ln(2)\,.
\end{equation}
This gauge choice reduces the constraint \eqref{eq:Ruu0} to 
\begin{equation}
\label{eq:FlatSpaced2psi}
    \partial_u^2 G =\frac{1}{4} \partial_u \gamma^{ij} \,\partial_u \gamma_{ij}\,,
\end{equation}
which we can solve formally as 
\begin{equation}
    \label{eq:FlatSpacepsiSolution}
    G =\frac{1}{4}\frac{1}{\partial_u^2}(\partial_u \gamma^{ij}\, \partial_u \gamma_{ij})\,.
\end{equation}
With this substitution, the action \eqref{eq:EHLCaction} depends only on the unit-determinant matrix $\gamma_{ij}$.
We will perturb around the flat metric by setting 
\begin{equation}
\label{eq:gammaPerturbation}
    \gamma_{ij} = \left(e^{\kappa H} \right)_{ij},
\end{equation}
where 
\begin{equation}
    H=
    \begin{pmatrix}
        h_{11} & h_{12}\\
        h_{12} & -h_{11}
    \end{pmatrix}.
\end{equation}
Note that $H$ being real, symmetric and traceless, corresponds to $\gamma$ being  real, symmetric and having unit determinant. The components of $H$ can be recast with 
\begin{equation}
\label{eq:handhbar}
    h=\frac{1}{\sqrt{2}}(h_{11}+i h_{12})\,, \qquad \bar{h}=\frac{1}{\sqrt{2}}(h_{11}-i h_{12})\,,
\end{equation}
which represent the positive and negative helicity degrees of freedom, respectively. Now, we can expand the action perturbatively in $\kappa$ to any desired order. We note that, from  \eqref{eq:FlatSpacepsiSolution}, we obtain 
\begin{equation}
\label{eq:psiInTermsOfh}
    G=-\frac{1}{\partial_u^2}(\partial_u h\, \partial_u \bar{h})\kappa^2 +\mathcal{O}(\kappa^4). 
\end{equation}
Keeping only terms linear in $\kappa$ in the action for brevity, we obtain
\begin{equation}
\label{eq:EHLCactionhhbar}
\begin{aligned}
     S_G =  \int d^4 x &\Bigg[ \frac{1}{4} \bar{h}\, \square\, h + \frac{\kappa}{\sqrt{2}}\partial_u^2\bar{h}\,\left(\frac{\partial_w}{\partial_u}h\,\frac{\partial_w}{\partial_u}h-h\frac{\partial_w^2}{\partial_u^2}h \right)  \\
     & + \frac{\kappa}{\sqrt{2}}\partial_u^2h\,\left( \frac{\partial_{\bar{w}}}{\partial_u}\bar{h}\,\frac{\partial_{\bar{w}}}{\partial_u}\bar{h}-\bar{h}\frac{\partial_{\bar{w}}^2}{\partial_u^2}\bar{h} \right)+ \mathcal{O}(\kappa^2)
     \Bigg],
\end{aligned}
\end{equation}
where \,$\square = \eta^{\mu\nu}\partial_\mu\partial_\nu$\,, and we remind the reader that
\begin{equation}
    \label{eq:DoubleNullCoords}
    w= x^1+ix^2\,,\quad \bar{w} = x^1-ix^2\,.
\end{equation}
This action agrees with (C-15) and (C-16) of \cite{Ananth:2006fh}. Higher-point terms in the light-cone action, starting with the four-point contribution at order $\kappa^2$, are messy.\footnote{Some improvements can be made with field redefinitions that do not affect the quadratic and cubic pieces \cite{Ananth:2006fh}. An alternative approach would be to start with the quadratic and cubic pieces, and try to construct higher-point contributions mandated by the Lorentz algebra; see e.g.~\cite{Ananth:2017pio,Metsaev:1991mt,Ponomarev:2016lrm} for related work.} Nevertheless, we verified that the complete action satisfies the basic properties:\footnote{To verify this, it is convenient to note that: \\
$$
\gamma_{ij} = \cosh(\kappa \sqrt{2h\bar h})\, \delta_{ij} + \frac{\sinh(\kappa \sqrt{2h\bar h})}{\sqrt{2h\bar h}}\, H_{ij} \,, \qquad
H_{ij} = \frac1{\sqrt{2}} \begin{pmatrix}
        h+\bar h & i(\bar h-h) \\
        i(\bar h-h) & -h-\bar h
    \end{pmatrix}\,.
$$
}
\begin{align}
    S_G |_{\bar h=0} & = 0\,, \\
    \frac{\delta S_G}{\delta\bar h} \left|_{\bar h=0} \right. & = \frac1{4}\square h + \frac{\kappa}{\sqrt{2}}\partial_u^2\left(\frac{\partial_w}{\partial_u}h\,\frac{\partial_w}{\partial_u}h-h\frac{\partial_w^2}{\partial_u^2}h \right) \,,
\end{align}
and analogously under complex conjugation. The relevance of these properties is that beyond the cubic vertices (order $\kappa$), the action has only vertices of the type $\bar h\bar h\cdots hh$, i.e.,~all four- and higher-point vertices have at least two fields of each helicity.  Therefore, considering the Feynman rules, a tree-level correlator of the type $\bar hhh\cdots h$ (with a single $\bar h$) is constructed solely with cubic vertices $\bar hhh$, and therefore lies in the self-dual sector of the theory. These basic properties apply also when we consider the perturbative action around AdS, to be studied below. Beyond the self-dual sector, a tree-level four-point correlator of the type $\bar h\bar h hh$ requires as usual all vertices up to four points ($\bar hhh$, $\bar h\bar hh$, $\bar h\bar h hh$), although tricks may be possible due to it being close to the self-dual sector; see e.g.~\cite{Monteiro:2011pc} for MHV amplitudes.

The action of self-dual gravity (here with $\Lambda=0$) is obtained by keeping only the terms linear in $\bar h$.  Redefining $h = \sqrt{2}\,\partial_u^2 \phi$ and $\bar h = \sqrt{2}\,\frac{1}{\partial_u^2} \bar\phi$, we can write
\begin{align}
S_{SDG}[\phi,\bar\phi] & = \frac{1}{2} \int d^4 x\, \bar \phi \Big( \Box \phi + 4\kappa (\partial_u \partial_{w}\phi\, \partial_u \partial_{w}\phi - \partial_u^2 \phi\, \partial_{w}^2\phi) \Big)\nonumber \\
& = \frac{1}{2} \int d^4 x\, \bar \phi \Big( \Box \phi - 4\kappa \{\{\phi,\phi\}\} \Big)\,,
\end{align}
where we use the notation
\begin{equation}
    \{\{f,g\}\}=\half \varepsilon^{\alpha \beta} \{\Pi_\alpha f,\Pi_\beta g\}\,,
\end{equation}
with 
\begin{equation}
    \{f,g\} = \half \varepsilon^{\alpha \beta}(\Pi_\alpha f {\Pi}_\beta g - \Pi_\alpha g {\Pi}_\beta f) = \partial_w f\, \partial_u g - \partial_u f\, \partial_w g
\end{equation}
and
\begin{equation}
    \Pi =(\Pi_v,\Pi_{\bar w})=({\partial}_{w},\partial_u)\,.
\end{equation}
This action agrees, up to conventions, with ref.~\cite{Siegel:1992wd}. The double-bracket notation here emphasises the double copy relating SDG to SDYM: the Fourier-space vertex in SDG (with $\Lambda=0$) is the square of the kinematic part of the SDYM vertex \cite{Monteiro:2011pc}.

\subsubsection*{Expansion around (A)dS}
\label{subsec:LCactiongrav}

For concreteness, we focus on AdS, but the dS case follows from the usual analytic continuation. To start with, we must make a choice. Considering the Poincar\'e patch, we must decide whether the `bulk' $z$ coordinate in the conformal factor is associated to the light-cone coordinates $(u,v)$, or is instead in $(x^1,x^2)$. The latter choice turns out to be simpler for deriving an action. However, we will follow here the former choice to match the conventions of \cite{Lipstein:2023pih}, and will leave the other choice to appendix~\ref{appendix:othercoordchoice} for completeness.

We want to expand about the AdS${}_4$ metric \eqref{eq:AdSmetricx30}, which we repeat here for convenience:
\begin{equation}
\label{eq:AdSmetricx3}
    ds^2_{AdS}= \frac{R^2}{z^2}\left(-du dv +(dx^1)^2 +(dx^2)^2\right)\,,
\end{equation}
where 
\begin{equation}
    z=\frac{u-v}{2}\,,
\end{equation}
and 
$R$ is related to the cosmological constant as
\begin{equation}
\label{eq:AdSLambdax3}
    \Lambda = -\frac{3}{R^2}\,.
\end{equation}
The coordinates above cover the Poincaré patch $z>0$, with $z=0$ defining the AdS boundary.\footnote{We note that the Poincaré patch covers the entire AdS spacetime in Euclidean signature, in contrast with Lorentzian signature.} Comparing to the metric \eqref{eq:LCMetric}, the AdS solution corresponds to
\begin{equation}
\label{eq:AdSZeroPerturbationsx3}
    F=\ln \left(\frac{R^2}{2z^2}\right)\,  , \,G=\ln \left(\frac{R^2}{z^2}\right)\,, \quad g_{vv}=g_{vi}=0 \,, \quad\gamma_{ij}=\delta_{ij}\,.
\end{equation}
Hence, we should ensure that when the metric perturbations are turned off, we retrieve these values. It is convenient to express
\begin{equation}
\label{eq:AdSpsiphix3x}
    G=\ln \left(\frac{R^2}{z^2}\right)+ \tilde{G}\,, \quad F= \ln \left(\frac{R^2}{2z^2}\right)+\tilde{F}\,.
\end{equation}
The constraint \eqref{eq:Ruu0} is now given by
\begin{equation}
\label{eq:giuuConditionAdSx3}
    \partial_u^2\tilde{G}-\partial_u \tilde{F}\partial_u \tilde{G}+\frac{1}{2}(\partial_u \tilde G)^2+\frac{2}{u-v}\partial_u \tilde{F} -\frac{1}{4} \partial_u \gamma^{ij} \partial_u \gamma_{ij}=0\,.
\end{equation}
Similarly to the flat space case, in order to simplify this constraint, we make now the final gauge choice by setting
\begin{equation}
\label{eq:AdSSpaceGaugeChoicex3}
    \tilde{F} =\half\, \tilde{G}\,.
\end{equation}
The constraint reduces to
\begin{equation}
    \partial_u^2\tilde{G}+\frac{1}{u-v}\partial_u \tilde{G} -\frac{1}{4} \partial_u \gamma^{ij} \partial_u \gamma_{ij}=0\,,
\end{equation}
which has the solution 
\begin{equation}
\label{eq:AdSSpacepsiSolutionx3}
     \tilde{G} =\frac{1}{4}\frac{1}{\partial_u}\left(\frac{1}{u-v}\frac{1}{\partial_u}\left((u-v)\partial_u \gamma^{ij} \partial_u \gamma_{ij}\right) \right)\,.
\end{equation}
For the unit-determinant matrix $\gamma_{ij}$, we proceed exactly as we did in flat space, by expanding it according to \eqref{eq:gammaPerturbation}-\eqref{eq:handhbar} in terms of positive and negative helicity fields, $h$ and $\bar h$. This leads now to
\begin{equation}
    \label{eq:psiInTermsOfhAdSx3}
    G=\ln \left(\frac{R^2}{z^2}\right)+\frac{1}{4}\frac{1}{\partial_u}\left(\frac{1}{u-v}\frac{1}{\partial_u}\left((u-v)\partial_u h \partial_u \bar{h}\right)\right)\kappa^2 +\mathcal{O}(\kappa^4)\,. 
\end{equation}
At this stage, we can fully express the action \eqref{eq:EHLCaction} in terms of $h$ and $\bar h$. Expanding the action in $\kappa$, we obtain for the first few orders, up to integration by parts,
\begin{equation}
S_G|_{\kappa^{-2}} = S_{AdS}\,, \quad S_G|_{\kappa^{-1}} = 0\,,
\end{equation}
and, recalling the coordinates \eqref{eq:DoubleNullCoords}, we have also
\begin{equation}
    \label{eq:EHLCactionhhbarAdSx3Kappa0}
    \begin{aligned}
    \hspace{0.0cm} S_G|_{\kappa^0}= \int d^4 x &\left[\frac{2R^2}{(u-v)^2}(\partial_u h \partial_v \bar{h}+\partial_v h \partial_u \bar{h})\right. \\
    &\left. -2R^2 (u-v)^2 \frac{1}{\partial_u}\left(\frac{1}{(u-v)^2}\partial_u \partial_{\bar{w}} h\right) \frac{1}{\partial_u}\left(\frac{1}{(u-v)^2}\partial_u \partial_{w} \bar{h}\right)  \right.\\
    &\left. -2R^2 (u-v)^2 \frac{1}{\partial_u}\left(\frac{1}{(u-v)^2}\partial_u \partial_{w} h\right) \frac{1}{\partial_u}\left(\frac{1}{(u-v)^2}\partial_u \partial_{\bar{w}} \bar{h}\right)  \right]\,,
    \end{aligned}
\end{equation}
\begin{equation}
    \label{eq:EHLCactionhhbarAdSx3Kappa1}
    \begin{aligned}
    \hspace{0.0cm} S_G|_{\kappa^1}= \frac{\kappa}{\sqrt{2}} \int d^4 x \, &\partial_u^2\bar{h}\left[\frac{4R^2}{(u-v)^2}\frac{\partial_w}{\partial_u}h\frac{\partial_w}{\partial_u}h \right.\\
    &-4R^2 (u-v)^2 h \frac{1}{\partial_u}\left(\frac{1}{(u-v)^4}\frac{\partial_w^2}{\partial_u}h\right)  \\
    &\left. \left. +16R^2 (u-v) \frac{\partial_w}{\partial_u}h\frac{1}{\partial_u}\left(\frac{1}{(u-v)^4}\frac{\partial_w}{\partial_u}h\right)\right.\right.\\
    &\left.+12R^2 (u-v)^4 \frac{1}{\partial_u}\left(\frac{1}{(u-v)^4}\frac{\partial_w}{\partial_u}h\right) \frac{1}{\partial_u}\left(\frac{1}{(u-v)^4}\frac{\partial_w}{\partial_u}h\right)  \right] \\
    & +C.C.\,,
    \end{aligned}
\end{equation}
where the last line corresponds to the terms $h\bar h\bar h$, which are the complex conjugate of the preceding terms $\bar hhh$. The comments about higher-point terms (order $\kappa^2$ and higher) that we made for the $\Lambda=0$ case apply here too.

The self-dual truncation of the theory is obtained by keeping the pieces linear in $\bar h$, which are $S_G|_{\kappa^0}$ and the $\bar hhh$ part of $S_G|_{\kappa^1}$. It is convenient to make the field redefinition
\begin{equation}
\label{eq:hphiAdS}
    h=\sqrt{2}\,\partial_u\left(\left(\partial_u -\frac{4}{u-v}\right)\phi\right)\,, \qquad \bar{h}=\sqrt{2}\,\frac{1}{\partial_u^2} \bar{\phi}\,.
\end{equation}
We then find the action for self-dual gravity with $\Lambda\neq 0$,
\begin{equation}
    \label{eq:EHLCactionSelfDualAdSBracketx3}
    S_{SDG}= 2\int d^4 x \left[  \sqrt{|g_{AdS}|}\, \bar{\phi}\left( \square_{AdS} +\frac{2}{R^2} \right) \phi-4\kappa \, \bar\phi \, \{\{\frac{R}{u-v}\phi,\frac{R}{u-v}\phi\}\}_*\right].
\end{equation}
This expression, which matches the one inferred in \cite{Lipstein:2023pih} from the self-dual equations of motion, employs the following notation:
we define
\begin{equation}
    \{\{f,g\}\}_*=\half \varepsilon^{\alpha \beta} \{\Pi_\alpha f,\Pi_\beta g\}_*
\end{equation}
where the deformed Poisson bracket is
\begin{equation}
    \{f,g\}_* = \half \varepsilon^{\alpha \beta}(\Pi_\alpha f \tilde{\Pi}_\beta g - \Pi_\alpha g \tilde{\Pi}_\beta f)
\end{equation}
with 
\begin{equation}
\label{eq:tracePi1}
    \Pi =(\Pi_v,\Pi_{\bar w})=({\partial}_{w},\partial_u)\,, \qquad    \tilde{\Pi} =(\tilde{\Pi}_v,\tilde{\Pi}_{\bar w})=({\partial}_{w},\partial_u-\frac{4}{u-v})\,.
\end{equation}
We also denote $\Box_{AdS}\phi= g^{-1/2}\partial_{\mu}\left(\sqrt{g}g^{\mu\nu}\partial_{\nu}\phi\right)$, where $g_{\mu \nu}$ is the AdS$_4$ background metric \eqref{eq:AdSmetricx3}. Note that the kinetic term corresponds to that of a conformally coupled scalar, which can be mapped to a massless scalar in half of flat space.\footnote{Notice the relation: \,$\sqrt{g}\, \bar \phi \left(\square_{AdS} +\frac{2}{R^2} \right) \phi = 4R^2\, \frac{\bar\phi}{u-v} (\partial_w\partial_{\wb}-\partial_u\partial_v) \frac{\phi}{u-v}  $\,. \label{eq:confflat}} Regarding the interaction term, its fully explicit form is
\begin{eqn}
\doublecomm{f, g}_* &= \frac{1}{2} \Big[\p_u^2 f\ \p_w^2 g + \p_u^2 g \ \p_w^2 f\Big] - \p_u \p_w f \ \p_u \p_w g\\
&+ \frac{1}{u - v} \Big[ \p_u\p_w f\ \p_w g + \p_w f\ \p_u \p_w g  - ( \p_w^2 f \ \p_u g + \p_u f \ \p_w^2 g )   \Big].
\end{eqn}
As explained in \cite{Lipstein:2023pih}, we can obtain this interaction from the one in SDYM by replacing the colour commutator with the deformed Poisson bracket,  which encodes the colour-kinematics duality.

Finally, notice that, in \eqref{eq:hphiAdS}, the leftmost field redefinition can be written as $h=\kappa\Pi_{\bar w}\tilde{\Pi}_{\bar w}\phi$.
We have checked that, by setting $\bar h=0$ in our metric ansatz, we obtain
\begin{equation}
    g_{\mu\nu}=g^{AdS}_{\mu\nu}+ \frac{4R^2}{(u-v)^2} \begin{pmatrix}
        0& 0&0&0\\
        0& -\frac{R^2}{(u-v)^2}g^{uu}&0&\kappa\Pi_{(v}\tilde{\Pi}_{{\bar w})}\phi\\
        0& 0&0&0\\
        0& \kappa\Pi_{(v}\tilde{\Pi}_{{\bar w})}\phi&0&\kappa\Pi_{({\bar w}}\tilde{\Pi}_{{\bar w})}\phi
    \end{pmatrix}. 
\end{equation}
This reproduces the metric ansatz used in \cite{Lipstein:2023pih}, except for $g^{uu}$, which is a Lagrange multiplier in our original action, to be fixed by the equations of motion such that we obtain $\kappa\Pi_{({\bar w}}\tilde{\Pi}_{{\bar w})}\phi$ for that component.

\section{Feynman rules} \label{feynmanrules}

We follow the conventions of \cite{Lipstein:2023pih} and work in Euclidean AdS$_4$ (EAdS):
\begin{eqn}
ds^2 = \frac{dt^2 + dz^2 + dx^2 + dy^2}{z^2}\,,
\end{eqn}
where $0<z<\infty$ is the radial coordinates in the Poincare patch and from now on, we set $R=1$, i.e.~$\Lambda=-3$. It will be convenient to introduce light-cone coordinates, which we define as 
 \begin{eqn}\label{LCcoord1}
 u &= i t + z\,, \qquad w = x+ i y\,, \\
  v &= i t - z\,, \qquad \bar w = x- i y\,.
 \end{eqn}
 In these coordinates, the metric becomes 
 \begin{eqn}\label{ads-LC}
 ds^2 = \frac{4(dw d\bar w - du dv)}{(u - v)^2}\,.
 \end{eqn}
We define the dot product between $k_\mu = (k_t, k_x, k_y,k_z)$ and $x^\mu = (t, x, y, z)$ as\footnote{Using \eqref{LCcoord1}, we can derive the relation between the momenta in light-cone variables and in Cartesian variables: 
 \begin{eqn}
 k_u = \frac12 (- i k_t + k_z) , \quad k_v = \frac12 (-i k_t - k_z), \quad 
 k_w = \frac12( k_x - i k_y),\quad k_{\wb} = \frac12( k_x + i k_y).
 \end{eqn}
 } 
 \begin{eqn}
 k \cdot x\, &= t k_t + z k_z + x k_x + y k_{y} \\
 &= u k_u + v k_v + w k_w + \wb k_{\wb}~.
 \end{eqn}

In section \ref{sec:correlators}, we will compute Witten diagrams of SDYM and SDG whose external legs end on the boundary of AdS ($z=0$). The purpose of this section is to present the Feynman rules for these computations. We consider Fourier modes labelled by 3-momenta $\vec{k}=(k_t,k_x,k_y)$. In order to make the comparison to massless fields in flat space more transparent, it is convenient to define associated 4-momenta $k_\mu = (k_t, k_x, k_y,k_z)$, chosen to be null. In EAdS, this corresponds to assigning
\begin{equation}
k_z = i |\vec k|\,, \qquad \text{where}
\quad |\vec k| = \sqrt{k_t^2 + k_x^2 + k_y^2}\,.
\end{equation}
This quantity $|\vec k|$ shall often be referred to as the energy of the particle. Notice that we have translational invariance along the boundary, resulting in conservation of the 3-momentum:
\begin{equation}
\sum_{i=1}^{n}\vec{k}_{i}=0 \,,
\end{equation}
where $n$ is the number of external legs. There is, however, no momentum conservation along the $z$-direction. We denote the sum of energies for a subset of legs as
\begin{equation}
k_{ij\ldots l}=k_i+k_j+\ldots+k_l\,,
\end{equation}
and the total energy of an $n$-point diagram as
\begin{equation}
k_{123 \ldots n} =\sum_{i=1}^{n}k_{i} \,, \qquad \text{with}
\quad k_{i}=|\vec{k}_{i}|\,.
\end{equation}
The flat space limit corresponds to taking $E\rightarrow0$ \cite{Raju:2012zr,Arkani-Hamed:2015bza}. In this limit, a Witten diagram will develop a pole in $k_{123 \ldots n}$, and the coefficient of the leading pole corresponds to a Feynman diagram in flat space.

\subsection{$\varphi^3$ theory} \label{phi3rules}

A useful toy model to consider is a massless $\varphi^3$ theory in 4-dimensional Euclidean flat space with a boundary at $z = 0$:\footnote{Alternatively, in view of the relation in footnote \ref{eq:confflat}, one can think of this setting as having a scalar $\tilde\varphi=z\,\varphi$ whose kinetic term is that of a conformally coupled scalar in AdS. Then, the potential term is $\sim \sqrt{g} \,z\, \tilde\varphi^3$. Under a conformal mapping to flat space, we obtain the theory \eqref{eq:phi3}. See e.g.~section 2 of \cite{Albayrak:2020isk}.}
\begin{equation}
\label{eq:phi3}
\mathcal{L}_{\varphi^{3}}=\frac12\partial_{\mu}\varphi\partial^{\mu}\varphi-\frac{1}{6}\varphi^{3}
\end{equation}
 As we will see later, the Witten diagrams of SDYM and SDG can be related to those of $\varphi^3$ theory in flat space. The propagators for this theory are given as \footnote{The derivation of scalar propagators in AdS momentum space can be found in \cite{Liu:1998ty,Raju:2011mp} so we refer the reader to those papers for more details.} 
\begin{eqn}\label{prop-phi3}
\mbox{Bulk-Boundary}:& \quad \varphi(z, k) = e^{- k z}\,, \\
\mbox{Bulk-Bulk}:& \quad G(z_1, z_2, k) = \frac{1}{2k} \left( e^{- k|z_1 - z_2|} - e^{ - k (z_1 + z_2)} \right)  \,,
\end{eqn}
where $k= |\vec k|=-ik_z$.  Here, we have dropped plane wave factors of the type $e^{i \vec{k} \cdot \vec{x} }$ corresponding to directions $\vec x=(t,x,y)$ along the boundary, which---due to translational invariance in those directions---will merely give rise to 3-momentum conservation in the correlator.\footnote{For each propagator, there is an implicit plane wave factor of the type $e^{i \vec k \cdot (\vec x_1 - \vec x_2)}$.} In contrast, the existence of the boundary at $z=0$ breaks momentum conservation along the $z$-direction, even in flat space, due to the boundary conditions. In particular, the bulk-bulk Green's function satisfies Dirichlet boundary conditions, vanishing as $z_1 \to 0$ and as $z_1\to\infty$ (similarly for $z_2$). This Green's function admits the following convenient representation:
\begin{equation}
   G(z_1, z_2, k) =\frac{1}{\pi} \intinf \frac{d\omega}{\omega^2 + k^2} \sin(\omega z_1) \sin(\omega z_2) \,.
\end{equation}

The propagators are diagrammatically expressed as 
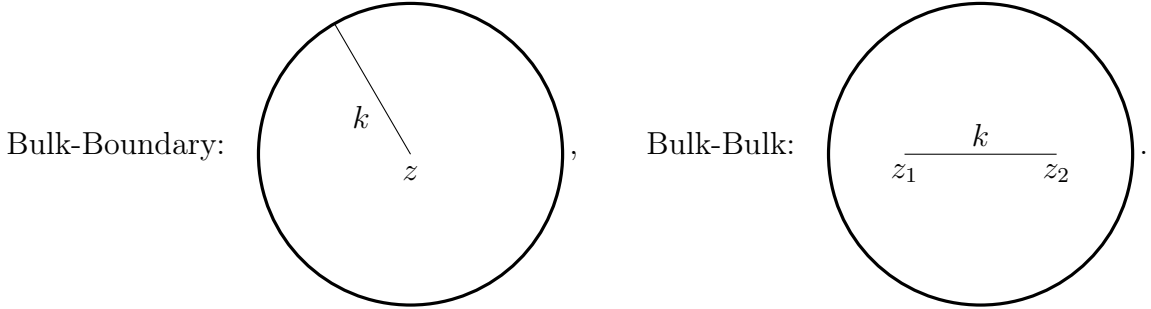
\begin{eqn*}
\mbox{Bulk-Boundary:}&\quad 
\begin{tikzpicture}[baseline]
  \draw[very thick] (0, 0) circle (2);
\draw (0, 0) -- ({2*cos(120)}, {2*sin(120)});
\node at (0, -0.25) {$z$};
\node at (-0.65, 0.5) {$k$};
\end{tikzpicture}\,, \qquad 
\mbox{Bulk-Bulk:}  \quad 
\begin{tikzpicture}[baseline]
\draw[very thick] (0, 0) circle (2);
\draw (-1, 0) -- (1, 0);
\node at (-1, -0.25) {$z_1$};
\node at (1, -0.25) {$z_2$};
\node at (0, 0.25) {$k$};
\end{tikzpicture}\,.
\end{eqn*}
We will sometimes suppress the explicit $z$-dependence in the diagrams. The interaction vertex is given as 
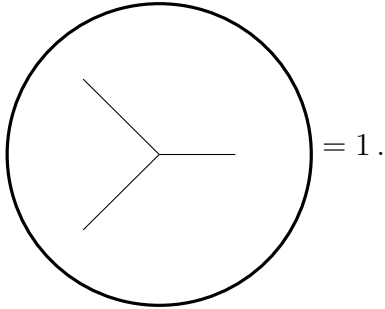
\begin{eqn*}
\begin{tikzpicture}[baseline]
\draw[very thick] (0, 0) circle (2);
\draw (-1, 1) -- (0,0);
\draw (1, 0) -- (0,0);
\draw (-1, -1) -- (0,0);
\end{tikzpicture}
 = 1\,.
\end{eqn*}
For each interaction vertex, we need an integration $\int_0^\infty dz(\cdots)$ over the $z$-location of the vertex. We will provide various examples in the next section. As alluded to previously, integrations over the boundary merely enforce the associated 3-momentum conservation.

Notice we have set the coupling constant to unity. This is because we are only going to consider tree-level diagrams, and the dependence on the coupling constant is straightforward to restore (coupling constant to the power $n-2$ at $n$ points).

\subsection{SDYM} \label{sec:SDYMfeynman}

The Lagrangian for SDYM, reviewed in section~\ref{sec:YMaction}, is given as 
\begin{eqn}\label{LSDYM}
\mathcal{L}_{SDYM} = \tr \big[ \bar\Phi \big( \Box \Phi + i [\p_u \Phi, \p_w \Phi] \big)\big]\,.
\end{eqn}
We remind the reader that it takes the same form as in flat space, because AdS is conformally flat and the action of SDYM is Weyl invariant.\footnote{Interestingly, while the latter property is broken by quantum effects in full Yang-Mills theory, results from the scattering amplitudes literature suggest that the property holds for SDYM also at quantum level \cite{Henn:2019mvc,Doran:2023cmj}.} Hence, we can trivially map the AdS problem to half of flat space with a boundary at $z=0$ using a Weyl transformation, as previously discussed.

Looking at the quadratic piece of the Lagrangian, we see that the propagators take the same form as in \eqref{prop-phi3}. The bulk-to-boundary propagators are
\begin{eqn}
\Phi(z, k) = e^{- k z},\quad 
\bar\Phi(z, k) = e^{- k z},
\label{bulktoboundaryprops}
\end{eqn}
but we should note that in SDYM the propagators see a different chirality from each end, i.e.
\begin{eqn}
\Phi(z, k)=\;\; & 
\begin{tikzpicture}[baseline]
\draw[very thick] (0,0) circle (2);
\draw ({2*cos(140)},{(2*sin(140)}) -- (-0, 0);
\node at ({2.2*cos(140)},{2.2*sin(140)}) {$+$};
\node at (0,-0.25) {$-$};
\node at (0, 0.25) {$z$};
\node at (-1, 0.5) {$k$};
\end{tikzpicture}\, ,\quad
\bar\Phi(z, k)=\;\; 
\begin{tikzpicture}[baseline]
\draw[very thick] (0,0) circle (2);
\draw ({2*cos(140)},{(2*sin(140)}) -- (-0, 0);
\node at ({2.2*cos(140)},{2.2*sin(140)}) {$-$};
\node at (0,-0.25) {$+$};
\node at (0, 0.25) {$z$};
\node at (-1, 0.5) {$k$};
\end{tikzpicture}\,.
\end{eqn}
Recalling the relation between scalar and gluon fields in \eqref{eq:defPhiPhibar} and the bulk-to-boundary propagators for scalar fields in \eqref{bulktoboundaryprops}, we can immediately read off the bulk-to-boundary propagators for components of the gluon field:
\begin{equation}
A_{\bar{w}}(z,k)=k_{u}\,e^{-kz}\,,\qquad A_{\bar{w}}(z,k)=\frac{1}{k_{u}}\,e^{-kz}\,,
\label{gaugeprops}
\end{equation}
where we recall that $k_u = \frac12 (- i k_t + k_z)=\frac{i}{2} (k-k_t)$.

The bulk-to-bulk propagator is again the Green's function
\begin{eqn}
G^{SDYM}(z_1, z_2, k)= \frac{1}{2k} \left( e^{- k|z_1 - z_2|} - e^{ - k (z_1 + z_2)} \right) = \frac{1}{\pi} \intinf \frac{d\omega \sin(\omega z_1) \sin(\omega z_2)}{\omega^2 + k^2}\,,
\label{btbulksdym}
\end{eqn}
which is now diagrammatically denoted as
\begin{eqn}
G^{SDYM}(z_1, z_2, k)= 
\begin{tikzpicture}[baseline]
\draw[very thick] (0,0) circle (2);
\draw (-1, 0) -- (1, 0);
\node at (-1, -0.25) {$+$};
\node at (1, -0.25) {$-$};
\node at (0, 0.25) {$k$};
\node at (1, 0.25) {$z_2$};
\node at (-1, 0.25) {$z_1$};
\end{tikzpicture}\,.
\end{eqn}

The interaction term in the Lagrangian \eqref{LSDYM} contains spacetime derivatives. In practice, we will consider colour-ordered correlators, so we suppress the colour data.\footnote{This means that we suppress the colour dependence $\delta^{a_1a_2}$ from the Green's function and $f^{a_1a_2a_3}$ from the cubic vertex, as usual. By a colour-ordered correlator at tree level with ordering $12\cdots n$, we mean the coefficient of $\tr (T^{a_1}T^{a_2}\cdots T^{a_n})$ in the complete correlator.} We will write the vertex as
\begin{eqn}\label{V-SDYM}
V^{SDYM}_z(P, Q) &= \p_u P \,\p_w Q - \p_w P \,\p_u Q \\
&= \scalebox{0.75}{\begin{tikzpicture}[baseline]
\draw[very thick] (0,0) circle (3);
\draw (0, 0) -- ({2*cos(60)}, {2*sin(60)});
\draw (0, 0) -- ({2*cos(-60)}, {2*sin(-60)});
\draw (0, 0) -- ({2*cos(180)}, {2*sin(180)});

\node at ({(2+0.3)*cos(60)}, {(2+0.3)*sin(60)}) {$P^+$};
\node at ({(2+0.3)*cos(-60)}, {(2+0.3)*sin(-60)}) {$Q^+$};
\node at ({(2+0.3)*cos(-180)}, {(2+0.3)*sin(-180)}) {$R^-$};

\node at (-0.3, -0.3) {$+$};
\node at (0.3, 0.3) {$-$};
\node at (0.3, -0.3) {$-$};
\node at (-0.3, 0.3) {$z$};
\end{tikzpicture}}\,.
\end{eqn}
Here, $P$ and $Q$ denote bulk-to-boundary or bulk-to-bulk propagators and the propagator $R$ will just multiply $V_z^{SDYM}(P, Q)$, so that we have the cases $V_z^{SDYM}(\Phi, \Phi)$, $V_z^{SDYM}(G, \Phi)$, and $V_z^{SDYM}(G, G)$. We suppress in our vertex notation a negative helicity leg $R$, which is not differentiated according to the Lagrangian in \eqref{LSDYM}. Given the translational invariance along the boundary direction, we can simply substitute boundary derivatives with boundary momenta:
\begin{eqn}
\label{eq:rule3Fourier}
\partial_u P = \frac12 (- i \partial_t + \partial_z) P \,\mapsto\, \frac12 (k_t+\partial_z) P\,, \qquad 
\partial_w P \,\mapsto\, i k_w P\,.
\end{eqn}
It is the $z$-dependence that requires more care. The subscript $z$ in $V_z$ indicates the $z$-coordinate of the interaction, and the rules include that we integrate over $z$ independently for each interaction. Again, this will become clearer in the explicit examples to be worked out later. Based on whether $P$ and $Q$ in \eqref{V-SDYM} are bulk-to-boundary or bulk-to-bulk propagators, and going into Fourier 3-space as in \eqref{eq:rule3Fourier}, we write down explicitly the three cases:
\begin{eqn}
\label{eq:V-SDYMphiphi}
V_{z}^{SDYM}(\Phi_1, \Phi_2) = - \Phi_1\, (X_{1,2})\, \Phi_2\,,
\end{eqn}
where we define $X_{i,j}=X(k_i,k_j)$, with
\begin{eqn}
\label{eq:Xij}
X(k_i,k_j)=k_{iu}k_{jw}-k_{iw}k_{ju} = \frac{i}{2} \left((k_i-k_{it})k_{jw} - k_{iw}(k_j-k_{jt})\right) \,;
\end{eqn}
\begin{equation}
\label{eq:V-SDYMGphi}
V^{SDYM}_{z}(\Phi_1,G(z,z_\ast,q))= -\Phi_1\left(X(k_1,q) +\frac{i}{2} k_{1w} (q + \partial_z)\right)G(z,z_\ast,q)\,;
\end{equation}
and
\begin{align}
\label{eq:V-SDYMGG}
 V^{SDYM}_{z}&(G(z,z_1,q_1),G(z,z_2,q_2))= \nonumber \\
 &-G(z,z_1,q_1)\left(X(q_1,q_2)+\frac{i}{2} \left(q_{1w}(q_2+\overrightarrow{\partial_z})-
 q_{2w} (q_1+\overleftarrow{\partial}\!_z)\right)\right)G(z,z_2,q_2)\,,
\end{align}
with the derivatives acting left or right as indicated by the arrows. We note that the operator acting on the two arguments of $ V^{SDYM}_{z}$ is always the same, but the result is simpler when it acts on a bulk-to-boundary propagator, because $(k+\partial_z)e^{-kz}=0$. That is, we can write in general 
\begin{eqn}
\label{eq:V-SDYMgen}
V_{z}^{SDYM}(P_{q_1},Q_{q_2}) = - P_{q_1} \Xd\!\!(q_1,q_2) \; Q_{q_2}\,,
\end{eqn}
where
\begin{eqn}
\label{eq:Xarrows}
\Xd\!\!(q_1,q_2) = X(q_1,q_2)+\frac{i}{2} \left(q_{1w}(q_2+\overrightarrow{\partial_z})-
 q_{2w} (q_1+\overleftarrow{\partial}\!_z)\right)\,.
\end{eqn}

\subsection{SDG} \label{sdgfeynman}

The Lagrangian for SDG, reviewed in section~\ref{subsec:LCactiongrav}, is 
\begin{eqn}\label{LSDG}
\mathcal{L}_{SDG} = \sqrt{g}\, \bar \phi (\Box_{AdS} + 2) \phi - 4\, \bar \phi \Bigg\{ \Bigg\{ \frac{\phi}{u - v}, \frac{\phi}{u - v} \Bigg\}\Bigg\}_*\,,
\end{eqn}
where we set the coupling $\kappa$ to unity, and recall that
\begin{eqn}\label{doublecomm1}
\doublecomm{f, g}_* &= \frac{1}{2} \Big[\p_u^2 f\ \p_w^2 g + \p_u^2 g \ \p_w^2 f\Big] - \p_u \p_w f \ \p_u \p_w g\\
&+ \frac{1}{u - v} \Big[ \p_u\p_w f\ \p_w g + \p_w f\ \p_u \p_w g  - ( \p_w^2 f \ \p_u g + \p_u f \ \p_w^2 g )   \Big].
\end{eqn}
While SDG is not conformally invariant, the kinetic term does correspond to that of a conformally coupled scalar, which can be mapped to a massless scalar in half of flat space (see footnote~\ref{eq:confflat}). For this reason, the propagators match those in \eqref{prop-phi3} up to a factor of the bulk coordinate $z$. In particular, the bulk-to-boundary propagators are given as
\begin{eqn}
\phi(z, k) &= z e^{- k z}\,, \quad 
 \bar \phi(z, k) = z e^{- k z}\,,
\label{sdgbulktoboundary}
\end{eqn}
which we will denote diagrammatically as 
\begin{eqn}
\phi(z, k) &= 
\begin{tikzpicture}[baseline]
\draw[very thick] (0,0) circle (2);
\draw[dashed] ({2*cos(140)},{(2*sin(140)}) -- (-0, 0);
\node at ({2.2*cos(140)},{2.2*sin(140)}) {$+$};
\node at (0,-0.25) {$-$};
\node at (0, 0.25) {$z$};
\node at (-1, 0.5) {$k$};
\end{tikzpicture}\, ,\quad
\bar\phi(z, k) = 
\begin{tikzpicture}[baseline]
\draw[very thick] (0,0) circle (2);
\draw[dashed] ({2*cos(140)},{(2*sin(140)}) -- (-0, 0);
\node at ({2.2*cos(140)},{2.2*sin(140)}) {$-$};
\node at (0,-0.25) {$+$};
\node at (0, 0.25) {$z$};
\node at (-1, 0.5) {$k$};
\end{tikzpicture}\,.
\end{eqn}

Plugging \eqref{sdgbulktoboundary} into the first relation in \eqref{eq:hphiAdS}, we can read off the associated bulk-to-boundary propagator for $h$:
\begin{equation}
h(k,z)=-\sqrt{2}\, k_{u}\left(i+k_{u}z\right)e^{-kz}\,,
\end{equation}
associated to the Bunch-Davies vacuum.
Similarly, consistency of the second relation in \eqref{eq:hphiAdS} with \eqref{sdgbulktoboundary} implies the following bulk-to-boundary propagator for $\bar{h}$:
\begin{equation}
\bar{h}(k,z)=-\sqrt{2}\,k_{u}^{-3}\left(i+k_{u}z\right)e^{-kz}.
\end{equation}
This indicates that we may wish to normalise the propagators differently to connect more readily to standard graviton correlators in the full (parity-invariant) theory, though we will not pursue this here.

The bulk-to-bulk propagator is given as 
\begin{eqn}
G^{SDG}(z_1, z_2, k)= \frac{z_1 z_2}{2k} \left( e^{- k|z_1 - z_2|} - e^{ - k (z_1 + z_2)} \right) = \frac{z_1 z_2}{\pi} \intinf \frac{d\omega \sin(\omega z_1) \sin(\omega z_2)}{\omega^2 + k^2}\,,
\label{btbulksdg}
\end{eqn}
and is diagrammatically denoted as 
\begin{eqn}
G^{SDG}(z_1, z_2, k) = \begin{tikzpicture}[baseline]
\draw[very thick] (0,0) circle (2);
\draw[dashed] (-1, 0) -- (1, 0);
\node at (-1, -0.25) {$+$};
\node at (1, -0.25) {$-$};
\node at (0, 0.25) {$k$};
\node at (1, 0.25) {$z_2$};
\node at (-1, 0.25) {$z_1$};
\end{tikzpicture}\,.
\end{eqn}

The interaction vertex of the action \eqref{LSDG} is given as
\begin{eqn}\label{V-SDG}
V_z^{SDG}(P, Q) &=  \{ \{ P/z,Q/z \} \}_*\,,\\
&\equiv  \scalebox{0.75}{\begin{tikzpicture}[baseline]
\draw[very thick] (0,0) circle (3);
\draw[dashed] (0, 0) -- ({2*cos(60)}, {2*sin(60)});
\draw[dashed] (0, 0) -- ({2*cos(-60)}, {2*sin(-60)});
\draw[dashed] (0, 0) -- ({2*cos(180)}, {2*sin(180)});

\node at ({(2+0.3)*cos(60)}, {(2+0.3)*sin(60)}) {$P^+$};
\node at ({(2+0.3)*cos(-60)}, {(2+0.3)*sin(-60)}) {$Q^+$};
\node at ({(2+0.3)*cos(-180)}, {(2+0.3)*sin(-180)}) {$R^-$};

\node at (-0.3, -0.3) {$+$};
\node at (0.3, 0.3) {$-$};
\node at (0.3, -0.3) {$-$};
\node at (-0.3, 0.3) {$z$};
\end{tikzpicture}\,,}
\end{eqn}
where $P,Q, R$ are propagators, and the integration over $z$ is done separately, just like for the SDYM case. Again, the rule \eqref{eq:rule3Fourier} can be used to take advantage of the translational invariance along the boundary directions. Analogously to the expression \eqref{eq:V-SDYMgen} for SDYM, we obtain for SDG
\begin{eqn}
\label{eq:V-SDGgen}
V_{z}^{SDG}(P_{q_1},Q_{q_2}) = \frac12\, \left(\frac{P_{q_1}}{z}\right) \left( \,\Xd\!\!(q_1,q_2)^2+ \frac{i\! \Xd\!\!(q_1,q_2)}{z}\,(q_{2w}-q_{1w})\right)
\left(\frac{Q_{q_2}}{z}\right)\,,
\end{eqn}
where the derivatives in $\frac{\Xd(q_1,q_2)}{z}$ do not act on the $z$ in that particular denominator.\footnote{To be fully explicit, $\frac{\Xd(q_1,q_2)}{z} = \frac{X(q_1,q_2)}{z}+\frac{i}{2} \left(q_{1w}\frac1{z}(q_2+\overrightarrow{\partial_z})-
 q_{2w} (q_1+\overleftarrow{\partial}\!_z )\frac1{z}\right)$, and we have also  $\Xd\!\!(q_1,q_2)^2 = X(q_1,q_2)^2 -\frac{1}{4} \left( \big(q_{1w}(q_2+\overrightarrow{\partial_z})\big)^2 +\big(q_{2w} (q_1+\overleftarrow{\partial}\!_z )\big)^2 - 2 q_{1w}
 q_{2w} (q_1+\overleftarrow{\partial}\!_z )(q_2+\overrightarrow{\partial_z})\right)$.} This is precisely the `deformed double copy relation' between the SDYM vertex \eqref{eq:V-SDYMgen} and the SDG vertex, observed in \cite{Lipstein:2023pih}. Notice that it applies at the level of $z$-integrands of Witten diagrams.

\section{Correlators}\label{sec:correlators}

In this section, we will compute three- and four-point correlators using the Feynman rules obtained in the previous section. We will show that they can be recast in terms of $\varphi^3$ correlators in flat space with a boundary. From the Feynman rules for SDYM and SDG, it is clear that only diagrams with a single minus leg can arise at tree-level, while at loop-level only one-loop all-plus diagrams can occur (where plus and minus refer to barred and un-barred scalar fields, respectively).

Before proceeding, let us recall that the flat space limit of Feynman diagrams in AdS is obtained by taking the total energy $k_{12...n}\rightarrow 0$ \cite{Raju:2012zr}. In particular, we find that the Feynman diagrams of SDYM have simple pole in the energy while those of SDG have a leading pole of order $n-1$, and the coefficients of these poles correspond to flat space Feynman diagrams. On the other hand, it is well-known that the scattering amplitudes of these theories vanish for $n>3$. This implies after summing over all AdS diagrams for a given correlator, the coefficient of $k_{12...n}^{-1}$ and $k_{12...n}^{n-1}$ must vanish in SDYM and SDG, respectively. We explicitly verify this property in Appendix \ref{3ptappendix}.

\subsection{$\varphi^3$ theory}\label{sec:phi3}

Let us begin by computing 3-point and 4-point correlators of $\varphi^3$ theory in half flat space with a boundary at $z=0$ using the Feynman rules in section \ref{phi3rules}. 

\subsection*{3 points:}
The 3-point correlator for this theory is obtained from a contact diagram:
\begin{eqn}\label{3pt-phi3}
A_3=
\begin{tikzpicture}[baseline]
\draw[very thick] (0, 0) circle (2);
\draw (0, 0) -- ({2*cos(60)}, {2*sin(60)});
\draw (0, 0) -- ({2*cos(-60)}, {2*sin(-60)});
\draw (0, 0) -- ({2*cos(180)}, {2*sin(180)});

\node at ({(2+0.3)*cos(60)}, {(2+0.3)*sin(60)}) {$1$};
\node at ({(2+0.3)*cos(-60)}, {(2+0.3)*sin(-60)}) {$2$};
\node at ({(2+0.3)*cos(-180)}, {(2+0.3)*sin(-180)}) {$3$};

\end{tikzpicture} = \intsinf dz e^{- (k_1 + k_2 + k_3) z} =\frac{1}{k_{123}}\,.
\end{eqn}
Notice that the integral over $z$ is cut-off at the boundary, $z = 0$. This results in the pole $1/k_{123}$. The residue of this pole is one, corresponding to the tree-level 3-point amplitude. Following \cite{Arkani-Hamed:2017fdk}, we will denote the 3-point graph above as
\begin{eqn}
\begin{tikzpicture}[baseline]
\draw[very thick] (0, 0) circle (2);
\draw (0, 0) -- ({2*cos(60)}, {2*sin(60)});
\draw (0, 0) -- ({2*cos(-60)}, {2*sin(-60)});
\draw (0, 0) -- ({2*cos(180)}, {2*sin(180)});

\node at ({(2+0.3)*cos(60)}, {(2+0.3)*sin(60)}) {$1$};
\node at ({(2+0.3)*cos(-60)}, {(2+0.3)*sin(-60)}) {$2$};
\node at ({(2+0.3)*cos(-180)}, {(2+0.3)*sin(-180)}) {$3$};

\end{tikzpicture}
= 
\begin{tikzpicture}[baseline]
\node at (0, 0) {\textbullet}; 
\node at (0, -0.25) {$k_{123}$}; 
\end{tikzpicture}\,.
\end{eqn}
In other words, the single bold dot \begin{tikzpicture}[baseline]
\node at (0,0) {\textbullet};
\node at (0,-0.25) {$x$};
\end{tikzpicture}
means $\frac{1}{x} $. We will make use of this stick graph representation later on in the paper.

\subsection*{4 points:}

Next, let us consider the following $s$-channel exchange diagram \footnote{Note that the full tree-level 4-point correlator is obtained by summing over $s$, $t$, and $u$ channels.}:
\begin{eqn}\label{4pt-phi3-1}
A_4^{(s)}=
 \begin{tikzpicture}[baseline]
\draw[very thick] (0, 0) circle (2);
\draw (-1, 0) -- ({2*cos(150)}, {2*sin(150)});
\draw (-1, 0) -- ({2*cos(150)}, {2*sin(-150)});

\draw (1, 0) -- ({2*cos(30)}, {2*sin(30)});
\draw (1, 0) -- ({2*cos(30)}, {2*sin(-30)});

\draw (-1, 0) -- (1, 0);

\node at ({(2+0.3)*cos(30)}, {(2+0.3)*sin(30)}) {$3$};
\node at ({(2+0.3)*cos(30)}, {(2+0.3)*sin(-30)}) {$4$};
\node at ({(2+0.3)*cos(150)}, {(2+0.3)*sin(150)}) {$2$};
\node at ({(2+0.3)*cos(150)}, {(2+0.3)*sin(-150)}) {$1$};

\node at (0, -0.35) {$\vec{k}_{12}$};
\end{tikzpicture}
= \intsinf dz dz' e^{- k_{12} z} G(z, z', k_I)  e^{- k_{34} z'}\,,
\end{eqn}
where $k_I=\big|\vec{k}_{12}\big|$. The full tree-level correlator is obtained by summing over all three channels. In the stick-graph notation this is denoted as 
\begin{tikzpicture}[baseline]
\node at (-1, 0) {\textbullet};
\node at (1, 0) {\textbullet};
\draw (-1, 0) -- (1, 0);
\node at (-1, -0.25) {$k_{12}$};
\node at (1, -0.25) {$k_{34}$};
\node at (0, 0.25) {$k_I$};
\end{tikzpicture}. We perform the integral in \eqref{4pt-phi3-1} by substituting the expression for the bulk-to-bulk propagator given in \eqref{prop-phi3}, and by first performing the $z$ integrals and then the $\omega$ integral. We obtain
\begin{eqn}\label{4pt-phi3-2}
\begin{tikzpicture}[baseline]
\node at (-1, 0) {\textbullet};
\node at (1, 0) {\textbullet};
\draw (-1, 0) -- (1, 0);
\node at (-1, -0.25) {$k_{12}$};
\node at (1, -0.25) {$k_{34}$};
\node at (0, 0.25) {$k_I$};
\end{tikzpicture}
&=  \frac{1}{\pi}\intinf \frac{d\omega \omega^2}{(\omega^2 + k_{12}^2)(\omega^2 +k_{34}^2) (\omega^2 +k_I^2)} = \frac{1}{k_{1234}E_{L}E_{R}}\,,
\end{eqn}
where $E_{L}=k_{12}+k_I$, and $E_{R}=k_{34}+k_I$. To obtain the final equality, we evaluated the residues of the $\omega$ integral at the poles in the upper-half plane. We have again suppressed the trivial dependence of the momentum conservation along the boundary, $\delta^3(\vec k_1 + \vec k_2 + \vec k_3 + \vec k_4)$.

The scattering amplitude can be obtained by taking the residue at the total energy pole $k_{1234}$, which results in 
\begin{eqn}
\lim_{k_{1234}\rightarrow0}A_{4}^{(s)}
= \frac{1}{k_I^{2}-k_{12}^{2}}\,,
\end{eqn}
where we recognise the usual four-dimensional Lorentz-invariant Mandelstam variable $s$ appearing in the denominator.

\subsection{SDYM}\label{SDYM-corr}

Now we turn to SDYM correlators using the Feynman rules presented in section \ref{sec:SDYMfeynman}. 

\subsection*{3 points:}

At three points, we have the following diagram:
\begin{eqn}\label{3ptSDYM1}
\begin{tikzpicture}[baseline]
\draw[very thick] (0, 0) circle (2);
\draw (0, 0) -- ({2*cos(60)}, {2*sin(60)});
\draw (0, 0) -- ({2*cos(-60)}, {2*sin(-60)});
\draw (0, 0) -- ({2*cos(180)}, {2*sin(180)});

\node at ({(2+0.3)*cos(60)}, {(2+0.3)*sin(60)}) {$1^+$};
\node at ({(2+0.3)*cos(-60)}, {(2+0.3)*sin(-60)}) {$2^+$};
\node at ({(2+0.3)*cos(-180)}, {(2+0.3)*sin(-180)}) {$3^-$};

\node at (-0.3, -0.3) {$+$};
\node at (0.3, 0.3) {$-$};
\node at (0.3, -0.3) {$-$};
\end{tikzpicture}
\end{eqn}
We restate the interaction vertex \eqref{eq:V-SDYMphiphi} here for convenience: 
\begin{eqn}
V_{z}^{SDYM}(\Phi_1, \Phi_2) = - X_{1,2}\, \Phi_1\,  \Phi_2\,,
\end{eqn}
where
\begin{equation}
X_{i,j}=k_{iu}k_{jw}-k_{iw}k_{ju}\,.
\end{equation}
With the bulk-to-boundary propagators \eqref{bulktoboundaryprops}, we obtain the following result for the (colour-stripped) correlator: 
\begin{eqn}
\mathcal{A}_{3}\left(1^{+},2^{+},3^{-}\right)=& \intsinf dz\,  V_z^{SDYM}(\Phi_1, \Phi_2) \bar\Phi_3 \\
=&  -\intsinf dz  e^{- (k_1 + k_2) z} (k_{1u} k_{2w} - k_{1w} k_{2u})e^{- k_3 z} \\
=&  - \frac{X_{1,2} }{k_{123}}=- X_{1,2} \begin{tikzpicture}[baseline]
\node at (0,0) {\textbullet};
\node at (0,-0.25) {$k_{123}$};
\end{tikzpicture}.
\label{sdym3pt}
\end{eqn}
The final result is proportional to the boundary three-point function of $\varphi^3$ theory in \eqref{3pt-phi3}. The flat space limit is obtained by taking the residue of the total energy pole:
\begin{equation}
\lim_{k_{123}\rightarrow 0} k_{123}\, \mathcal{A}_3 = -X_{1,2}\,, 
\label{sdym3ptflat}
\end{equation}
which is the expected result; see e.g.~\cite{Monteiro:2011pc}.

In Appendix \ref{3ptappendix}, we explain how to lift the scalar 3-point function in \eqref{sdym3pt} to a spinning correlator, and show that it agrees with the 3-point function of full YM in light-cone gauge up to a boundary term.


\subsection*{4 points:}

Next, let us consider the following 4-point exchange diagram \footnote{Note that the full tree-level color-ordered correlator is obtained by summing over $s$ and $t$ channels.}:
\begin{eqn}
\mathcal{A}_4^{(s)} = \begin{tikzpicture}[baseline]
\draw[very thick] (0, 0) circle (2);
\draw (-1, 0) -- ({2*cos(150)}, {2*sin(150)});
\draw (-1, 0) -- ({2*cos(150)}, {2*sin(-150)});

\draw (1, 0) -- ({2*cos(30)}, {2*sin(30)});
\draw (1, 0) -- ({2*cos(30)}, {2*sin(-30)});

\draw (-1, 0) -- (1, 0);

\node at ({(2+0.3)*cos(30)}, {(2+0.3)*sin(30)}) {$3^+$};
\node at ({(2+0.3)*cos(30)}, {(2+0.3)*sin(-30)}) {$4^-$};
\node at ({(2+0.3)*cos(150)}, {(2+0.3)*sin(150)}) {$2^+$};
\node at ({(2+0.3)*cos(150)}, {(2+0.3)*sin(-150)}) {$1^+$};

\node at (0, -0.35) {$\vec{k}_I$};

\node at (-1.3, 0.25) {$-$};
\node at (-1.3, -0.25) {$-$};
\node at (-1+0.15, 0.15) {$+$};

\node at (1.3, 0.25) {$-$};
\node at (1.3, -0.25) {$+$};
\node at (1-0.15, 0.15) {$-$};
\end{tikzpicture}
\end{eqn}
This diagram is given by (stripping off the colour factor)
\begin{equation}
\mathcal{A}_4^{(s)}(1^+,2^+,3^+,4^-)= \intsinf dz_1 dz_2 \,V^{SDYM}_{z_1}(\Phi_1, \Phi_2) V^{SDYM}_{z_2}(G, \Phi_3) e^{- k_4 z_2}\,.
\end{equation}
Note that the vertex on the left acts on two bulk-to-boundary propagators, as in \eqref{eq:V-SDYMphiphi}, whereas the vertex on the right acts also on a bulk-to-bulk propagator, as in \eqref{eq:V-SDYMGphi}. Using the representation of the bulk-to-bulk propagator in \eqref{btbulksdym} then gives
\begin{align}
V^{SDYM}_{z_2}(G_{12}, \Phi_3)= \; & \frac{1}{2\pi} \intinf d\omega  \; \frac{\sin(\omega z_1)}{\omega^2 + k_I^2}  \nonumber \\
&
\cdot \Big(  (- 2 X_{3,4} - i k_{3w} k_{34}) \sin(\omega z_2) + i k_{3w} \, \omega \cos(\omega z_2) \Big)  e^{- k_{3}z_2}\,,
\end{align}
where $k_I$ was defined below \eqref{4pt-phi3-1}. Integrating over the $z$ variables, and then performing the $\omega$ integral by summing over residues in the upper half plane, finally gives\footnote{One important difference with respect to flat space is that we now have $X_{1+2,3}\neq X_{3,4}$, namely $$ X_{1+2,3}=X_{1,3}+X_{2,3}=X_{3,4} + \frac1{2}(k_{1z}+k_{2z}+k_{3z}+k_{4z})k_{3w}=X_{3,4} +\frac{i}{2}k_{1234}\,k_{3w}\,, $$ due to the lack of momentum conservation along the $z$-direction.}
\begin{eqn}\label{SDYM-4pt}
\mathcal{A}_4^{(s)}(1^+,2^+,3^+,4^-)  = \frac{ X_{1,2}X_{3,4}  }{k_{1234} E_L E_R }
=X_{1,2}X_{3,4}   \begin{tikzpicture}[baseline]
\node at (-1, 0) {\textbullet};
\node at (1, 0) {\textbullet};
\draw (-1, 0) -- (1, 0);
\node at (-1, -0.25) {$k_{12}$};
\node at (1, -0.25) {$k_{34}$};
\node at (0, 0.25) {$k_I$};
\end{tikzpicture}\,,
\end{eqn}
where $E_{L,R}$ are defined below \eqref{4pt-phi3-2}.

Note that the final result is obtained by dressing the $\varphi^3$ correlator in \eqref{4pt-phi3-2} with kinematic structure constants \eqref{eq:Xij} associated with each vertex. Moreover, the residue at the total energy pole matches with the result expected from the flat space Feynman rules. However, when summing over the other channels (see Appendix \ref{3ptappendix}), the flat space amplitude vanishes due to on-shell kinematic cancellations, so the full correlator has no energy pole. Computations at five and six points for SDYM can be found in Appendix~\ref{app:5ptSDG}.

\subsection{SDG}

Finally, in this section we will compute three and four-point SDG correlators using the Feynman rules in section \ref{sdgfeynman}. 

\subsection*{3 points:}

We must compute the following diagram at three points:
\begin{eqn}\label{3ptSDG1}
\mathcal{M}_3=
\begin{tikzpicture}[baseline]
\draw[very thick] (0, 0) circle (2);
\draw[dashed] (0, 0) -- ({2*cos(60)}, {2*sin(60)});
\draw[dashed] (0, 0) -- ({2*cos(-60)}, {2*sin(-60)});
\draw[dashed] (0, 0) -- ({2*cos(180)}, {2*sin(180)});

\node at ({(2+0.3)*cos(60)}, {(2+0.3)*sin(60)}) {$1^+$};
\node at ({(2+0.3)*cos(-60)}, {(2+0.3)*sin(-60)}) {$2^+$};
\node at ({(2+0.3)*cos(-180)}, {(2+0.3)*sin(-180)}) {$3^-$};

\node at (-0.3, -0.3) {$+$};
\node at (0.3, 0.3) {$-$};
\node at (0.3, -0.3) {$-$};
\end{tikzpicture}
\end{eqn}
The 3-point vertex in SDG can be represented as in \eqref{eq:V-SDGgen}, which leads in the present case to
\begin{eqn}
V_z^{SDG}(\phi_1, \phi_2) 
=  \frac{e^{-k_{12}z}}{z} X_{1,2} \Big(z X_{1,2} - i (k_{1w} - k_{2w})  \Big) \,.
\end{eqn}
We are left with the following $z$ integral:\footnote{It is convenient to perform these integrals using integration by parts, as we demonstrate this for other examples in Appendix \ref{app:5ptSDG}. } 
\begin{eqn}
\mathcal{M}_{3}\left(1^{+},2^{+},3^{-}\right)=& \intsinf dz\,  V_z^{SDG}(\phi_1, \phi_2) \bar\phi_3 \\
&= X_{1,2} \intsinf dz \Big(z X_{1,2} - i (k_{1w} - k_{2w})  \Big)  e^{- k_{123} z} \\
&= X_{1,2} \left( \frac{X_{1,2}}{k_{123}^2} - \frac{i}{k_{123}}(k_{1w} - k_{2w})\right) \\
&=- X_{1,2}\hat{\mathcal{D}}_{12}\left(\frac{1}{k_{123}}\right) \equiv - X_{1,2}\Big(\mathcal{\hat{D}}_{1,2} 
\begin{tikzpicture}[baseline]
\node at (0,0) {\textbullet};
\node at (0,-0.25) {$k_{123}$};
\end{tikzpicture}\Big)\,,
\label{sdg3pt}
\end{eqn}
where $\mathcal{\hat{D}}_{1,2}= X_{1,2}\,\p_{k_{123}} + i ( k_{1w} - k_{2w} )$ is a differential operator acting on the three-point $\varphi^3$ correlator in \eqref{3pt-phi3}. Note that the derivative is with respect to the sum of external energies entering the vertex, which in this case is $k_{123}$. We will find similar differential operators appearing at higher points, where the derivative will once again be with respect to a sum of external energies.

The flat space limit is obtained from the residue of the leading pole in the energy:
\begin{equation}
\label{eq:sgdres3pt}
\lim_{k_{123}\rightarrow 0} k_{123}^2 \,\mathcal{M}_{3}=X_{1,2}^2 \,,
\end{equation}
which is the expected flat space result. Comparing to \eqref{sdym3ptflat}, we verify that in flat space the double copy is realised by squaring $X_{12}$, which can be thought of as the structure constant of a kinematic algebra. On the other hand, comparing \eqref{sdg3pt} to the SDYM correlator in \eqref{sdym3pt}, we see that the double copy in AdS is realised by inserting the differential operator $\mathcal{\hat{D}}_{1,2}$, which acts on the $\varphi^3$ correlator. This reduces to introducing a second copy of $X_{1,2}$ when we take the residue at the leading pole, as in \eqref{eq:sgdres3pt}.

It is interesting to note that the differential operator satisfies a Jacobi identity. If we define the following object,  
\begin{eqn}\label{3ptJacobi}
M[1,2,3] &= \hat{\mathcal D}_{1, 2}  \hat{\mathcal D}_{1+2, 3}\\
&= \big( X_{1,2} \p_{k_{123}} + i (k_{1w} - k_{2w})\big)  \big( X_{1+2, 3} \p_{k_{123}} + i (k_{1w} + k_{2w} - k_{3w})\big)\,,
\end{eqn}
where $X_{1+2,3} = X_{1,3} + X_{2,3}$, it is easily checked that for any function $f(k_{123})$, we have 
\begin{eqn}
&\Big( M[1, 2,3] +  M[2, 3, 1] + M[3, 1, 2] \Big) f(k_{123}) = 0\,.
\end{eqn}
This encodes the Jacobi identity pointed out in \cite{Lipstein:2023pih}. We will find, however, that this does not apply directly to the 4-point correlator.

\subsubsection*{4 points:}

Next we look at the following $s$-channel exchange diagram \footnote{Note that the full tree-level 4-point correlator can be obtained by summing this diagram over all permutations of the positive helicity legs. Since the left vertex is symmetric under exchange of the two external legs (which follows from \eqref{doublecomm1}), this simply reduces to a sum over $s$, $t$, and $u$ channels.}:
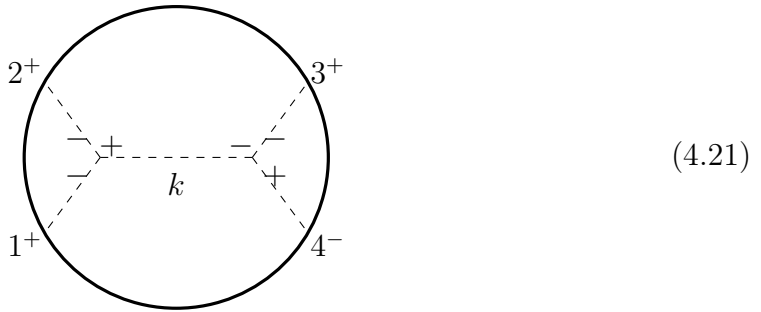
\begin{eqn}
\begin{tikzpicture}[baseline]
\draw[very thick] (0, 0) circle (2);
\draw[dashed] (-1, 0) -- ({2*cos(150)}, {2*sin(150)});
\draw[dashed] (-1, 0) -- ({2*cos(150)}, {2*sin(-150)});

\draw[dashed] (1, 0) -- ({2*cos(30)}, {2*sin(30)});
\draw[dashed] (1, 0) -- ({2*cos(30)}, {2*sin(-30)});

\draw[dashed] (-1, 0) -- (1, 0);

\node at ({(2+0.3)*cos(30)}, {(2+0.3)*sin(30)}) {$3^+$};
\node at ({(2+0.3)*cos(30)}, {(2+0.3)*sin(-30)}) {$4^-$};
\node at ({(2+0.3)*cos(150)}, {(2+0.3)*sin(150)}) {$2^+$};
\node at ({(2+0.3)*cos(150)}, {(2+0.3)*sin(-150)}) {$1^+$};

\node at (0, -0.35) {$k$};

\node at (-1.3, 0.25) {$-$};
\node at (-1.3, -0.25) {$-$};
\node at (-1+0.15, 0.15) {$+$};

\node at (1.3, 0.25) {$-$};
\node at (1.3, -0.25) {$+$};
\node at (1-0.15, 0.15) {$-$};

\end{tikzpicture}
\end{eqn}
This is given by
\begin{eqn}
&\mathcal{M}_{4}^{(s)}\left(1^{+},2^{+},3^{+},4^{-}\right)\\
&= \intsinf dz_1 dz_2 V^{SDG}_{z_1}(\phi_1, \phi_2) V_{z_2}^{SDG}(G_{12}, \phi_3) (z_2 e^{- k_4 z_2})\\
&= \frac{1}{32\pi} 
X_{12} \intinf \frac{d\omega}{\omega^2 + k^2}
\intsinf dz_1 dz_2  e^{- k_{12} z_1 - k_{34} z_2 } \big( z_1 X_{12} - i (k_{1w} - k_{2w}) \big) \sin(\omega z_1) \\
&\times  \Bigg[ \sin(\omega z_2) \big\{ 2 \omega ^2 z_2 k_{3 w}^2-2 \left( -2 i X_{34} + k_{3w} k_{34} \right) \big(z_2 (-2 i X_{34} + k_{3w} k_{34} ) +2 ( k_{4w}+2k_{3 w})\big) \big\} \\
&\qquad\qquad  + 4  k_{3 w} \omega \cos(\omega z_2) \Big\{ z_2 \big(-2 i X_{34} + k_{3w} k_{34} \big) +k_{4w}+2k_{3 w} \Big\} \Bigg]\,.
\end{eqn}
After performing the $z_1$ and $z_2$ integrals, we obtain 
\begin{eqn}\label{SDG-4pt1}
&\mathcal{M}_{4}^{(s)}\left(1^{+},2^{+},3^{+},4^{-}\right)\\
&= \frac{X_{1,2} X_{3,4}}{4\pi} \intinf \frac{d\omega \omega^2}{\omega^2 + k_I^2}
\Bigg\{  \frac{2i k_{12} X_{1,2}}{\omega^2 + k_{12}^2} +  (k_{1w}-k_{2w})\Bigg\} 
 \Bigg\{ \frac{- 2i k_{34} X_{3,4}}{\omega^2 + k_{34}^2} +(3k_{3w} + k_{4w}) \Bigg\} \\
 &\hspace{2cm}
 \times\frac{1}{(\omega^2 + k_{12}^2)(\omega^2 + k_{34}^2)}\\
&= \frac{X_{1,2} X_{3,4}}{4\pi} \intinf \frac{d\omega \omega^2}{\omega^2 + k_I^2} \hat{\mathcal{D}}_{1,2}\hat{\mathcal{D}}_{3,4} \left(\frac{1}{(\omega^{2}+k_{12}^{2})(\omega^{2}+k_{34}^{2})}\right) \\
&= \frac{1}{4} X_{1,2}X_{3,4}\hat{\mathcal{D}}_{1,2}\hat{\mathcal{D}}_{3,4}\left(\frac{1}{k_{1234}E_{L}E_{R}}\right)\\
&= \frac{1}{4} X_{1,2} X_{3,4} \hat{\mathcal D}_{1,2}\hat{\mathcal D}_{3,4}
\begin{tikzpicture}[baseline]
\node at (-1, 0) {\textbullet};
\node at (1, 0) {\textbullet};
\draw (-1, 0) -- (1,0);
\node at (-1, -0.25) {$k_{12}$};
\node at (1, -0.25) {$k_{34}$};
\node at (0, +0.25) {$k$};
\end{tikzpicture}\,,
\end{eqn}
where $k_I$ was defined below \eqref{4pt-phi3-1}, $\hat{\mathcal{D}}_{1,2}=X_{1,2} \p_{k_{12}} + i (k_{1w}-k_{2w} )$, and $\hat{\mathcal{D}}_{3,4}=X_{3,4} \p_{k_{34}} -  i (3k_{3w}+k_{4w})$. Notice that the derivatives in \eqref{SDG-4pt1} are with respect to external energies at each vertex, which we take to be independent of the exchanged energy $k_I$. This can be formally justified by considering an off-shell correlator where the external energies are not equal to the magnitudes of boundary momenta\cite{Salcedo:2022aal}. To obtain the third equality, we pulled the derivative operators out of the integrand and identified the remaining integral with the 4-point $\varphi^3$ exchange diagram evaluated in \eqref{4pt-phi3-2}. We note that the precise form of our derivative operators is not universal in the way that we presented them, i.e.~they are not obtained from each other (and from those seen earlier at 3 points) by trivial relabelling. This is an issue that deserves further study. The representation we find is reminiscent of the differential representation for AdS correlators in \cite{Diwakar:2021juk}. It would be interesting to relate those differential operators to the much simpler ones arising here.

In the flat space limit, there is only a contribution from the term with two derivatives in \eqref{SDG-4pt1}, which gives cubic pole in the total energy:
\begin{eqn}\label{SDG-4pt-flat1}
\lim_{k_{1234}\to 0} k_{1234}^3 \mathcal{M}_{4}^{(s)} = - \frac{1}{2} \frac{X_{1,2}^2 X_{3,4}^2}{k_I^{2}-k_{12}^{2}}\,,
\end{eqn}
which is consistent with the expected flat space limit, the denominator becoming the usual Mandelstam variable $s$. Looking at the SDYM result in \eqref{SDYM-4pt}, we see that the SDG correlator can be obtained by inserting differential operators at each vertex. In the flat space limit, this reduces to squaring the kinematic numerator $X_{1,2} X_{3,4}$. It would be interesting to derive \eqref{SDG-4pt1} from a systematic double copy prescription. Note, however, that \eqref{SDG-4pt-flat1} corresponds to a single channel, and the complete SDG correlator at four points does not have a cubic energy pole, as we show in  Appendix \ref{3ptappendix}. A computation at five points in SDG can be found in Appendix~\ref{app:5ptSDG}.

\section{Soft limits} \label{softlimits}

In flat space, the scattering amplitudes of gluons and gravitons have the following universal soft limits found long ago by Weinberg \cite{Weinberg:1965nx}:
\begin{equation}
\lim_{k_{n+1}^{\mu}\rightarrow0}\mathcal{\mathcal{A}}_{n+1}=\left(\sum_{h=1,n}\frac{\epsilon\cdot k_{h}}{k_{n+1}\cdot k_{h}}+...\right)\mathcal{\mathcal{A}}_{n}\,,
\end{equation} 
\begin{equation}
\lim_{k_{n+1}^{\mu}\rightarrow0}\mathcal{M}_{n+1}=\left(\sum_{h=1}^{n}\frac{\left(\epsilon\cdot k_{h}\right)^{2}}{k_{n+1}\cdot k_{h}}+...\right)\mathcal{M}_{n}\,,
\end{equation}
where $\mathcal{A}_n$ and $\mathcal{M}_n$ refer to $n$-point gluon and graviton amplitudes, respectively, and ... are subleading in the soft momentum.
These soft theorems have been extended to higher orders in the soft momentum  \cite{Cachazo:2014fwa} and more recently interpreted as Ward identities of asymptotic symmetries \cite{Strominger:2017zoo}.

In this section, we will analyse soft limits of SDYM and SDG Feynman diagrams in AdS$_4$ and show that they exhibit structure similar to the soft limits of YM and GR in AdS recently found in \cite{Chowdhury:2024wwe, Chowdhury:2024snc}. In particular, the soft limits give rise to Weinberg-like soft factors, where the soft pole is replaced by an energy derivative. We expect that this similarity can be turned into a direct identification once we work out the precise uplift of our correlators to those studied in these references.
The universal-looking structure we describe suggests an underlying symmetry analogous to the asymptotic symmetries that give rise to the Weinberg soft theorems \cite{Strominger:2013jfa}. It would be interesting to explore whether it is possible to adapt $\Lambda$-BMS symmetry \cite{Compere:2019bua} or cosmological soft theorems \cite{Creminelli:2012ed,Hinterbichler:2013dpa, McFadden:2014nta} to self-dual theories in AdS.

We first consider the toy example of $\varphi^3$ theory in flat space and then show how the soft limits generalize to SDYM and SDG in AdS. For each of these theories we will first illustrate how the soft limits work at three and four points and then compute the soft limit for the following general class of diagrams:
\begin{eqn}
\mbox{Class I:}\quad & \lim_{\vec{k}\to 0}
\scalebox{0.9}{\begin{tikzpicture}[baseline]
\node at (-2.5, 0.15) {$\vdots$};
\draw[fill=lightgray!50] (-1.5, 0) circle (0.5);
\draw ({3*cos(160)}, {3*sin(160)}) -- (-1.9, 0.35);
\draw ({3*cos(200)}, {3*sin(200)}) -- (-1.9, -0.35);

\draw (-1, 0) -- (1, 0);
\draw ({3*cos(20)}, {3*sin(20)}) -- (1, 0);
\draw ({3*cos(-20)}, {3*sin(-20)}) -- (1, 0);

\node at (1.75, 1) {$ \vec{k}$};

\draw[very thick] (0,0) circle (3);
\end{tikzpicture}} \;.
\end{eqn}
We refer to these as class I diagrams. One also also consider other types of diagrams such as class II diagrams:
\begin{eqn}
\mbox{Class II:}\quad &\lim_{\vec{k} \to 0}
\scalebox{0.9}{\begin{tikzpicture}[baseline]
\node at (-2.5, 0.15) {$\vdots$};
\node at (2.5, 0.15) {$\vdots$};
\draw[fill=lightgray!50] (-1.5, 0) circle (0.5);
\draw ({3*cos(160)}, {3*sin(160)}) -- (-1.9, 0.35);
\draw ({3*cos(200)}, {3*sin(200)}) -- (-1.9, -0.35);

\draw[fill=lightgray!50] (1.5, 0) circle (0.5);
\draw ({3*cos(20)}, {3*sin(20)}) -- (1.9, 0.35);
\draw ({3*cos(-20)}, {3*sin(-20)}) -- (1.9, -0.35);

\draw (-1, 0) -- (1, 0);

\draw (0, 0)-- (0, 3);
\node at (0.25, 1.25) {$\vec{k}$};

\draw[very thick] (0,0) circle (3);
\end{tikzpicture} }\,.
\end{eqn}
Note that class I diagrams give rise to the Weinberg soft theorems for scattering amplitudes in flat space whereas class II diagrams contribute to the subleading soft theorem \cite{Cachazo:2014fwa, White:2014qia, Sen:2017xjn}. In this paper we explore the soft limit class I diagrams in AdS. \footnote{See \cite{Chowdhury:2024wwe, Chowdhury:2024snc} for an analysis of Class 2 diagrams in pure Yang Mills and Yang-Mills coupled to matter.}

Before discussing the soft limits of $n$-point diagrams we pause to discuss the nature of two-point functions, since these will arise from the soft limit of three-point functions. In particular, they can be obtained from a bulk perspective by plugging two bulk-to-boundary propagators into the kinetic term of the action \cite{Maldacena:2002vr}, or from the boundary perspective using the conformal Ward identities in momentum space \cite{Bzowski:2013sza,Bzowski:2015pba}. Using either point view, one finds that the two-point functions of scalar operators with scaling dimension $\Delta$ in $d$ dimensions is proportional to $k^{2\Delta - d}$, where $k$ is the energy of either operator. In this paper, all the theories we consider are described at the level of the kinetic term by conformally coupled scalars in AdS$_4$, which corresponds to $\Delta=2$ and $d=3$. Hence, the two-point functions in all cases will scale linearly in the energy. Moreover, looking at the kinetic terms for SDYM and SDG in \eqref{LSDYM} and \eqref{LSDG}, respectively, we see that the 2-point functions must connect scalars of opposite chirality which lift to spinning fields of opposite helicity. While this is consistent with the usual conventions for propagators used in the scattering amplitude literature, this is different than the convention used for cosmological correlators in \cite{Maldacena:2011nz}, where 2-point functions are nonzero for fields of the same helicity. We explain the origin of this difference in Appendix \ref{3ptappendix}.

\subsection{$\varphi^3$ theory}\label{phi3class1}
Let us first demonstrate how the soft limit works for some simple examples in $\varphi^3$ theory.

\subsubsection*{3 $\to$ 2}
Taking the limit $\vec{k}_3 \to 0$ of $A_3$ in \eqref{3pt-phi3} gives
\begin{eqn}
\label{phi3322}
\lim_{\vec{k}_3\to 0} A_3 = \frac{1}{k_{12}}=\frac{1}{2k_{2}}=\frac{1}{2k_{2}}\partial_{k_{2}} A_{2}\,,
\end{eqn}
where we noted that in the soft limit $k_1$ becomes equal to $k_2$ (since $\vec{k}_1$ becomes $-\vec{k}_2$ by boundary momentum conservation), and fixed the normalisation of the 2-point function such that $A_2= k_2$. 

\subsubsection*{4 $\to$ 3}
Now consider the taking the limit $\vec{k}_4 \to 0$ in the 4-point exchange diagram in \eqref{4pt-phi3-2}. This gives 
\begin{eqn}
\lim_{\vec{k}_{4}\rightarrow0} A_{4}=\frac{1}{2k_{3}k_{123}^{2}}=-\frac{1}{2k_{3}}\partial_{k_{3}} A_{3}\,,
\label{phi3423}
\end{eqn}
where we recall the 3-point function given in \eqref{3pt-phi3} and noted that in the soft limit $k_{I}=\left|\vec{k}_{34}\right|\rightarrow k_{3}$.
Let us now pause to discuss the physical interpretation of \eqref{phi3423}. First recall that when we take the flat space limit of the 4-point diagram in \eqref{4pt-phi3-2} we get
\begin{equation}
\lim_{k_{1234}\rightarrow0} A_{4}=\frac{1}{k_{1234}}\frac{1}{\vec k_{34}^2 - k_{34}^2}\,,
\end{equation} 
where we have the usual pole in the total energy and recognize the  is the usual 
standard Lorentz-invariant Mandelstam variable $k_{34}^{\mu}k_{34\mu}$ in the denominator. If we now take the soft limit $k_4^\mu \rightarrow 0$, we obtain 
\begin{equation}
\lim_{\vec{k}_{4}^{\mu}\rightarrow0}\lim_{k_{1234}\rightarrow0} A_{4}=\frac{1}{k_{123}}\frac{1}{2k_{3}\cdot k_{4}}\,,
\end{equation}
where we now have a three-point energy pole, $k_{123}$, and $k_3\cdot k_4$ is the usual Weinberg soft pole. Now let us take the soft limit followed by the flat space limit of \eqref{4pt-phi3-2}.   Using \eqref{phi3423}, we see that 
\begin{equation}
\lim_{k_{1234}\rightarrow0}\lim_{\vec{k}_{4}\rightarrow0} A_{4}=\frac{1}{2k_{3}k_{123}^{2}}\,.
\end{equation}
Hence, we now get a double pole in $k_{123}$ which arises from acting with $\partial_{k_3}$ on the the three-point function in \eqref{phi3423}. Using energy conservation at four points which arises from taking the flat space limit, we may then write the double pole in $k_{123}$ as a single pole in $k_{123}$ times a pole in the energy of the soft leg $k_4$. This shows that the flat space limit does not commute with the soft limit. From this we also see that the analogue of a Weinberg soft factor in AdS is an energy derivative which gives a double pole in the energy.

We will now generalise the relations in \eqref{phi3322} and \eqref{phi3423} to all multplicity for class I diagrams. 

\subsubsection*{Class I}
Consider a general class I diagram in the $\varphi^3$ theory: 
\begin{equation}\label{CCS-class1}
F^{\varphi^3(\rm{I})}_{n+1} \equiv
\scalebox{1}{\begin{tikzpicture}[baseline]
\draw[fill=lightgray!50] (-1.5, 0) circle (0.75);
\draw ({3*cos(160)}, {3*sin(160)}) -- (-1.9, 0.35);
\draw ({3*cos(200)}, {3*sin(200)}) -- (-1.9, -0.35);
\node at (-2.65, 0.15) {$\vdots$};

\draw (-1, 0) -- (1, 0);
\draw ({3*cos(20)}, {3*sin(20)}) -- (1, 0);
\draw ({3*cos(-20)}, {3*sin(-20)}) -- (1, 0);

\node at (1.75, 1) {$ \vec{k}_s$};
\node at (1.75, -1) {$ \vec{k}$};
\node at (-0, -0.25) {$\vec{y}$};

\draw[fill=lightgray!50] (-1.5, 0) circle (0.75);
\draw[very thick] (0,0) circle (3);
\node at (-1.5, 0) {$F_n^{\varphi^3}$};
\end{tikzpicture} }\,,
\end{equation}
where $\vec{k}_s$ is the boundary momentum of the leg we will take soft, $\vec{k}$ is the boundary momentum of an adjacent hard leg, and by momentum conservation $\vec{y}=\vec{k}+\vec{k}_s$, and the gray blob denoted by $F_n^{\varphi^3}$ is a generic $n$-point sub-diagram. From the Feynman rules given in \eqref{prop-phi3} we have 
\begin{eqn}
    F^{\varphi^3(\rm{I})}_{n+1} = \intsinf dz_1 dz_2 F_n^{\varphi^3}(z_1) G(z_1, z_2,  y) e^{-(k + k_s) z_2}~.
\end{eqn}
In the limit $\vec k_s \to 0$, $\vec y = \vec k$ and we can expand the Green functions up to the leading order in $k_s$:
\begin{eqn}
\lim_{\vec{k}_s \to 0} F^{\varphi^3(\rm{I})}_{n+1} &= \intsinf dz_1 dz_2 G(z_1, z_2, k) F_n^{\varphi^3}(z_1)  e^{- k z_2}~.
\end{eqn}
We can then perform the integral over $z_2$ by using the following identity:
\begin{eqn}
\intsinf dz_2 G(z_1, z_2, k) e^{- k z_2} =  \frac{ z_1}{2k} e^{- kz_1}~.
\end{eqn}
We then obtain
\begin{eqn}
\lim_{\vec{k}_s \to 0} F^{\varphi^3(\rm{I})}_{n+1}  &= -\frac{1}{2k} \p_k \intsinf dz_1 F_n^{\varphi^3}(z_1) e^{- kz_1}~.
\end{eqn}
This equation can be diagrammatically expressed as
\begin{eqn}
\lim_{\vec{k}_s \to 0}
\scalebox{0.9}{\begin{tikzpicture}[baseline]
\draw ({3*cos(160)}, {3*sin(160)}) -- (-1.9, 0.35);
\draw ({3*cos(200)}, {3*sin(200)}) -- (-1.9, -0.35);
\node at (-2.65, 0.15) {$\vdots$};

\draw (-1, 0) -- (1, 0);
\draw ({3*cos(20)}, {3*sin(20)}) -- (1, 0);
\draw ({3*cos(-20)}, {3*sin(-20)}) -- (1, 0);

\node at (1.75, 1) {$ \vec{k}_s$};
\node at (1.75, -1) {$ \vec{k}$};
\node at (-0, -0.25) {$\vec{y}$};

\draw[fill=lightgray!50] (-1.5, 0) circle (0.75);

\draw[very thick] (0,0) circle (3);
\node at (-1.5, 0) {$F_n^{\varphi^3}$};
\end{tikzpicture} }
= - \frac{1}{2k} \p_k 
\scalebox{0.9}{\begin{tikzpicture}[baseline]
\draw ({3*cos(160)}, {3*sin(160)}) -- (-1.9, 0.35);
\draw ({3*cos(200)}, {3*sin(200)}) -- (-1.9, -0.35);
\node at (-2.65, 0.15) {$\vdots$};

\draw (-1, 0) -- (3, 0);

\node at (0.5, -0.25) {$\vec{y} = \vec{k}$};

\draw[fill=lightgray!50] (-1.5, 0) circle (0.75);
\draw[very thick] (0,0) circle (3);
\node at (-1.5, 0) {$F_n^{\varphi^3}$};
\end{tikzpicture} }\,.
\end{eqn}

In the following sections, we will see that similar structure arises in the soft limit of SDYM and SDG diagrams. Note that the energy derivative also arises in the soft limit of full YM and GR boundary correlators \cite{Chowdhury:2024wwe} and plays the role of the Weinberg soft pole for scattering amplitudes.

\subsection{SDYM}
Let us proceed as we did in the $\varphi^3$ theory and first look at some simple examples. For simplicity we shall study the colour-stripped correlator and denote the rest of the correlator as $\mathcal{A}_n$ with $n$ denoting the number of external legs.

\subsubsection*{3 $\to$ 2}

Consider the 3-point function given in \eqref{sdym3pt}. By taking the soft limit of leg 2, i.e, $\vec k_2 \to 0$ we obtain 
\begin{eqn}
\label{softym1}
\lim_{\vec k_{2}\rightarrow0}\mathcal{A}_{3}(1^{+},2^{+},3^{-})=\frac{X_{1,2}}{2k_{1}}=-\frac{X_{1,2}}{2k_{1}}\partial_{k_{1}}\mathcal{A}_{2}(1^{+},3^{-})\,,
\end{eqn}
where we have normalised the 2-point function such that $\mathcal{A}_{2}(1^{+},3^{-})= k_1$. It is also natural to ask what happens upon taking the soft limit of a minus helicity field. After taking $\vec k_3 \to 0$ in \eqref{sdym3pt} we obtain 
\begin{eqn}\label{3ptminussoft}
\lim_{\vec k_{3} \rightarrow0}\mathcal{A}_{3}(1^{+},2^{+},3^{-}) = \frac{X_{1, 2}}{ 2k_{1}} \,.
\end{eqn}
In this case, we do not obtain a correlator in SDYM, as it corresponds to a two-point function with both helicities positive. In Appendix \ref{3ptappendix}, we will show that after lifting the three-point SDYM scalar correlator to a spinning correlator, the soft limit in \eqref{softym1} becomes finite, while the one in \eqref{3ptminussoft} (which gives rise to a correlator that does not occur in in SDYM) goes to zero. 

\subsubsection*{4 $\to$ 3}

Let us recall the 4-point result in \eqref{SDYM-4pt}:
\begin{eqn}
\mathcal{A}_{4}^{(s)}\left(1^{+},2^{+},3^{+},4^{-}\right)=\frac{X_{1,2}X_{3,4}}{k_{1234}E_{L}E_{R}}\,.
\end{eqn}
Taking the soft limit $\vec{k}_3 \to 0$ then gives
\begin{eqn}
\lim_{\vec{k}_{3}\rightarrow0}\mathcal{A}_{4}^{(s)}\left(1^{+},2^{+},3^{+},4^{-}\right)=\frac{X_{1,2}X_{3,4}}{2k_{4}k_{124}^{2}}=-\frac{X_{3,4}}{2k_{4}}\partial_{k_{4}}\mathcal{A}_{3}\left(1^{+},2^{+},4^{-}\right)\,,
\label{sdym423}
\end{eqn}
where we noted that $E_{L}=k_{12}+\left|\vec{k}_{34}\right|\rightarrow k_{124}$ and $E_{R}=k_{34}+\left|\vec{k}_{34}\right|\rightarrow2k_{4}$ Hence, in the soft limit we obtain the same type of structure that we found in the $3\rightarrow 2$ soft limit, notably a kinematic structure constant $X_{i,j}$ times an energy derivative acting on a lower-point correlator. An energy derivative also arises in soft limit of the $\varphi^3$ theory, as shown in the previous subsection. 

We get a similar result when taking one of the other positive helicity legs soft. For example if we take leg 1 soft we obtain,
\begin{eqn}
\lim_{\vec{k}_1 \to 0} \mathcal A_4^{(s)}\left(1^{+},2^{+},3^{+},4^{-}\right) = -\frac{X_{1,2}}{k_2} \p_{k_2} \mathcal A_3(2^+, 3^-, 4^+)\,.
\end{eqn} 
On the other hand, we take the soft limit of the leg with negative helicity we obtain 
\begin{eqn}
\lim_{\vec k_4 \to 0} \mathcal A_4^{(s)}(1^+, 2^+, 3^+, 4^-) = \frac{X_{1,2} X_{3, 4}}{k_{123}^2 k_{3}}\,,
\end{eqn}
which gives an all-plus three-point correlator which does not exist in SDYM theory.    

\subsubsection*{class I}

We now compute the soft limit of a general class I diagram:
\begin{eqn}
F^{SDYM\rm{(I)}}_{n+1} = \begin{tikzpicture}[baseline]
\node at (-2.65, 0.15) {$\vdots$};

\draw ({3*cos(160)}, {3*sin(160)}) -- (-1.9, 0.35);
\draw ({3*cos(200)}, {3*sin(200)}) -- (-1.9, -0.35);

\draw (-1, 0) -- (1, 0);
\draw ({3*cos(20)}, {3*sin(20)}) -- (1, 0);
\draw ({3*cos(-20)}, {3*sin(-20)}) -- (1, 0);

\node at (1.75, 1) {$\vec{k}_s$};
\node at (1.75, -1) {$ \vec{k}$};
\node at (-0, -0.35) {$\vec{y}$};

\node at ({3.2*cos(20)}, {3.2*sin(20)}) {$+$};
\node at ({3.2*cos(-20)}, {3.2*sin(-20)}) {$-$};

\node at ({3.2*cos(200)}, {3.2*sin(200)}) {$+$};
\node at ({3.2*cos(160)}, {3.2*sin(160)}) {$+$};

\draw[very thick] (0,0) circle (3);

\draw[fill=lightgray!50] (-1.5, 0) circle (0.75);
\node at (-1.5, 0) {\footnotesize $F_n^{SDYM}$};

\end{tikzpicture} \,.
\end{eqn}
where the blob is a generic subdiagram in SDYM. Using the Feynman rules in section \ref{sec:SDYMfeynman} we have the following expression for the diagram above:
\begin{eqn}
F_{n+1}^{SDYM(\rm{I})} = \intsinf dz_1 dz_2 F_n^{SDYM}(z_1)  V_{z_2}^{SDYM}(G, \Phi_s) e^{-k z_2}\,,
\end{eqn}
where $V_z^{SDYM}(G, \Phi_s)$ is obtained from \eqref{V-SDYM} and is given as 
\begin{eqn}
 V_{z_2}^{SDYM}(G, \Phi_s)  = \frac{1}{ \pi} \intinf \frac{d\omega \sin(\omega z_1)}{\omega^2 + y^2} \Big[ f_1 \sin(\omega z_2) + i (k_s)_w\omega \cos(\omega z_2) \Big] e^{- k_s z_2}\,,
\end{eqn}
where $f_1 = - 2 X_{k, k_s}  - i k (k_s)_w$~.

In the limit $\vec{k}_s \to 0$ we have $\vec y = \vec k$ and obtain 
\begin{eqn}
&\lim_{\vec{k}_s \to 0}F_{n+1}^{SDYM(\rm{I})} \\
&= \frac1\pi \intsinf dz_1 F_n^{SDYM}(z_1) \intinf \frac{d\omega \sin(\omega z_1)}{\omega^2 + k^2} \intsinf dz_2  \Big[ f_1 \sin(\omega z_2) + i (k_s)_w \omega \cos(\omega z_2) \Big] e^{- k z_2}\,.
\end{eqn}
We now evaluate the $z_2$ integrals by using 
\begin{eqn}
\intsinf dz_2 \sin(\omega z_2) e^{- k z_2} &= \frac{\omega}{\omega^2  + k^2}, \qquad
  \intsinf dz_2 \omega \cos(\omega z_2) e^{- k z_2} = \frac{k \omega}{\omega^2  + k^2}\,.
\end{eqn}
and are left with the following $\omega$-integral:
\begin{eqn}
\lim_{\vec{k}_s \to 0}F_{n+1}^{SDYM(\rm{I})} &= \frac1\pi \intsinf dz_1 F_n^{SDYM}(z_1) \intinf \frac{d\omega }{(\omega^2 + k^2)^2}  \big[ f_1  +i (k_s)_w k \big]   \omega\sin(\omega z_1)\,.
\end{eqn}
The factor in the square bracket simplifies to $ f_1  + i (k_s)_w k = - 2 X_{k, k_s}$~.

Finally, the $\omega$ integral can be performed by the method of residues resulting in 
\begin{eqn}
\label{sdymclass1gensoft}
\lim_{\vec{k}_s \to 0}F_{n+1}^{SDYM(\rm{I})} &= \frac{1}{2 k_1}  X_{k_s, k} \intsinf dz_1 z_1 F_n^{SDYM}(z_1) e^{- k z_1} \\
&=- \frac{1}{2 k} X_{k_s, k}\p_k \intsinf dz_1  F_n^{SDYM}(z_1) e^{- k z_1} \\
&=- \frac{X_{k, k_s}}{2}  \p_{k} F_n^{SDYM}\,.
\end{eqn}
For $n =3$, \eqref{sdymclass1gensoft} reduces to \eqref{sdym423}. This relation can be diagrammatically summarized as 
\begin{eqn}\label{class1-SDYM1}
\lim_{\vec k_s \to 0}
\scalebox{0.9}{
\begin{tikzpicture}[baseline]
\node at (-2.65, 0.15) {$\vdots$};

\draw ({3*cos(160)}, {3*sin(160)}) -- (-1.9, 0.35);
\draw ({3*cos(200)}, {3*sin(200)}) -- (-1.9, -0.35);

\draw (-1, 0) -- (1, 0);
\draw ({3*cos(20)}, {3*sin(20)}) -- (1, 0);
\draw ({3*cos(-20)}, {3*sin(-20)}) -- (1, 0);

\node at (1.75, 1) {$\vec{k}_s$};
\node at (1.75, -1) {$ \vec{k}$};
\node at (-0, -0.35) {$\vec{y}$};

\node at ({3.2*cos(20)}, {3.2*sin(20)}) {$+$};
\node at ({3.2*cos(-20)}, {3.2*sin(-20)}) {$-$};

\node at ({3.2*cos(200)}, {3.2*sin(200)}) {$+$};
\node at ({3.2*cos(160)}, {3.2*sin(160)}) {$+$};

\draw[very thick] (0,0) circle (3);

\draw[fill=lightgray!50] (-1.5, 0) circle (0.75);
\node at (-1.5, 0) {\footnotesize $F_n^{SDYM}$};

\end{tikzpicture}
}
= - \frac{X_{k_s, k}}{2k}  \p_k \scalebox{0.9}{\begin{tikzpicture}[baseline]
\node at (-2.65, 0.15) {$\vdots$};

\draw ({3*cos(160)}, {3*sin(160)}) -- (-1.9, 0.35);
\draw ({3*cos(200)}, {3*sin(200)}) -- (-1.9, -0.35);

\draw (-1, 0) -- (3, 0);

\node at (0.5, -0.35) {$\vec{y} = \vec{k}$};

\node at ({3.2*cos(0)}, {3.2*sin(0)}) {$-$};

\node at ({3.2*cos(200)}, {3.2*sin(200)}) {$+$};
\node at ({3.2*cos(160)}, {3.2*sin(160)}) {$+$};

\draw[very thick] (0,0) circle (3);

\draw[fill=lightgray!50] (-1.5, 0) circle (0.75);
\node at (-1.5, 0) {\footnotesize $F_n^{SDYM}$};
\end{tikzpicture} }\,.
\end{eqn}
There is a similar formula when the soft leg is adjacent to a leg with positive helicity: 
\begin{eqn}
\lim_{\vec k_s \to 0}\scalebox{0.9}{\begin{tikzpicture}[baseline]
\draw ({3*cos(160)}, {3*sin(160)}) -- (-2, 0);
\draw ({3*cos(200)}, {3*sin(200)}) -- (-2, 0);

\draw ({3*cos(20)}, {3*sin(20)}) -- (1.9, 0.35);
\draw ({3*cos(-20)}, {3*sin(-20)}) -- (1.9, -0.35);

\draw (-2, 0) -- (2, 0);

\node at (2.65, 0) {$\vdots$};
\node at (-0.5, -0.25) {$\vec{y}$};

\node at ({(3+0.25)*cos(200)}, {(3+0.25)*sin(200)}) {$+$};
\node at ({(3+0.25)*cos(160)}, {(3+0.25)*sin(160)}) {$+$};

\node at ({(3+0.25)*cos(20)}, {(3+0.25)*sin(20)}) {$+$};
\node at ({(3+0.25)*cos(-20)}, {(3+0.25)*sin(-20)}) {$-$};

\node at (-2.2, 0.75) {$\vec{k}$};
\node at (-2.2, -0.75) {$\vec{k}_s$};

\draw[very thick] (0,0) circle (3);

\draw[fill=lightgray!50] (1.5, 0) circle (0.75);
\node at (1.5, 0) {\footnotesize $F_n^{SDYM}$};

\end{tikzpicture}}
=
 - \frac{X_{k_s k} }{2k} \p_k
 \scalebox{0.9}{\begin{tikzpicture}[baseline]


\draw ({3*cos(20)}, {3*sin(20)}) -- (1.9, 0.35);
\draw ({3*cos(-20)}, {3*sin(-20)}) -- (1.9, -0.35);

\draw (-3, 0) -- (2, 0);

\node at (2.65, 0) {$\vdots$};


\node at ({(3+0.25)*cos(20)}, {(3+0.25)*sin(20)}) {$+$};
\node at ({(3+0.25)*cos(-20)}, {(3+0.25)*sin(-20)}) {$-$};

\node at (-0.75, 0.25) {$\vec{y} = \vec{k}$};
\node at (-3+.25, -0.25) {$+$};
\draw[very thick] (0,0) circle (3);

\draw[fill=lightgray!50] (1.5, 0) circle (0.75);
\node at (1.5, 0) {\footnotesize $F_n^{SDYM}$};

\end{tikzpicture}}\,.
\end{eqn}

\subsection{SDG}

Let us begin by computing the soft limit of three and four-point SDG diagrams evaluated in section \ref{sec:correlators}.

\subsection*{3 $\to$ 2}
The soft limit of the 3-point function given in \eqref{sdg3pt} is given by
\begin{eqn}
\lim_{\vec{k_{2}}\rightarrow0}\mathcal{M}_{3}(1^{+},2^{+},3^{-})=-i \frac{X_{1,2}k_{1w}}{2k_{1}}=-\frac{X_{1,2}k_{1w}}{2k_{1}}\partial_{k_{1}}\mathcal{M}_{2}(1^{+},3^{-})\,,
\end{eqn}
where we normalise the 2-point function such that $\mathcal{M}_{2}(1^{+},3^{-}) = k_1$. 

\subsection*{4 $\to$ 3}
We can similarly evaluate the soft limit of the four-point function in \eqref{SDG-4pt1} and obtain
\begin{eqn}
\lim_{\vec{k}_{3}\rightarrow0}\mathcal{\mathcal{M}}_{4}^{(s)}\left(1^{+},2^{+},3^{+},4^{-}\right)=\frac{X_{1,2}X_{3,4}k_{4w}}{2k_{4}}\hat{\mathcal{D}}_{1,2}\left(\frac{1}{k_{124}^{2}}\right)=-\frac{X_{3,4}k_{4w}}{2k_{4}}\partial_{k_{4}}\mathcal{M}_{3}\left(1^{+},2^{+},4^{-}\right)\,,
\end{eqn}
where we recall that $E_{L}=k_{12}+\left|\vec{k}_{34}\right|\rightarrow k_{124}$, $E_{R}=k_{34}+\left|\vec{k}_{34}\right|\rightarrow2k_{4}$,  and the three-point correlator in \eqref{sdg3pt}. Hence, in the soft limits we once again find an energy derivative acting on a lower-point correlator as we found for SDYM in \eqref{sdym423} and the $\varphi^3$ theory in \eqref{phi3423}. As we explained in section \ref{phi3class1}, this encodes the analogue of the Weinberg soft factor in AdS.

In the following section we examine the soft limit for a general $n$-point function and show that this structure generalizes.

\subsubsection*{class I}
Now we consider a general class I diagram:
\begin{eqn}
F_{n+1}^{SDG (\rm{I})}=
\begin{tikzpicture}[baseline]
\draw[fill=lightgray!50] (-1.5, 0) circle (0.5);
\draw[dashed] ({3*cos(160)}, {3*sin(160)}) -- (-1.9, 0.35);
\draw[dashed] ({3*cos(200)}, {3*sin(200)}) -- (-1.9, -0.35);
\node at (-2.5, 0.15) {$\vdots$};

\draw[dashed] (-1, 0) -- (2, 0);
\draw[dashed] ({3*cos(20)}, {3*sin(20)}) -- (2, 0);
\draw[dashed] ({3*cos(-20)}, {3*sin(-20)}) -- (2, 0);

\node at (1.75, 1) {$ \vec{k}_s$};
\node at (1.75, -1) {$ \vec{k}$};
\node at (-0, -0.35) {$\vec{y}$};

\draw[very thick] (0,0) circle (3);

\node at ({3.2*cos(20)}, {3.2*sin(20)}) {$+$};
\node at ({3.2*cos(-20)}, {3.2*sin(-20)}) {$-$};

\node at ({3.2*cos(200)}, {3.2*sin(200)}) {$+$};
\node at ({3.2*cos(160)}, {3.2*sin(160)}) {$+$};

\draw[fill=lightgray!50] (-1.5, 0) circle (0.75);
\node at (-1.5, 0) {\small $F_n^{SDG}$};

\end{tikzpicture} \,,
\end{eqn}
where the $F_n^{SDG}$ is a generic $n$-point subdiagram in SDG. The expression for this diagram is given as 
\begin{eqn}
F_{n+1}^{SDG (\rm{I})} =  \frac{1}{8}\int dz_1 F_n^{SDG}(z_1) z_1 \intsinf dz_2  \Big[ \tilde{\tilde G} z_2 g_1 + \tilde G (z_2 g_2  + g_3 ) + G (z_2 g_4 + g_5) \Big] e^{- k_2 z_2} \,,\\
\end{eqn}
where
\begin{eqn}\label{gdefn1}
g_1 &= (k_{s})_w^2~, \\
g_2 &= 4 i (k_{s})_w \big(y_{w} (k_{s})_u - y_{u} (k_{s})_w \big)~,\\
g_3 &= 2 (k_{s})_w \big(- y_{w} + (k_{s})_w \big) ~, \\
g_4 &= 4 \big(y_{u} (k_{s})_w - y_{w} (k_{s})_w \big)^2~, \\
g_5 &= 4i \big((k_{s})_w - y_{w})(q_{u} (k_{s})_w - y_{w} (k_{s})_u \big)\,,
\end{eqn}
with the functions
\begin{eqn*}
G &= \frac1\pi \int \frac{d\omega}{\omega^2 + y^2} \sin(\omega z_1) \sin(\omega z_2)\,, \\
\tilde G &= \frac1\pi \int \frac{\omega d\omega}{\omega^2 + y^2} \sin(\omega z_1) \cos(\omega z_2)\,, \\
\tilde{\tilde G} &= \frac1\pi  \int \frac{\omega^2 d\omega}{\omega^2 + y^2} \sin(\omega z_1) \sin(\omega z_2)\,.
\end{eqn*}
There is a huge simplification that occurs upon taking the soft limit $\vec k_s \to 0$, where we see that $g_1, g_2, g_3 \sim O(k_s^2)$ and $g_3, g_5 \sim O(k_s)$. Using this simplification and various manipulations similar to the SDYM case, the contribution to $F_{n+1}^{SDG(\rm{I})}$ in the soft limit takes a fairly simple form, 
\begin{eqn}
\lim_{\vec{k}_s \to 0} F_{n+1}^{SDG(\rm{I})} = - \frac{1}{2k} k_w  X_{k, k_s}\p_k   \intsinf dz_1 F_n^{SDG}(z_1) z_1 e^{- k z_1}
=- \frac{1}{2k} k_w  X_{k, k_s}\p_k F_{n}^{SDG}
\,.
\end{eqn}
This can be diagrammatically summarized as 
\begin{eqn}\label{class1-SDG1}
&\lim_{\vec k_s \to 0}
\scalebox{0.85}{\begin{tikzpicture}[baseline]
\draw[dashed] ({3*cos(160)}, {3*sin(160)}) -- (-1.9, 0.35);
\draw[dashed] ({3*cos(200)}, {3*sin(200)}) -- (-1.9, -0.35);
\node at (-2.65, 0.15) {$\vdots$};

\draw[dashed] (-1, 0) -- (1, 0);
\draw[dashed] ({3*cos(20)}, {3*sin(20)}) -- (1, 0);
\draw[dashed] ({3*cos(-20)}, {3*sin(-20)}) -- (1, 0);

\node at (1.75, 1) {$ \vec{k}_s$};
\node at (1.75, -1) {$ \vec{k}$};
\node at (-0, -0.35) {$\vec{y}$};

\draw[very thick] (0,0) circle (3);

\node at ({3.2*cos(20)}, {3.2*sin(20)}) {$+$};
\node at ({3.2*cos(-20)}, {3.2*sin(-20)}) {$-$};

\node at ({3.2*cos(200)}, {3.2*sin(200)}) {$+$};
\node at ({3.2*cos(160)}, {3.2*sin(160)}) {$+$};

\draw[fill=lightgray!50] (-1.5, 0) circle (0.75);
\node at (-1.5, 0) {\small $F_n^{SDG}$};

\end{tikzpicture} } 
= -  \frac{k_w  X_{k, k_s}}{2k}\p_k \scalebox{0.85}{\begin{tikzpicture}[baseline]
\draw[fill=lightgray] (-1.5, 0) circle (0.5);
\draw[dashed] ({3*cos(160)}, {3*sin(160)}) -- (-1.9, 0.35);
\draw[dashed] ({3*cos(200)}, {3*sin(200)}) -- (-1.9, -0.35);
\node at (-2.65, 0.15) {$\vdots$};

\draw[dashed] (-1, 0) -- (3, 0);

\node at (0.5, -0.35) {$\vec{y} = \vec{k}$};

\draw[very thick] (0,0) circle (3);

\node at ({3.2*cos(0)}, {3.2*sin(0)}) {$-$};

\node at ({3.2*cos(200)}, {3.2*sin(200)}) {$+$};
\node at ({3.2*cos(160)}, {3.2*sin(160)}) {$+$};

\draw[fill=lightgray!50] (-1.5, 0) circle (0.75);
\node at (-1.5, 0) {\small $F_n^{SDG}$};

\end{tikzpicture} }\,.
\end{eqn}
One can perform a similar analysis when the soft leg is is adjacent to another positive-helicity leg to obtain
\begin{eqn}
\lim_{\vec k_s \to 0}\scalebox{0.85}{\begin{tikzpicture}[baseline]
\draw[dashed] ({3*cos(160)}, {3*sin(160)}) -- (-2, 0);
\draw[dashed] ({3*cos(200)}, {3*sin(200)}) -- (-2, 0);

\draw[dashed] ({3*cos(20)}, {3*sin(20)}) -- (1.9, 0.35);
\draw[dashed] ({3*cos(-20)}, {3*sin(-20)}) -- (1.9, -0.35);

\draw[dashed] (-2, 0) -- (2, 0);

\draw[fill=lightgray] (1.5, 0) circle (0.5);
\node at (2.65, 0) {$\vdots$};
\node at (-0.5, -0.25) {$y$};

\node at ({(3+0.25)*cos(200)}, {(3+0.25)*sin(200)}) {$+$};
\node at ({(3+0.25)*cos(160)}, {(3+0.25)*sin(160)}) {$+$};

\node at ({(3+0.25)*cos(20)}, {(3+0.25)*sin(20)}) {$+$};
\node at ({(3+0.25)*cos(-20)}, {(3+0.25)*sin(-20)}) {$-$};

\node at (-2.2, 0.75) {$\vec{k}$};
\node at (-2.2, -0.75) {$\vec{k}_s$};

\draw[very thick] (0,0) circle (3);

\draw[fill=lightgray!50] (1.5, 0) circle (0.75);
\node at (1.5, 0) {\small $F_n^{SDG}$};

\end{tikzpicture}}
=
 - \frac{k_w X_{k_s k} }{2k} \p_k
 \scalebox{0.85}{\begin{tikzpicture}[baseline]


\draw[dashed] ({3*cos(20)}, {3*sin(20)}) -- (1.9, 0.35);
\draw[dashed] ({3*cos(-20)}, {3*sin(-20)}) -- (1.9, -0.35);

\draw[dashed] (-3, 0) -- (2, 0);

\draw[fill=lightgray] (1.5, 0) circle (0.5);
\node at (2.65, 0) {$\vdots$};


\node at ({(3+0.25)*cos(20)}, {(3+0.25)*sin(20)}) {$+$};
\node at ({(3+0.25)*cos(-20)}, {(3+0.25)*sin(-20)}) {$-$};

\node at (-0.75, 0.25) {$\vec{y} = \vec{k}$};
\node at (-3+.25, -0.25) {$+$};
\draw[very thick] (0,0) circle (3);

\draw[fill=lightgray!50] (1.5, 0) circle (0.75);
\node at (1.5, 0) {\small $F_n^{SDG}$};

\end{tikzpicture}}\,.
\end{eqn}

It has previously been shown that soft limits of A(dS) correlators can be derived from the Ward identities corresponding to asymptotic symmetries.\footnote{CC would like to thank Radu Moga and Kostas Skenderis for useful discussions on this subject.} For example, Ward identities for large diffeomorphisms have been used to used to derive the soft limits of cosmological correlators \cite{Maldacena:2002vr,Creminelli:2012ed,  Hinterbichler:2013dpa,McFadden:2014nta}. In these approaches, it was found that the soft limit gives lower-point correlators acted on by a derivative with respect to a hard momentum and similar structures have recently been derived for AdS correlators from a diagrammatic perspective \cite{Chowdhury:2024wwe,Chowdhury:2024snc}. In this paper, we have found similar structure arising in the soft limit of Witten diagrams of SDYM and SDG in AdS$_4$, so it would be interesting to explore if this can be derived from some underlying symmetry. 

Indeed, it has recently been shown that the scattering amplitudes of SDYM and SDG enjoy an infinite dimensional symmetry known as a $Lw_{1+\infty}$ symmetry that is encoded in their soft and collinear limits \cite{Strominger:2021mtt} and recent work has demonstrated that this symmetry has an analogue in AdS$_4$ \cite{Lipstein:2023pih,Taylor:2023ajd,Bittleston:2024rqe}, although its implications for correlation functions have yet to be understood. It is therefore natural to expect that the elegant structures we find in the SDYM and SDG correlators can at least be partial explained by such symmetries, and we leave this investigation for future work. Indeed, in the soft limit of SDYM diagrams \eqref{class1-SDYM1} we see the appearance of $X_{k_a,k_b}$ factors which correspond to the structure constants of a kinematic algebra dual to the color algebra of the gauge theory, which is intimately related to $Lw_{1+\infty}$ symmetry \cite{Monteiro:2022lwm}. Moreover, these structure constants also arise in the soft limit of SDG diagrams now multiplied by a $k_w$ (see \eqref{class1-SDG1}). Naively, we would expect to have square of a kinematic structure constant, but in curved background the additional structure constant gets deformed \cite{Lipstein:2023pih} (as one can see from the 3-point correlator in \eqref{sdg3pt}) and only the $k_w$ term survives in the soft limit. 

Note that the kinematic structure constants appearing the soft limit of SDYM and SDG diagrams actually vanish in the soft limit. In Appendix \ref{3ptappendix} we explain how to lift the three-point SDYM scalar correlator to a spinning correlators. After doing so, we find that the soft limit of a positive helicity leg gives a nonzero result while the soft limit of a negative helicity leg vanishes, as expected. We expect similar results to hold after lifting scalar SDG correlators to spinning ones.

\section{Conclusion} \label{conclusion}

In this article, we have made progress towards a more systematic understanding of the recent formulation of SDG in an AdS$_4$ background \cite{Lipstein:2023pih}, using also the analogous case of SDYM in AdS$_4$ as guidance. Firstly, we described how the lightcone actions for SDYM and SDG in AdS$_4$ are part of the full lightcone actions for YM and gravity, written in terms of positive and negative helicity fields. In particular, guided by past work in flat background \cite{Ananth:2006fh}, we showed how the introduction of the cosmological constant modifies the structure of the lightcone graviton vertices, matching the results of \cite{Lipstein:2023pih} in the self-dual sector. No such modification occurs in the YM case, due to conformal symmetry. We did not study in detail the boundary terms that arise due to the conformal boundary, and only kept track of such terms in an Appendix in the case of YM. Nevertheless, our results provide a starting point to systematically perturb away from the self-dual sector. Since boundary terms can give non-trivial contributions to correlation functions, a very important direction for the future would be to carry out a systematic analysis of boundary terms in SDYM, SDG, as well as more general theories.

Secondly, we took the first steps in the computation of boundary correlators in the self-dual sector, showing that they can be recast in terms of flat-space boundary correlators built from diagrams with cubic vertices. This leads to major simplifications and offers glimpses of an underlying double-copy structure. There has been a great deal of recent progress in the study of the double copy of (A)dS correlators \cite{Farrow:2018yni,Armstrong:2020woi,Albayrak:2020fyp,Diwakar:2021juk,Cheung:2022pdk,Herderschee:2022ntr,Armstrong:2023phb,Lee:2022fgr}. Due to the simplicity of SDYM and SDG correlators, we believe that they may provide the simplest setting for studying the double copy in these backgrounds and we hope to develop a more systematic understanding in the future. At this stage, we do not have a systematic understanding of how to map the scalar SDYM and SDG correlators to spinning ones so this is another important future direction. In the case of gluons, we discuss this mapping in Appendix \ref{3ptappendix}, which highlights the importance of boundary terms and appropriate translation between gauges. 

Finally, we computed soft limits of the SDYM/SDG correlators at the diagrammatic level, and noticed they have a very similar structure to those of full YM and gravity given in \cite{Chowdhury:2024wwe}, a work that was partly motivated by our early investigations. Again, the explicit relation requires us to translate our results for the `scalar' correlators into those of gluons and gravitons. The ultimate hope is that SDG in AdS could be a powerful toy model, capable of revealing hidden structures in the computation of correlators, mimicking the successes of its flat-space counterpart. 

There are a number of directions open for future work, apart from the ones already mentioned. The soft limits of the correlators, some of which we have computed, may have an interpretation as Ward identities of some suitably defined asymptotic symmetries \cite{Strominger:2013jfa,He:2014laa}. Indeed, in flat space, an infinite tower of such symmetries, constructed recursively, was described in \cite{Campiglia:2021srh} for SDYM and SDG in the lightcone formalism. Related to this, the $Lw_{\infty+1}$ algebra of soft-collinear limits which has played a central role in celestial holography \cite{Strominger:2021mtt,Guevara:2021abz} was shown to follow directly from the Poisson structure of the colour-kinematics duality in flat-space SDG \cite{Monteiro:2022lwm,Adamo:2021lrv}. The deformation of this Poisson structure presented in \cite{Lipstein:2023pih}, based on the lightcone formalism, provides a natural counterpart, and it would be important to understand how it relates to the deformation of $Lw_{\infty+1}$ later described in \cite{Taylor:2023ajd,Bittleston:2024rqe}. In addition, it would be important to understand how these relate to the soft/collinear limits of correlators.

The computation of graviton correlators in de Sitter backgrounds is of natural interest to cosmology, with recent progress at four and five points \cite{Bonifacio:2022vwa,Armstrong:2023phb,Mei:2024abu}. It would thus be interesting to generalise our results to in-in correlators following \cite{Sleight:2021plv,DiPietro:2021sjt}, or other de Sitter observables recently proposed in \cite{Melville:2024ove}. Indeed, it has recently been shown that in-in correlators of certain scalar theories have simpler structure than wavefunction coefficients which arise from Wick rotation of Witten diagrams for boundary correlators \cite{Chowdhury:2023arc}, so it would be interesting to see if simpler simplifications arise for SDYM and SDG correlators. Another interesting direction would be to see if the correlators in this paper can be expressed geometrically as cosmological polytopes \cite{Arkani-Hamed:2017fdk, Benincasa:2024leu} or simplices \cite{Bzowski:2019kwd}, where previous work has focused on scalar fields. In this regard, our work could provide a bridge to spinning theories like YM and gravity since the scalar theories we consider encode a subsector of the dynamics of full YM and GR. We can also consider SDYM and SDG correlators in more general FLRW spacetimes, for which a similar self-dual formulation based on Jacobi brackets was given in \cite{CarrilloGonzalez:2024sto}. This may in turn provide provide an interesting new arena for generalising the differential equations for cosmological correlators recently presented in \cite{Arkani-Hamed:2023kig}.

Beyond these specific applications, there is the general aim---which has been central for scattering amplitudes---to provide the `simplest formulas' for correlators, exhibiting as manifestly as possible their analytic structure and symmetries. We expect that twistor theory will play a role, and indeed, there is a close connection to the lightcone formalism. Ref.~\cite{Baumann:2024ttn} recently took some steps in applying twistor methods to correlators. There is significant related work using either twistor techniques and/or lightcone methods that would be important to understand in a unified framework, e.g.~\cite{Adamo:2012nn,Adamo:2013cra,Krasnov:2016emc,Adamo:2016rtr,Albrychiewicz:2021ndv,Miller:2024oza}. There has also been progress in all-multiplicity formulae for certain scalar correlators in the framework of scattering equations\cite{Roehrig:2020kck,Eberhardt:2020ewh,Gomez:2021qfd}. We note that we have included in an appendix calculations up to six points in SDYM and up to five points in SDG, which will hopefully prove to be useful data points for unveiling general structures and for constructing a holographic dual to SDG and it higher spin extensions in AdS \cite{Skvortsov:2018uru,Jain:2024bza,Aharony:2024nqs}.

Finally, the lightcone formalism is well suited to compute correlators at loop level, due to the absence of ghost contributions. SDYM and SDG in (A)dS$_4$ are expected from the Feynman rules to be perturbatively one-loop exact, because one cannot draw higher-loop diagrams with the vertex of the self-dual sector. We can try to extend the lightcone flat-space computations of \cite{Boels:2013bi} (see also \cite{Brandhuber:2006bf}) using the Feynman rules we developed. Loop effects in (A)dS are a thorny subject, and the self-dual theories may provide an insightful toy model. In fact, motivated by recent flat-space results \cite{Costello:2021bah,Bittleston:2022nfr,Monteiro:2022nqt}, a conjecture was made in \cite{Doran:2023cmj}---in particular (5.6) and (5.7) in that article---that would suggest a relatively simple form of the self-dual one-loop (A)dS correlators.

\subsection*{Acknowledgements}

We are thankful to Tim Adamo, Graham Brown, Wei Bu, Sachin Jain, Kirill Krasnov, Alok Laddha, Lionel Mason, Scott Melville, Radu Moga, David Skinner and Evgeny Skvortsov for discussions. CC is supported by the STFC consolidated grant (ST/X000583/1) ``New Frontiers in Particle Physics, Cosmology and Gravity''. RM is supported by the Royal Society via a University Research Fellowship. GD is supported by the Royal Society via a studentship grant. A.L. and S.N. are
supported by an STFC Consolidated Grant ST/T000708/1.

\begin{appendix}

\section{Boundary terms}\label{app:boundary}
In section \ref{sec:lightcone}, we reviewed how to derive a  lightcone action for YM written in terms of positive and negative helicity fields. While performing integrations by parts, one accumulates several boundary terms at $z = 0$. Together with the bulk action, these form the definition of the theory. In this appendix, we list the set of boundary terms one accumulates from the original action in lightcone gauge, namely expression \eqref{eq:F2lightcone}, to the simplified expression \eqref{eq:YMlightcone}. One can perform a similar (but much more complicated) analysis for gravity. 

The set of terms accumulated at the boundary of the Poincare patch, $z = 0$, in the case of YM, is:
\begin{eqn}\label{bndySDYM}
S_{\p} &= \frac12  \tr \int_{z = 0} d^3 x \Bigg[ A_w \p_z A_{\bar w} + A_v (\p_w A_{\bar w} + \p_{\bar w} A_w) - i A_v \Big( [\p_u A_w, \frac{1}{\p_u} A_{\bar w}] + [\p_u A_{\bar w}, \frac{1}{\p_u} A_{w}]  \Big) \\
& \qquad -  \frac{1}{\p_u^2} \big( [A_{\bar w}, \p_u A_w] + [A_{w}, \p_u A_{\bar w}] \big) \frac{1}{\p_u} \big( [A_{\bar w}, \p_u A_w] + [ A_{w}, \p_u A_{\bar w}] \big)\\
&\qquad -  [A_w, A_{\bar w}] \frac{1}{\p_u} [A_w, A_{\bar w}] -  \frac{1}{\p_u} [A_{\bar w}, \p_u A_w] \frac{1}{\p_u^2} [A_w, \p_u A_{\bar w}] +  \frac{1}{\p_u^2} [A_{\bar w}, \p_u A_w] \frac{1}{\p_u} [A_w, A_{\bar w}] \\
&\qquad -4i \Big( \p_w A_{\bar w} [A_{\bar w}, \p_u A_w] + \p_{\bar w} A_{w} [A_{w}, \p_u A_{\bar w}] \Big)\Bigg]\,,
\end{eqn}
where $d^3 x = dt\, dx\, dy$. Notice that none of the terms depend on $A_w A_w A_w$ or $A_{\wb} A_{\wb} A_{\wb}$, and hence we do not obtain any contribution to the all-plus (or all-minus) helicity amplitude, even from the boundary terms, with our definition of helicity. In order to use the effective action \eqref{eq:YMlightcone} for YM---and therefore the associated action for SDYM---, we need to add a boundary term that exactly cancels \eqref{bndySDYM}. Together with the original action for YM \eqref{eq:F2lightcone}, this set of boundary terms forms the definition of the theory which results in the effective action \eqref{eq:YMlightcone}. We note that, as usual in light-cone gauge, we have performed certain operations up to boundary terms, such as \,$\int f\frac{1}{\partial_u}g = -\int g\frac{1}{\partial_u}f $\,.

\section{Alternative light-cone coordinates}
\label{appendix:othercoordchoice}

In section~\ref{subsec:LCactiongrav}, we derived the action for Einstein gravity in AdS in terms of positive and negative helicity fields, $h$ and $\bar h$. There, in order to match the conventions of \cite{Lipstein:2023pih}, we took the lightcone directions $(u,v)$ to depend on the bulk coordinate $z$. Here, we will instead consider an alternative choice where $(u,v)$ are boundary coordinates:
\begin{equation}
\label{eq:AdSmetric}
    ds^2_{AdS}= \frac{R^2}{z^2}(-du dv + (dx^1)^2 + dz^2).
\end{equation}
As a result, the lightcone gauge condition in \eqref{eq:LCgauge} will set to zero boundary components of the metric rather than bulk components. In the choice we made previously, there is a non-trivial interplay between the inverse derivatives with respect to $u$, on the one hand, and the non-translationally-invariant direction $z$ of the Poincar\'e coordinates. That is the main motivation for considering in this Appendix the other choice, where $z$ and $u$ are not intertwined. Then inverse derivatives with respect to $u$ present no difficulties (e.g.~considering Fourier space), as this is a translationally-invariant direction in AdS. This is likely to be a more convenient choice for future work. In fact, the derivation of the AdS light-cone action simplifies considerably.

Starting with the AdS metric given by \eqref{eq:AdSmetric}, the expression \eqref{eq:AdSZeroPerturbationsx3} for AdS still holds, now with $x^i=(x^1,z)$, and the fields $G$ and $F$ are still defined to deviate from the AdS values as in \eqref{eq:AdSpsiphix3x}. However, in the coordinates we use here, the constraint \eqref{eq:Ruu0} is given by
\begin{equation}
\label{eq:giuuConditionAdS}
     \partial_u \tilde{F} \partial_u \tilde{G} - \half (\partial_u \tilde{G})^2 - \partial_u^2 \tilde{G} +\frac{1}{4} \partial_u \gamma^{ij} \partial_u \gamma_{ij}=0\,.
\end{equation}
This is the same expression as for $\Lambda=0$, and it should be contrasted with \eqref{eq:giuuConditionAdSx3}. Hence, in the coordinates we now use, the gauge choice
\begin{equation}
    \tilde{F} =\half\, \tilde{G}
\end{equation}
leads to
\begin{equation}
\label{eq:AdSSpacepsiSolution}
     \tilde{G} =\frac{1}{4}\frac{1}{\partial_u^2}(\partial_u \gamma^{ij} \partial_u \gamma_{ij}).
\end{equation}

The action \eqref{eq:EHLCaction} can now be expanded in $\kappa$. Integrating by parts several times, we find
\begin{equation}
    \label{eq:EHLCactionhhbarAdS0}
    \begin{aligned}
     S_G|_{\kappa^0}= \int d^4 x & \sqrt{|g_{AdS}|}\left( -g^{\mu \nu}_{AdS} \partial_\mu h \partial_\nu \bar{h} +\frac{2}{R^2} h \bar{h}\right)
    \end{aligned}
\end{equation}
and
\begin{equation}
    \label{eq:EHLCactionhhbarAdS1}
    \begin{aligned}
     S_G|_{\kappa^1}= \frac{\kappa}{\sqrt{2}} \int d^4 x \;\; & \partial_u^2\bar{h}\; \frac{R^2}{z^2}\, \left[ \left(\frac{\partial_w}{\partial_u}h\frac{\partial_w}{\partial_u}h-h\frac{\partial_w^2}{\partial_u^2}h \right) + \frac{2i}{z}\left(\frac{1}{\partial_u}h\frac{\partial_w}{\partial_u}h-h\frac{\partial_w}{\partial_u^2}h \right) \right.\\
      &\left. + \frac{1}{z^2}\left(h\frac{1}{\partial_u^2}h-\frac{3}{4}\frac{1}{\partial_u}h\frac{1}{\partial_u}h\right) \right]
      + C.C.\,,
    \end{aligned}
\end{equation}
where $C.C.$ denotes the complex conjugate contribution $h\bar h\bar h$. Notice that we now defined
\begin{equation}
    \label{eq:DoubleNullCoordsA1}
    w= x^1+iz\,,\quad \bar{w} = x^1-iz\,.
\end{equation}
The terms of order $\kappa^2$ and higher obey the same properties as in section~\ref{subsec:LCactiongrav}, i.e.,~only $\bar h\bar h \cdots hh$ contributions occur.

Restricting now to the self-dual sector, where we keep the terms linear in $\bar h$, the expression simplifies if we employ the field redefinition
\begin{equation}
    h=\sqrt{2}\,\partial_u^2 \phi\,, \qquad \bar{h}=\sqrt{2}\,\frac{1}{\partial_u^2} \bar\phi\,.
\end{equation}
This is the same field redefinition as for $\Lambda=0$, and it should be contrasted with \eqref{eq:hphiAdS}. Notwithstanding these differences to the coordinate choice made in section~\ref{subsec:LCactiongrav}, we still find the action for self-dual gravity to be given as
\begin{equation}
    S_{SDG}= 2\int d^4 x \left[  \sqrt{|g_{AdS}|} \bar{\phi}\left( \square_{AdS} +\frac{2}{R^2} \right) \phi-4\kappa  \bar\phi  \{\{\frac{R}{u-v}\phi,\frac{R}{u-v}\phi\}\}_*\right],
\end{equation}
just like \eqref{eq:EHLCactionSelfDualAdSBracketx3},
with the important distinction that now the double-bracket $ \{\{\cdot,\cdot\}\}_*$ is defined in terms of 
\begin{equation}
    \Pi =(\Pi_v,\Pi_{\bar w})=({\partial}_{w},\partial_u)\,, \qquad    \tilde{\Pi} =(\tilde{\Pi}_v,\tilde{\Pi}_{\bar{w}})=(\partial_{w}-\frac{4}{w-\bar{w}},\partial_u)\,,
\end{equation}
as opposed to \eqref{eq:tracePi1}.

Similarly, the action for self-dual Yang-Mills theory in AdS still takes the form \eqref{lag_sdym} if we use the coordinate/gauge choice in this appendix, but now with the definition \eqref{eq:DoubleNullCoordsA1}.

\section{Higher points}\label{app:5ptSDG}

In this Appendix we extend the calculations in section \ref{sec:correlators} to higher points. In particular, we go up to six points in SDYM and five points in SDG.

\subsection{SDYM}

Let us begin with SDYM.

\subsubsection*{5 points}

Two types of diagrams can arise at 5 points, which are depicted in \eqref{5ptSDYM1} and \eqref{5ptSDYM2}. The first type of diagram has a negative helicity field at one of the external vertices. This can be evaluated using the same methods as done in the main text and the final answer is given as 
\begin{eqn}\label{5ptSDYM1}
\begin{tikzpicture}[baseline]
\draw[very thick] (0, 0) circle (2.21);
\draw (-2, 1) -- (-1, 0);
\draw (-2, -1) -- (-1, 0);
\draw (-1, 0) -- (0, 0);
\draw (0,0) -- (0,2.21);
\draw (0, 0) -- (1, 0);
\draw (2, 1) -- (1, 0);
\draw (2, -1) -- (1, 0);

\node at (-2.2, -1.25) {$1^+$};
\node at (-2.2, 1.25) {$2^+$};
\node at (0, 2.5) {$3^+$};
\node at (2.2, 1.25) {$4^+$};
\node at (2.2, -1.25) {$5^-$};
\end{tikzpicture}
&= 
X_{1,2} X_{4,5} \Big[2 (X_{3, 1} + X_{3, 2}) \begin{tikzpicture}[baseline]
\draw (-0.5, 0) -- (0.5, 0);
\node at (-0.5, 0) {\textbullet};
\node at (0, 0) {\textbullet};
\node at (0.5, 0) {\textbullet};
\end{tikzpicture}
+ i k_{3w}
\begin{tikzpicture}[baseline]
\draw (0.5, 0) -- (0, 0);
\node at (-0.5, 0) {\textbullet};
\node at (0, 0) {\textbullet};
\node at (0.5, 0) {\textbullet};
\end{tikzpicture}\Big],
\end{eqn}
where we use the same shorthand notation introduced in the main text. The first term in the bracket on the right-hand-side can be expressed in terms of the lower-point functions via the following recursion relation \cite{Arkani-Hamed:2017fdk}:
\begin{eqn*}
 \begin{tikzpicture}[baseline]
\draw (-0.5, 0) -- (0.5, 0);
\node at (-0.5, 0) {\textbullet};
\node at (0, 0) {\textbullet};
\node at (0.5, 0) {\textbullet};
\end{tikzpicture} = \frac{1}{k_{12345}} \Big[ \begin{tikzpicture}[baseline]
\draw (0.5, 0) -- (0, 0);
\node at (-0.5, 0) {\textbullet};
\node at (0, 0) {\textbullet};
\node at (0.5, 0) {\textbullet};
\end{tikzpicture}
+ 
\begin{tikzpicture}[baseline]
\draw (0, 0) -- (0.5, 0);
\node at (-0.5, 0) {\textbullet};
\node at (0, 0) {\textbullet};
\node at (0.5, 0) {\textbullet};
\end{tikzpicture}
\Big].
\end{eqn*}
This representation makes the flat space limit manifest and we note that only the first term on the right-hand-side in \eqref{5ptSDYM1} will contribute to the flat space limit, which is given as
\begin{eqn}
\lim_{k_{12345}\rightarrow 0} k_{12345} \begin{tikzpicture}[baseline]
\draw[very thick] (0, 0) circle (2.21);
\draw (-2, 1) -- (-1, 0);
\draw (-2, -1) -- (-1, 0);
\draw (-1, 0) -- (0, 0);
\draw (0,0) -- (0,2.21);
\draw (0, 0) -- (1, 0);
\draw (2, 1) -- (1, 0);
\draw (2, -1) -- (1, 0);

\node at (-2.2, -1.25) {$1^+$};
\node at (-2.2, 1.25) {$2^+$};
\node at (0, 2.5) {$3^+$};
\node at (2.2, 1.25) {$4^+$};
\node at (2.2, -1.25) {$5^-$};
\end{tikzpicture}
&= \frac{2 X_{3, 1 + 2} X_{1,2} X_{4,5}}{s_{12} s_{123}},
\end{eqn}
where $s_{12}$ and $s_{123}$ are the Mandelstam's in flat space\footnote{$s_{i\cdots j} = (k_i + \cdots + k_j)^2 - (\vec k_i + \cdots + \vec k_j)^2$}. This matches the expected result in flat space which can be found in equation (41) of \cite{Monteiro:2011pc}. 

The second type of diagram that can appear at 5 points has a negative helicity field at the middle vertex. Such a diagram evaluates to
\begin{eqn}
\label{5ptSDYM2}
\begin{tikzpicture}[baseline]
\draw[very thick] (0, 0) circle (2.21);
\draw (-2, 1) -- (-1, 0);
\draw (-2, -1) -- (-1, 0);
\draw (-1, 0) -- (0, 0);
\draw (0,0) -- (0,2.21);
\draw (0, 0) -- (1, 0);
\draw (2, 1) -- (1, 0);
\draw (2, -1) -- (1, 0);

\node at (-2.2, -1.25) {$1^+$};
\node at (-2.2, 1.25) {$2^+$};
\node at (0, 2.5) {$3^-$};
\node at (2.2, 1.25) {$4^+$};
\node at (2.2, -1.25) {$5^+$};
\end{tikzpicture}
&= - X_{1,2} X_{4,5} \Big[X_{1+2, 4+5} \begin{tikzpicture}[baseline]
\node at (-0.5, 0) {\textbullet};
\node at (0, 0) {\textbullet};
\node at (0.5, 0) {\textbullet};
\draw (-0.5, 0) -- (0.5, 0);
\end{tikzpicture} 
+ \frac{i}{2}  y_{2w} 
\begin{tikzpicture}[baseline]
\node at (-0.5, 0) {\textbullet};
\node at (0, 0) {\textbullet};
\node at (0.5, 0) {\textbullet};
\draw (0, 0) -- (0.5, 0);
\end{tikzpicture}
+ \frac{i}{2} y_{1w}
\begin{tikzpicture}[baseline]
\node at (-0.5, 0) {\textbullet};
\node at (0, 0) {\textbullet};
\node at (0.5, 0) {\textbullet};
\draw (-0.5, 0) -- (0, 0);
\end{tikzpicture} \Big] 
\end{eqn}
This expression again makes the flat space limit manifest as the only first term contributes to it, resulting in
\begin{eqn}
\lim_{k_{12345}\rightarrow 0} k_{12345}
\begin{tikzpicture}[baseline]
\draw[very thick] (0, 0) circle (2.21);
\draw (-2, 1) -- (-1, 0);
\draw (-2, -1) -- (-1, 0);
\draw (-1, 0) -- (0, 0);
\draw (0,0) -- (0,2.21);
\draw (0, 0) -- (1, 0);
\draw (2, 1) -- (1, 0);
\draw (2, -1) -- (1, 0);

\node at (-2.2, -1.25) {$1^+$};
\node at (-2.2, 1.25) {$2^+$};
\node at (0, 2.5) {$3^-$};
\node at (2.2, 1.25) {$4^+$};
\node at (2.2, -1.25) {$5^+$};
\end{tikzpicture}
&= \frac{-2 X_{1+2, 4+5} X_{1,2} X_{4,5}}{s_{12} s_{123}}~,
\end{eqn}
which matches with the result in \cite{Monteiro:2011pc} 


\subsubsection*{Six points}\label{APP_6_PT}

Higher point diagrams have a recursive structure in terms of stick graphs. For example, the following 6-point diagram in SDYM can be expressed as
\begin{eqn}
\begin{tikzpicture}[baseline]
\draw[very thick] (0, 0) circle (2.21);

\draw (-2, 1) -- (-1.5, 0);
\draw (-2, -1) -- (-1.5, 0);
\draw (-1.5, 0) -- (0, 0);
\draw (-0.5,0) -- (-0.5,2.18);
\draw (0.5,0) -- (0.5,2.18);
\draw (0, 0) -- (1.5, 0);
\draw (2, 1) -- (1.5, 0);
\draw (2, -1) -- (1.5, 0);

\node at (-2.2, -1.25) {$1^+$};
\node at (-2.2, 1.25) {$2^+$};
\node at (-0.5, 2.5) {$3^+$};
\node at (0.5, 2.5) {$4^+$};
\node at (2.2, 1.25) {$5^+$};
\node at (2.2, -1.25) {$6^-$};

\node at (-1, -0.25) {$y_1$};
\node at (-0, -0.25) {$y_2$};
\node at (1, -0.25) {$y_3$};
\end{tikzpicture} 
= X_{1,2} X_{5,6} \Bigg\{ f_1 f_3 F_6 + f_2 f_3 F_6^{(1)}  + f_1 f_4 F_6^{(2)} + f_2 f_4 F_6^{(3)} \Bigg\},
\end{eqn}
where
\begin{eqn}
F_6 &= 
\begin{tikzpicture}[baseline]
\node at (-2, 0) {\textbullet};
\node at (-1, 0) {\textbullet};
\node at (0, 0) {\textbullet};
\node at (1, 0) {\textbullet};
\draw (-2, 0) -- (1, 0);
\end{tikzpicture} \\
F^{(1)}_6 &= - k_{12} F_6 + 
\begin{tikzpicture}[baseline]
\node at (-2, 0) {\textbullet};
\node at (-1, 0) {\textbullet};
\node at (0, 0) {\textbullet};
\node at (1, 0) {\textbullet};
\draw (-1, 0) -- (1, 0);
\end{tikzpicture} \\
F^{(2)}_6 &= - k_3 F_6 + F_6^{(1)} + 
\begin{tikzpicture}[baseline]
\node at (-2, 0) {\textbullet};
\node at (-1, 0) {\textbullet};
\node at (0, 0) {\textbullet};
\node at (1, 0) {\textbullet};
\draw (-2, 0) -- (-1, 0);
\draw (0, 0) -- (1, 0);
\end{tikzpicture} \\
F^{(3)}_6 &= k_{12} F_6^{(2)} - (k_3 + y_1)
\begin{tikzpicture}[baseline]
\node at (-2, 0) {\textbullet};
\node at (-1, 0) {\textbullet};
\node at (0, 0) {\textbullet};
\node at (1, 0) {\textbullet};
\draw (-1, 0) -- (1, 0);
\end{tikzpicture} 
+ 
\begin{tikzpicture}[baseline]
\node at (-2, 0) {\textbullet};
\node at (-1, 0) {\textbullet};
\node at (0, 0) {\textbullet};
\node at (1, 0) {\textbullet};
\draw (0, 0) -- (1, 0);
\end{tikzpicture}, 
\end{eqn}
where $f_1 = - i y_{1t} k_{3w} - 2 y_{1w} k_{3u}$, $f_2 = i k_{3w}$, $f_3 = - i y_{2t} k_{4w} - 2 y_{2w} k_{4u}$, $f_4 = i k_{4w} $. In general, any diagram admits a representation where each term is obtained from a stick graph and its sub-graphs.

\subsection{SDG}\label{APP_5_PT_GRAV}

Now let us consider tree-level 5-point diagrams in SDG. Once again there are two types of diagrams, which are depicted in \eqref{5ptSDG} and \eqref{5ptSDG2}. A diagram with a negative helicity leg at an outer vertex is given by
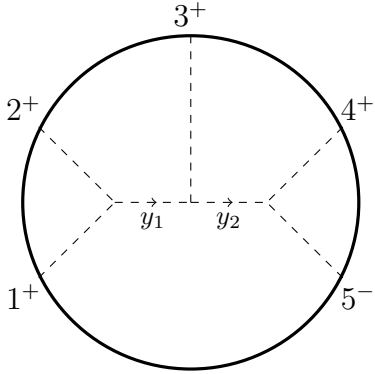
\begin{eqn}\label{5ptSDG}
&\begin{tikzpicture}[baseline]
\draw[very thick] (0, 0) circle (2.21);
\draw[dashed] (-2, 1) -- (-1, 0);
\draw[dashed] (-2, -1) -- (-1, 0);
\draw[fermion, black, dashed] (-1, 0) -- (0, 0);
\draw[dashed] (0,0) -- (0,2.21);
\draw[fermion, black, dashed] (0, 0) -- (1, 0);
\draw[dashed] (2, 1) -- (1, 0);
\draw[dashed] (2, -1) -- (1, 0);

\node at (-2.2, -1.25) {$1^+$};
\node at (-2.2, 1.25) {$2^+$};
\node at (0, 2.5) {$3^+$};
\node at (2.2, 1.25) {$4^+$};
\node at (2.2, -1.25) {$5^-$};

\node at (-0.5, -0.25) {\footnotesize $y_1$};
\node at (0.5, -0.25) {\footnotesize $y_2$};
\end{tikzpicture}\\
&=4i X_{1,2} X_{3, 1 + 2} X_{4,5} \hat{\mathcal D}_{1,2} \hat{\mathcal D}_{3} \hat{\mathcal D}_{4,5} 
\begin{tikzpicture}[baseline]

\node at (-1, 0) {\textbullet};
\node at (0, 0) {\textbullet};
\node at (1, 0) {\textbullet};
\draw (-1, 0) -- (1, 0);
\node at (-1, -0.25) {$k_{12}$};
\node at (1, -0.25) {$k_{45}$};
\node at (0, -0.25) {$k_{3}$};
\node at (-0.5, 0.25) {$y_1$};
\node at (0.5, 0.25) {$y_2$};
\end{tikzpicture} \\
&+ X_{1,2} X_{4,5} \hat{\mathcal D}_{1,2}  \hat{\mathcal D}_{4,5}  k_{3w} \Big[  \big\{ 4i X_{3, 1 + 2} + k_{3w} (y_1 - k_{12})\big\} \p_{k_3} - 2 (y_{1w} + k_{3w}) \Big]
\begin{tikzpicture}[baseline]
\node at (-1, 0) {\textbullet};
\node at (0, 0) {\textbullet};
\node at (1, 0) {\textbullet};
\draw (0, 0) -- (1,0);
\node at (-1, 0.25) {$k_{12} + y_1$};
\node at (1, -0.25) {$k_{45}$};
\node at (0, -0.25) {$k_{3} + y_1$};

\node at (0.5, 0.25) {$y_2$};
\end{tikzpicture},
\end{eqn}
where $X_{3, 1 + 2} = X_{3, 1} + X_{3, 2}$ 
 and the differential operators are 
\begin{eqn}\label{5ptSFG-der}
\hat{\mathcal D}_{1,2} &=  X_{1, 2} \p_{k_{12}} + i (k_{1w} -k_{2w})\\
\hat{\mathcal D}_{3} &= X_{3, 1+2} \p_{k_{3}} + i (k_{3w} + y_{1w})\\
\hat{\mathcal D}_{4,5} &= X_{4, 5} \p_{k_{45}} + i  (2k_{4w} + y_{2w}),
\end{eqn}
with boundary momentum conservation implying: \begin{eqn*}
\vec k_1 + \vec k_2 = - \vec y_1, \quad 
\vec y_1 = \vec k_3 + \vec y_2 , \quad 
\vec y_2 = \vec k_4 + \vec k_5.
\end{eqn*}
By comparing with equation \eqref{5ptSDYM1} we see that the term containing the total energy pole admits a similar double copy structure to the 3- and 4-point graphs discussed in section \ref{sec:correlators}. It would be interesting to explore how this structure generalizes to the other subgraphs in \eqref{5ptSDG}. 
The other type of graph which can arise at 5 points has a negative helicity leg in the middle: 
\begin{eqn}\label{5ptSDG2}
&\begin{tikzpicture}[baseline]
\draw[very thick] (0, 0) circle (2.21);
\draw[dashed] (-2, 1) -- (-1, 0);
\draw[dashed] (-2, -1) -- (-1, 0);
\draw[fermion, black, dashed] (-1, 0) -- (0, 0);
\draw[dashed] (0,0) -- (0,2.21);
\draw[fermion, black, dashed] (0, 0) -- (1, 0);
\draw[dashed] (2, 1) -- (1, 0);
\draw[dashed] (2, -1) -- (1, 0);

\node at (-2.2, -1.25) {$1^+$};
\node at (-2.2, 1.25) {$2^+$};
\node at (0, 2.5) {$3^-$};
\node at (2.2, 1.25) {$4^+$};
\node at (2.2, -1.25) {$5^+$};

\node at (-0.5, -0.25) {\footnotesize $y_1$};
\node at (0.5, -0.25) {\footnotesize $y_2$};
\end{tikzpicture}\\
&= \hat{\mathcal D}_{1,2} \hat{\mathcal D}_{4,5} \Bigg( -4 X_{1+2,4+5}\Big[ X_{1+2,4+5} \p_{k_3} - i (y_{1w} + y_{2w})  \Big] \begin{tikzpicture}[baseline]
\node at (-1, 0) {\textbullet};
\node at (0, 0) {\textbullet};
\node at (1, 0) {\textbullet};
\draw (-1, 0) -- (1, 0); 
\end{tikzpicture}\\
&+ y_{1w} \Big[2 (y_{1w} + y_{2w}) +  \big\{ 4i  X_{1+2,4+5} + y_{2w} k_{12345} \big\} \p_{k_3}\Big]
\begin{tikzpicture}[baseline]
\node at (-1, 0) {\textbullet};
\node at (0, 0) {\textbullet};
\node at (1, 0) {\textbullet};
\draw (0, 0) -- (1, 0); 
\end{tikzpicture} \\
&+ y_{2w} \Big[2 (y_{1w} + y_{2w}) +  \big\{ 4i  X_{1+2,4+5} + y_{1w}k_{12345}  \big\} \p_{k_3}\Big]
\begin{tikzpicture}[baseline]
\node at (-1, 0) {\textbullet};
\node at (0, 0) {\textbullet};
\node at (1, 0) {\textbullet};
\draw (-1, 0) -- (0, 0); 
\end{tikzpicture}  \Bigg)
\end{eqn}

The compact results presented above were obtained using an integration by parts technique, which we spell out below. Let us illustrate how this works for the diagram in \eqref{5ptSDG} (similar comments apply to the diagram in \eqref{5ptSDG2}). Using the Feynman rules in section \ref{sdgfeynman}, the digram in \eqref{5ptSDG} is given as
\begin{eqn}
&\begin{tikzpicture}[baseline]
\draw[very thick] (0, 0) circle (2.21);
\draw[dashed] (-2, 1) -- (-1, 0);
\draw[dashed] (-2, -1) -- (-1, 0);
\draw[fermion, black, dashed] (-1, 0) -- (0, 0);
\draw[dashed] (0,0) -- (0,2.21);
\draw[fermion, black, dashed] (0, 0) -- (1, 0);
\draw[dashed] (2, 1) -- (1, 0);
\draw[dashed] (2, -1) -- (1, 0);

\node at (-2.2, -1.25) {$1^+$};
\node at (-2.2, 1.25) {$2^+$};
\node at (0, 2.5) {$3^+$};
\node at (2.2, 1.25) {$4^+$};
\node at (2.2, -1.25) {$5^-$};

\node at (-0.5, -0.25) {\footnotesize $y_1$};
\node at (0.5, -0.25) {\footnotesize $y_2$};
\end{tikzpicture} \\
&= \intsinf dz_1 dz_2 dz_3 V_{z_1} (\phi_1, \phi_2) V_{z_2}(G_{12}, \phi_3) V_{z_3}(G_{23}, \phi_4) z_3 e^{- k_5 z_3} \equiv F_{5, 1}^{SDG}.
\end{eqn}
Momentum conservation gives us 
\begin{eqn}
\vec k_1 + \vec k_2 = - \vec y_1, \quad 
\vec y_1 = \vec k_3 + \vec y_2 , \quad 
\vec y_2 = \vec k_4 + \vec k_5.
\end{eqn}

\subsection*{Vertex Factors}

The vertex factors are given by
\begin{eqn}
V_{z_1}(\phi_1, \phi_2) &= \frac{e^{i k_1 \cdot x_1} e^{i k_2 \cdot x_2}}{2z_1} X_{1,2}\, \big(z_1 X_{1,2} - i (k_{1w} - k_{2w})  \big),
\end{eqn}

\begin{eqn}
&V_{z_2}(G_{12}, \phi_3) \\
&= \doublecomm{ \frac{G_{12}}{z_2}, \frac{\phi_3}{z_2}}_*\\
 &= z_1  e^{i \vec y_1 \cdot (\vec x_1 - \vec x_2)} e^{- k_3 z_2 + i \vec k_3 \cdot \vec x_2} \intinf \frac{d\omega}{\omega^2 + y_1^2} \sin(\omega z_1) \\
&\times \frac{1}{16 z_2} \Bigg[ \sin(\omega z_2) \big\{ 2 \omega^2 z_2 k_{3 w}^2-2 \left(y_{1t} k_{3 w}-2 i k_{3 u} y_{1w}\right) \left(z_2 y_{1t} k_{3 w}-2 i z_2 k_{3 u} y_{1w} +2 y_{1w}+2 k_{3 w}\right) \big\} \\
&\qquad + \omega \cos(\omega z_2) \big\{ 4  k_{3 w} \left(z_2 y_{1t} k_{3 w}-2 i z_2 k_{3 u} y_{1w} + y_{1w} +k_{3 w}\right) \big\} \Bigg] \\
&= z_1  e^{i \vec y_1 \cdot (\vec x_1 - \vec x_2)} e^{- k_3 z_2 + i \vec k_3 \cdot \vec x_2} \intinf \frac{d\omega}{\omega^2 + y_1^2} \sin(\omega z_1) \\
&\times \frac{1}{8 z_2} \Bigg[ \sin(\omega z_2) \big\{ \omega^2 z_2 k_{3w}^2 - c_1(z_2 c_1 + 2 c_2) \big\}  + 2k_{3w} \omega \cos(\omega z_2) \big\{ z_2 c_1 + c_2 \big\} \Bigg], 
\end{eqn}
where $c_1= y_{1t} k_{3w} - 2 i k_{3u} y_{1w} = (2 i X_{3, 1 + 2} - k_{3w} k_{12} )$ and $c_2=  k_{3w} + y_{1w} $,

\begin{eqn}
&V_{z_3}(G_{23}, \phi_4) \\
 &= z_2  e^{i \vec y_2 \cdot (\vec x_2 - \vec x_3)} e^{- k_4 z_3 + i \vec k_4 \cdot \vec x_3} \intinf \frac{d\omega}{\omega^2 + y_2^2} \sin(\omega z_2) \\
&\times \frac{1}{8 z_3} \Bigg[ \sin(\omega z_3) \big\{ \omega^2 z_3 k_{4w}^2 - c_3(z_3 c_3 + 2 c_4) \big\}  + 2k_{4w} \omega \cos(\omega z_3) \big\{ z_3 c_3 + c_4 \big\} \Bigg], 
\end{eqn}
where $c_3= (-2 i X_{4,5} + k_{4w} k_{45} )$ and $c_4=( k_{5w}+2k_{4 w})$.

\subsection*{$z$ integrals}
By directly performing the $z_1$ and $z_2$ integrals, we obtain the following (up to overall numerical factors): 
\begin{eqn}
F_{5, 1}^{SDG} &= X_{1, 2} X_{4, 5}\intsinf dz_1 dz_3 e^{- k_{12} z_1} e^{- k_{45} z_3}  \\
&\times \intinf \frac{d\omega d\omega' \omega \omega'}{(\omega^2 + y_1^2) (\omega^2 + y_2^2)} \Bigg\{ \frac{2i k_{12} X_{1,2}}{\omega^2 + k_{12}^2} + (2 k_{1w} + y_{1w}) \Bigg\} \Bigg\{ \frac{2i k_{45} X_{4,5}}{\omega'^2 + k_{45}^2} + (2 k_{4w} + y_{2w}) \Bigg\} \\
&\times \frac{1}{(\omega^2 + k_{12}^2)(\omega'^2 + k_{45}^2)} \\
&\times\intsinf  dz_2 e^{- k_3 z_2}  \Bigg[ \sin(\omega z_2) \big\{ \omega^2 z_2 k_{3w}^2 - c_1(z_2 c_1 + 2 c_2) \big\}  + 2k_{3w} \omega \cos(\omega z_2) \big\{ z_2 c_1 + c_2 \big\} \Bigg]  \sin(\omega' z_2) 
\end{eqn}
The left most and the right most vertex can be easily expressed in terms of the differential operators, and having done that we are left with 
\begin{eqn}
F_{5, 1}^{SDG} &= X_{1,2} X_{4, 5} \hat{\mathcal D}_{1,2}  \hat{\mathcal D}_{4, 5}  \intsinf dz_1 dz_2 dz_3e^{- k_{12} z_1} e^{- k_3 z_2} e^{- k_{45} z_3} \\
&\times G_{23} \left\{  \Big( - z_2 k_{3w}^2 \p_{z_2}^2 - c_1(z_2 c_1 + 2 c_2) \Big)  + 2k_{3w} \Big( z_2 c_1 + c_2 \Big) \p_{z_2}  \right\} G_{12},   
\end{eqn}
where the differential operators are defined in \eqref{5ptSFG-der}. We thus are left with the following five integrals, which are conveniently evaluated using IBP:
\begin{enumerate}
\item 
\begin{eqn}
& \intsinf dz_1 dz_2 dz_3e^{- k_{12} z_1} e^{- k_3 z_2} e^{- k_{45} z_3}  z_2 (- k_{3w}^2 \p_{z_2}^2 G_{12}) G_{23} \\
&= k_{3w}^2 \p_{k_3} \intsinf dz_1 dz_2 dz_3 e^{- k_{12} z_1} e^{- k_3 z_2} e^{- k_{45} z_3} (\p_{z_2}^2 G_{12} )G_{23}  \\
&= k_{3w}^2 \p_{k_3} \Big[k_{12}^2  \begin{tikzpicture}[baseline]
\node at (-1, 0) {\textbullet};
\node at (0, 0) {\textbullet};
\node at (1, 0) {\textbullet};
\draw (-1, 0) -- (1,0);
\end{tikzpicture}
- (k_{12} + y_1) 
\begin{tikzpicture}[baseline]
\node at (-1, 0) {\textbullet};
\node at (0, 0) {\textbullet};
\node at (1, 0) {\textbullet};
\draw (0, 0) -- (1,0);
\end{tikzpicture}
\Big],
\end{eqn}
where the last step follows from the relation 
\begin{eqn*}
&\intsinf dz_1 dz_2 dz_3 e^{- k_{12}z_1} e^{- k_{3}z_2} e^{- k_{45}z_3} \p_{z_2}^2 G_{12} G_{23} \\
&= k_{12}^2 \begin{tikzpicture}[baseline]
\node at (-1, 0) {\textbullet};
\node at (0, 0) {\textbullet};
\node at (1, 0) {\textbullet};
\draw (-1, 0)-- (1,0);
\end{tikzpicture}
- (k_{12} + y_1) 
\begin{tikzpicture}[baseline]
\node at (-1.5, 0) {\textbullet};
\node at (-0.5, 0) {\textbullet};
\node at (0.5, 0) {\textbullet};
\draw (-0.5, 0)-- (0.5,0);
\end{tikzpicture}
\end{eqn*}

\item 
\begin{eqn}
& \intsinf dz_1 dz_2 dz_3 e^{- k_{12} z_1} e^{- k_3 z_2} e^{- k_{45} z_3}  z_2 (- c_1^2) G_{12} G_{23} = c_1^2 \p_{k_3} 
\begin{tikzpicture}[baseline]
\node at (-1, 0) {\textbullet};
\node at (0, 0) {\textbullet};
\node at (1, 0) {\textbullet};
\draw (-1, 0) -- (1,0);
\end{tikzpicture}
\end{eqn}

\item 
\begin{eqn}
& \intsinf dz_1 dz_2 dz_3 e^{- k_{12} z_1} e^{- k_3 z_2} e^{- k_{45} z_3}  (- 2 c_1 c_2) G_{12} G_{23} = -2c_1 c_2 \begin{tikzpicture}[baseline]
\node at (-1, 0) {\textbullet};
\node at (0, 0) {\textbullet};
\node at (1, 0) {\textbullet};
\draw (-1, 0) -- (1,0);
\end{tikzpicture}
\end{eqn}

\item 
\begin{eqn}
& \intsinf dz_1 dz_2 dz_3 e^{- k_{12} z_1} e^{- k_3 z_2} e^{- k_{45} z_3}  (2 k_{3w}) (c_1 z_2) (\p_{z_2} G_{12} )G_{23} \\
&= - (2 k_{3w} c_1 \p_{k_3}) \intsinf dz_1 dz_2 dz_3 e^{- k_{12} z_1} e^{- k_3 z_2} e^{- k_{45} z_3}   (\p_{z_2} G_{12} )G_{23}  \\
&=- 2k_{3w} c_1  \p_{k_3} \Big[ - k_{12}  \begin{tikzpicture}[baseline]
\node at (-1, 0) {\textbullet};
\node at (0, 0) {\textbullet};
\node at (1, 0) {\textbullet};
\draw (-1, 0) -- (1,0);
\end{tikzpicture}
+ 
\begin{tikzpicture}[baseline]
\node at (-1, 0) {\textbullet};
\node at (0, 0) {\textbullet};
\node at (1, 0) {\textbullet};
\draw (0, 0) -- (1,0);
\end{tikzpicture}
\Big],
\end{eqn}
where the last equality follows from 
\begin{eqn*}
&\intsinf dz_1 dz_2 dz_3 e^{- k_{12} z_1} e^{- k_{3} z_2} e^{- k_{45} z_3} \p_{z_1} G_{12} G_{23} =- k_{12}\begin{tikzpicture}[baseline]
\node at (-1, 0) {\textbullet};
\node at (0, 0) {\textbullet};
\node at (1, 0) {\textbullet};
\draw (-1, 0)-- (1,0);
\end{tikzpicture}
+ 
\begin{tikzpicture}[baseline]
\node at (-1.5, 0) {\textbullet};
\node at (-0.5, 0) {\textbullet};
\node at (0.5, 0) {\textbullet};
\draw (-0.5, 0)-- (0.5,0);
\end{tikzpicture}
\end{eqn*}

\item 
\begin{eqn}
& \intsinf dz_1 dz_2 dz_3 e^{- k_{12} z_1} e^{- k_3 z_2} e^{- k_{45} z_3}  (2 k_{3w}) (c_2) (\p_{z_2} G_{12} )G_{23} \\
&= 2 k_{3w} c_2 \Big[ - k_{12} \begin{tikzpicture}[baseline]
\node at (-1, 0) {\textbullet};
\node at (0, 0) {\textbullet};
\node at (1, 0) {\textbullet};
\draw (-1, 0) -- (1,0);
\end{tikzpicture}
+ \begin{tikzpicture}[baseline]
\node at (-1, 0) {\textbullet};
\node at (0, 0) {\textbullet};
\node at (1, 0) {\textbullet};
\draw (0, 0) -- (1,0);
\end{tikzpicture}
\Big]
\end{eqn}

\end{enumerate}

Combining everything (1 + 2 + 3 + 4 + 5) we get 
\begin{eqn}
&=  (c_1 + k_{3w} k_{12}) \big\{(c_1 +  k_{3w} k_{12})\p_{k_3} - 2 c_2 \big\} \begin{tikzpicture}[baseline]
\node at (-1, 0) {\textbullet};
\node at (0, 0) {\textbullet};
\node at (1, 0) {\textbullet};
\draw (-1, 0) -- (1,0);
\end{tikzpicture}\\
&- k_{3w} \Big[ \big\{ k_{3w} (k_{12} + y_1) + 2 c_1 \big\} \p_{k_3} - 2 c_2 \Big]
\begin{tikzpicture}[baseline]
\node at (-1, 0) {\textbullet};
\node at (0, 0) {\textbullet};
\node at (1, 0) {\textbullet};
\draw (0, 0) -- (1,0);
\end{tikzpicture},
\end{eqn}
where $c_1 = 2 i X_{3, 1+2} - k_{3w} k_{12}$ and $c_2 = k_{3w} + y_{1w}$. Substituting this we obtain a simplified form
\begin{eqn}
&= 4i  X_{3, 1+ 2}\big\{i X_{3, 1 + 2}\p_{k_3} -   (y_{1w} + k_{3w}) \big\} \begin{tikzpicture}[baseline]
\node at (-1, 0) {\textbullet};
\node at (0, 0) {\textbullet};
\node at (1, 0) {\textbullet};
\draw (-1, 0) -- (1,0);
\end{tikzpicture}\\
&- k_{3w} \Big[ \big\{ 4i X_{3, 1 + 2} + k_{3w} (y_1 - k_{12})\big\} \p_{k_3} - 2 (y_{1w} + k_{3w}) \Big]
\begin{tikzpicture}[baseline]
\node at (-1, 0) {\textbullet};
\node at (0, 0) {\textbullet};
\node at (1, 0) {\textbullet};
\draw (0, 0) -- (1,0);
\end{tikzpicture}
\end{eqn}
In the first line above we can recognize the term inside the curly bracket $\{i X_{3, 1 + 2}\p_{k_3} -  (y_{1w} + k_{3w})\}$ as $\hat{ \mathcal D}_{3}$ which appears in \eqref{5ptSDG} and is given in \eqref{5ptSFG-der}. Hence the final answer for 5-point graphs can be expressed in terms of differential operators given in \eqref{5ptSFG-der} acting on scalar graphs, in a similar manner as the four-point graphs in \ref{sec:correlators}.

\section{Spinorial correlators} \label{3ptappendix}

In this Appendix, we will explain how to convert scalar correlators computed in section \ref{sec:correlators} to spinor variables and lift a subset of them to spinning correlators. In particular, we will show that lifting the three-point scalar correlator of SDYM to a spinning one gives the three-point correlator of full YM plus a boundary term. Extending this story to higher points and to SDG will be left for future work. 


\subsection*{Spinor-helicity in lightcone gauge}

First let us describe the spinor-helicity formalism for AdS. Whereas the spinor-helicity formalism is usually implemented in axial gauge (where components along the $z$ direction are set to zero) \cite{Maldacena:2011nz,Raju:2012zs}, we will adapt it to light-cone gauge (where components along a null direction are set to zero). To our knowledge, this has not been done before.

Noting that the 4d Lorentz group is locally $SU(2)\times SU(2)$, a 4d momentum can be written in terms of spinor indices as follows:
\begin{equation}
p^{\alpha\dot{\beta}}=q^{\mu}\sigma_{\mu}^{\alpha\dot{\beta}}=\left(\begin{array}{cc}
iq^{t}+q^{z} & q^{x}+iq^{y}\\
q^{x}-iq^{y} & iq^{t}-q^{z}
\end{array}\right)=\left(\begin{array}{cc}
q^{u} & q^{w}\\
q^{\bar{w}} & q^{v}
\end{array}\right),
\end{equation}
where $\sigma_\mu$  are the Pauli matrices, and $\alpha, \dot{\alpha}$ are spinor indices. In AdS$_{4}$, the $z$ direction is special since momentum is not
conserved along that direction. Converting the unit vector along this
direction to spinor indices gives
\begin{equation}
T_{\,\,\,\,\dot{\beta}}^{\beta}=\sigma_{3}^{\beta\dot{\alpha}}\epsilon_{\dot{\alpha}\dot{\beta}}=\left(\begin{array}{cc}
0 & 1\\
1 & 0
\end{array}\right).\label{eq:T}
\end{equation}
This object can be used to convert dotted indices to undotted indices,
breaking the 4d bulk Lorentz group to the 3d Lorentz group in the boundary \cite{Lipstein:2012kd}. 
To illustrate how this works,
let us consider a generic 4-vector $q^{\alpha\dot{\alpha}}$. After
converting it to undotted indices, its components are explicitly given
by
\begin{equation}
q^{\alpha\beta}=q^{\alpha\dot{\beta}}T_{\,\,\,\,\dot{\beta}}^{\beta}=\left(\begin{array}{cc}
q^{w} & q^{u}\\
q^{v} & q^{\bar{w}}
\end{array}\right)=\left(\begin{array}{cc}
q^{w} & iq^{t}\\
iq^{t} & q^{\bar{w}}
\end{array}\right)+q^{z}\left(\begin{array}{cc}
0 & 1\\
-1 & 0
\end{array}\right).\label{eq:spinorparam}
\end{equation}
From this, we see that the antisymmetric part encodes the $z$ component
while the symmetric part encodes boundary components. 

Recall that a 4d null momentum can be written in bispinor form as
follows:
\begin{equation}
\label{bispinor}
k^{\alpha\dot{\beta}}=\lambda^{\alpha}\bar{\lambda}^{\dot{\beta}}.
\end{equation}
Using \eqref{eq:T} to convert dotted to undotted indices gives 
\begin{equation}
k^{\alpha\beta}=\lambda^{\alpha}\bar{\lambda}^{\beta}=\lambda^{(\alpha}\bar{\lambda}^{\beta)}+\lambda^{[\alpha}\bar{\lambda}^{\beta]}=\lambda^{(\alpha}\bar{\lambda}^{\beta)}+k\epsilon^{\alpha\beta},\label{spinormomentum}
\end{equation}
where $k=\big|\vec{k}\big|=-\frac{1}{2}\epsilon^{\alpha\beta}\lambda_{\alpha}\bar{\lambda}_{\beta}=-\frac{1}{2}\left\langle \lambda\bar{\lambda}\right\rangle $. From this, we see that the boundary momentum is given by $\lambda^{(\alpha}\bar{\lambda}^{\beta)}$
while the energy (which is how we refer to the $z$-component) is given by $k$. Recalling that boundary momentum
is conserved for an $n$-point correlator, we also find that 
\begin{equation}
\sum_{i=1}^{n}\lambda_{i}^{\alpha}\bar{\lambda}_{i}^{\beta}=-k_{1...n}\epsilon^{\alpha\beta}.\label{energynon}
\end{equation}

Let us now turn to polarisation vectors for gluons. In a general gauge, these
are given by
\begin{equation}
\epsilon_{+}^{\alpha\beta}=\frac{\mu^{\alpha}\bar{\lambda}^{\beta}}{\left\langle \mu\lambda\right\rangle },\,\,\,\epsilon_{-}^{\alpha\beta}=\frac{\lambda^{\alpha}\bar{\mu}^{\beta}}{\left\langle \bar{\lambda}\bar{\mu}\right\rangle },\label{polarisations}
\end{equation}
where the momentum of the gluon is $\lambda^{\alpha}\bar{\lambda}^{\dot{\beta}}$ and $\mu$ and $\bar{\mu}$ are reference spinors which encode the gauge choice. A standard choice is axial gauge for which $\mu^{\alpha}=\bar{\lambda}^{\alpha}$
 and $\bar{\mu}^{\alpha}=\lambda^{\alpha}$ \cite{Maldacena:2011nz}. In this case the polarisations are symmetric in the spinor indices,
so they have no $z$-component. In this paper, we instead use light-cone gauge
$0=A_{u}=q^{\mu}A_{\mu}$, where $q^{\mu}=\left(q^{u},q^{v},q^{w},q^{\bar{w}}\right)=\left(1,0,0,0\right)$.
Since $q$ is a null vector, it can be written in bispinor form. Using \eqref{eq:spinorparam}, we can read off the reference spinors:
\begin{equation}
q^{\alpha\beta}=\mu^{\alpha}\bar{\mu}^{\beta}=\left(\begin{array}{cc}
0 & 1\\
0 & 0
\end{array}\right)\rightarrow\mu^{\alpha}=(1,0),\,\,\,\bar{\mu}^{\alpha}=(0,1)\label{referencespinors}\,.
\end{equation}
We shall label external momenta and polarisations with the index index
$i$, i.e.~$k_{i}^{\alpha\beta}=\lambda_{i}^{\alpha}\bar{\lambda}_{i}^{\beta}$.
We then see that in light-cone gauge $\epsilon_{\pm i}\cdot\epsilon_{\pm j}=0$,
so all-plus and all-minus correlators must vanish. This is in contrast to axial gauge, where such correlators are
nonzero. In contrast to
scattering amplitudes, correlators are not field-redefinition
invariant. Under field redefinitions they shift by boundary contact
terms in position space, which can take a nontrivial form in momentum
space. This indicates that all-plus and all-minus correlators are
boundary contact terms. This was previously observed at three points in \cite{Maldacena:2011nz,Jain:2024bza}.

Let us now apply this formalism to the correlators computed above.

\subsection*{2 points}

First let us recall the form of two and three-point correlators for YM and GR in AdS \cite{Maldacena:2011nz,Farrow:2018yni}:
\begin{equation}
\mathcal{A}_{{\rm YM}}^{2{\rm pt}}\propto k_{1}\epsilon_{1}\cdot\epsilon_{2},\,\,\,\mathcal{A}_{{\rm GR}}^{2{\rm pt}}\propto k_{1}^{3}\left(\epsilon_{1}\cdot\epsilon_{2}\right)^{2},
\end{equation}
where we will not be concerned with numerical prefactors. We will
follow the same convention usually followed for scattering amplitudes in flat space,
where the propagator connects particles with opposite helicity. Using \eqref{polarisations}, we see that in lightcone gauge
\begin{equation}
\epsilon_{i}^{+}\cdot\epsilon_{j}^{-}=\frac{\left\langle \mu j\right\rangle \left\langle \bar{i}\bar{\mu}\right\rangle }{\left\langle \mu i\right\rangle \left\langle j\bar{\mu}\right\rangle },
\label{polarasationproduct}
\end{equation}
where the reference spinors are given in \eqref{referencespinors}.
At two points, boundary momentum conservation implies that
\begin{equation}
p_{1}^{\alpha\beta}=\lambda_{1}^{(\alpha}\bar{\lambda}_{1}^{\beta)}=-p_{2}^{\alpha\beta}=-\lambda_{2}^{(\alpha}\bar{\lambda}_{2}^{\beta)},
\label{2ptkinematics}
\end{equation}
so we set $\lambda_{1}=\lambda_{2}$ and $\bar{\lambda}_{2}=-\bar{\lambda}_{2}$. plugging this into \eqref{polarasationproduct}, then implies that $\epsilon_{1}^{+}\cdot\epsilon_{2}^{-}=1$
so we find that
\begin{equation}
\mathcal{A}_{{\rm YM}}^{+-}\propto k_{1},\,\,\,\mathcal{A}_{{\rm GR}}^{+-}\propto k_{1}^{3}.
\label{2ptymgr}
\end{equation}
Note that \cite{Maldacena:2011nz} makes a different choice for the 2-point kinematics, notably $\lambda_{1}=\bar{\lambda}_{2}$
and $\lambda_{2}=-\bar{\lambda}_{1}$. As a result, in axial gauge nonzero two point functions must have the same helicity rather than opposite helicities. Indeed, in axial gauge we have
\begin{equation}
\epsilon_{i}^{+,\text{axial}}\cdot\epsilon_{j}^{+,\text{axial}}=\frac{\left\langle \bar{i}\bar{j}\right\rangle ^{2}}{\left\langle \bar{i}i\right\rangle \left\langle \bar{j}j\right\rangle },\qquad \epsilon_{i}^{+,\text{axial}}\cdot\epsilon_{j}^{-,\text{axial}}=\frac{\left\langle \bar{i}j\right\rangle ^{2}}{\left\langle \bar{i}i\right\rangle \left\langle j\bar{j}\right\rangle },
\label{eq:axialgauge}
\end{equation}
so we get zero 2-point functions for opposite helicities when $\lambda_{1}=\bar{\lambda}_{2}$ and $\lambda_{2}=-\bar{\lambda}_{1}$.

Let us now consider how to lift the scalar 2-point correlators of SDYM and SDG to
spinning ones. Following the approach in flat space \cite{Boels:2013bi}, we multiply them
by certain factors which encode external helicities:
\begin{equation}
\tilde{\mathcal{A}}_{{\rm SDYM}}^{+-}=\varepsilon_{1}^{+}\varepsilon_{2}^{-}\mathcal{A}_{{\rm SDYM}}^{+-},\,\,\,\tilde{\mathcal{A}}_{{\rm SDG}}^{+-}=\left(\varepsilon_{1}^{+}\varepsilon_{2}^{-}\right)^{2}k_{1}^{2}\mathcal{A}_{{\rm SDG}}^{+-},\label{dressed2pt}
\end{equation}
where 
\begin{equation}
\varepsilon_{i}^{+}=\frac{\left\langle \bar{\mu}\bar{i}\right\rangle }{\left\langle \mu i\right\rangle },\,\,\,\varepsilon_{i}^{-}=\left(\varepsilon_{i}^{+}\right)^{-1},
\label{scalarpolarfactors}
\end{equation}
and a factor of $k_{1}^{2}$ was included in the second expression since the gravity 2-point function scales like $k_{1}^{3}$ \footnote{We can already see from this 2-point example that lifting from SDG
scalar correlators to gravitational correlators is nontrivial because the scalars describing SDG are conformally
coupled while graviton states can be treated like massless
scalars dressed with polarisation tensors. }.
Using the 2-point kinematics
described below \eqref{2ptkinematics}, we find that
\begin{equation}
\varepsilon_{1}^{\pm}\varepsilon_{2}^{\mp}=1
\end{equation}
from which we immediately find that the dressed scalar 2-point correlators
defined in \eqref{dressed2pt} are indeed proportional to those of YM and GR in \eqref{2ptymgr}.

\subsection*{3 points}

Let us now see what form three-point correlators take in spinor notation. Recall that they are expressed in terms of kinematic structure constants 
\begin{equation}
X_{i,j}=k_{iw}k_{ju}-k_{iu}k_{jw}.
\end{equation}
Using \eqref{eq:spinorparam} and \eqref{referencespinors}, we see that 
\begin{equation}
k_{w}=k^{\bar{w}}=\mu_{\alpha}k^{\alpha\beta}\mu_{\beta},\,\,\,k_{u}=-k^{v}=\mu_{\alpha}k^{\alpha\beta}\bar{\mu}_{\beta},\label{componentsp}
\end{equation}
where
\begin{equation}
\mu_{\alpha}=\epsilon_{\alpha\beta}\bar{\mu}^{\beta}=(0,1),\,\,\,\bar{\mu}_{\alpha}=\epsilon_{\alpha\beta}\bar{\mu}^{\beta}=(-1,0).
\end{equation}
Hence, the kinematic structure constant can be written in spinor form
as follows:
\begin{align}\label{Xijspinor}
X_{i,j}=&\left\langle \mu i\right\rangle \left\langle \mu j\right\rangle \left(\left\langle \bar{i}\mu\right\rangle \left\langle \bar{j}\bar{\mu}\right\rangle -i\leftrightarrow j\right) \nonumber \\
=&\left\langle \mu i\right\rangle \left\langle \mu j\right\rangle \mu^{\alpha}\bar{\mu}^{\beta}\bar{\lambda}_{i}^{[\alpha}\bar{\lambda}_{j}^{\beta]} \nonumber \\
=&\left\langle \mu i\right\rangle \left\langle \mu j\right\rangle \left\langle \bar{i}\bar{j}\right\rangle ,
\end{align}
where we noted that $\bar{\lambda}_{i}^{[\alpha}\bar{\lambda}_{j}^{\beta]}=\epsilon^{\alpha\beta}\left\langle \bar{i}\bar{j}\right\rangle $
and $\left\langle \mu\bar{\mu}\right\rangle $=1. Hence, we find that
the SDYM scalar 3-point function computed in \eqref{sdym3pt} is given by
\begin{equation}
\mathcal{A}_{3}(1^+,2^+,3^-)=\frac{X_{1,2}}{k_{123}}=\frac{\left\langle \mu1\right\rangle \left\langle \mu2\right\rangle \left\langle \bar{1}\bar{2}\right\rangle }{k_{123}}.
\label{sdym3ptspinor}
\end{equation}
Similarly, the SDG 3-point function computed in \eqref{sdg3pt} is given by
\begin{align}
\mathcal{M}_{3}(1^+,2^+,3^-)&=\frac{X_{1,2}}{k_{123}}\left(\frac{X_{1,2}}{k_{123}}-i\left(k_{1w}-k_{2w}\right)\right)\\
=&\frac{\left\langle \mu1\right\rangle \left\langle \mu2\right\rangle \left\langle \bar{1}\bar{2}\right\rangle }{k_{123}}\left(\frac{\left\langle \mu1\right\rangle \left\langle \mu2\right\rangle \left\langle \bar{1}\bar{2}\right\rangle }{k_{123}}-i\left(\left\langle \mu1\right\rangle \left\langle \bar{1}\mu\right\rangle -\left\langle \mu2\right\rangle \left\langle \bar{2}\mu\right\rangle \right)\right),
\end{align}
where we have used \eqref{componentsp} and \eqref{Xijspinor}.

Let us now consider how to lift the scalar SDYM correlator in \eqref{sdym3ptspinor} to a spinning correlator. We will leave the analogous calculation in gravity for future work. We follow the same procedure used in flat space \cite{Boels:2013bi} and dress it with the following factors: 
\begin{equation}
\mathcal{\tilde{\mathcal{A}}}_{3}^{{\rm SDYM}}(1^+,2^+,3^-)=\varepsilon_{1}^{+}\varepsilon_{2}^{+}\varepsilon_{3}^{-}\frac{k_{3u}}{k_{1u}k_{2u}}\mathcal{A}_{3}^{{\rm SDYM}},\label{3ptdressing}
\end{equation}
where $\varepsilon_i^{\pm}$ are defined in \eqref{scalarpolarfactors}.
The factors of $k_u$ are implied by\eqref{gaugeprops} and can be converted to spinor notation using \eqref{bispinor} and \eqref{componentsp}. Putting everything together, we
find that
\begin{equation}
\mathcal{\tilde{\mathcal{A}}}_{3}^{{\rm SDYM}}(1^+,2^+,3^-)=\frac{1}{k_{123}}\frac{\left\langle \bar{1}\bar{2}\right\rangle \left\langle \mu3\right\rangle ^{2}}{\left\langle \mu1\right\rangle \left\langle \mu2\right\rangle }.\label{sdym3ptsp}
\end{equation}

Let us compare this result to the three-point correlator for full YM in AdS$_{4}$. The calculation is essentially the same as in flat space, except that the integral over $z$ runs from $0$ to $\infty$, which gives a pole in the total energy rather than a delta function: 
\begin{equation}
\mathcal{A}_{3}^{{\rm YM}}(1^+,2^+,3^-)=\frac{1}{2k_{123}}\left(\epsilon_{1}\cdot\epsilon_{2}\epsilon_{3}\cdot k_{1}+{\rm cyclic}\right)-\left(1\leftrightarrow2\right).
\end{equation}
Recalling that in lightcone gauge $\epsilon_{i}^{\pm}\cdot\epsilon_{j}^{\pm}=0$ and
using the spinorial expressions in \eqref{polarisations} we see that 
\begin{equation}
\epsilon_{i}^{+}\cdot\epsilon_{j}^{-}=\frac{\left\langle \mu j\right\rangle \left\langle \bar{i}\bar{\mu}\right\rangle }{\left\langle \mu i\right\rangle \left\langle \bar{\mu}\bar{j}\right\rangle },\,\,\,p_{i}\cdot\epsilon_{j}^{+}=\frac{\left\langle i\mu\right\rangle \left\langle \bar{i}\bar{j}\right\rangle }{\left\langle \mu j\right\rangle },\,\,\,p_{i}\cdot\epsilon_{j}^{-}=\frac{\left\langle ij\right\rangle \left\langle \bar{i}\bar{\mu}\right\rangle }{\left\langle \bar{j}\bar{\mu}\right\rangle }.
\end{equation}
Using these formulae, we find that
\begin{align}
\epsilon_{1}^{+}\cdot\epsilon_{2}^{+}\epsilon_{3}^{-}\cdot k_{1}+{\rm cyclic}=&\frac{\left\langle \mu3\right\rangle }{\left\langle \mu1\right\rangle \left\langle \mu2\right\rangle \left\langle \bar{\mu}\bar{3}\right\rangle }\left(\left\langle \bar{2}\bar{\mu}\right\rangle \left\langle 2\mu\right\rangle \left\langle \bar{2}\bar{1}\right\rangle +\left\langle \bar{\mu}\bar{1}\right\rangle \left\langle \bar{3}\bar{2}\right\rangle \left\langle \mu3\right\rangle \right) \nonumber \\
=&\frac{\left\langle \mu3\right\rangle }{\left\langle \mu1\right\rangle \left\langle \mu2\right\rangle \left\langle \bar{\mu}\bar{3}\right\rangle }\left\{ \left\langle \bar{2}\bar{\mu}\right\rangle \left\langle 2\mu\right\rangle \left\langle \bar{2}\bar{1}\right\rangle -\left(\left\langle \bar{\mu}\bar{3}\right\rangle \left\langle \bar{2}\bar{1}\right\rangle +\left\langle \bar{\mu}\bar{2}\right\rangle \left\langle \bar{1}\bar{3}\right\rangle \right)\left\langle \mu3\right\rangle \right\} \nonumber \\
=&\frac{\left\langle \bar{1}\bar{2}\right\rangle \left\langle \mu3\right\rangle ^{2}}{\left\langle \mu1\right\rangle \left\langle \mu2\right\rangle }+k_{123}\frac{\left\langle \mu3\right\rangle \left\langle \mu\bar{1}\right\rangle \left\langle \bar{\mu}\bar{2}\right\rangle }{\left\langle \mu1\right\rangle \left\langle \mu2\right\rangle \left\langle \bar{\mu}\bar{3}\right\rangle },
\end{align}
where we obtained the second line using the Schouten identity:
\begin{equation}
\bar{\lambda}_{i}^{\alpha}\left\langle \bar{j}\bar{k}\right\rangle +\bar{\lambda}_{j}^{\alpha}\left\langle \bar{k}\bar{i}\right\rangle +\bar{\lambda}_{k}^{\alpha}\left\langle \bar{i}\bar{j}\right\rangle =0,
\end{equation}
and we obtained the third line using \eqref{energynon}. Hence we find that 
\begin{align}
\mathcal{A}_{3}^{{\rm YM}}(1^+,2^+,3^-)=&\frac{1}{k_{123}}\frac{\left\langle \bar{1}\bar{2}\right\rangle \left\langle \mu3\right\rangle ^{2}}{\left\langle \mu1\right\rangle \left\langle \mu2\right\rangle }+\frac{\left\langle \mu3\right\rangle }{2\left\langle \mu1\right\rangle \left\langle \mu2\right\rangle \left\langle \bar{\mu}\bar{3}\right\rangle }\left(\left\langle \mu\bar{1}\right\rangle \left\langle \bar{\mu}\bar{2}\right\rangle -\left\langle \mu\bar{2}\right\rangle \left\langle \bar{\mu}\bar{1}\right\rangle \right) \nonumber \\
=&\left(1-\frac{k_{123}}{2k_{3u}}\right)\mathcal{\tilde{A}}_{3}^{{\rm SDYM}}(1^+,2^+,3^-),\label{fullYM3pt}
\end{align}
where we noted that $\left\langle \mu\bar{1}\right\rangle \left\langle \bar{\mu}\bar{2}\right\rangle -\left\langle \mu\bar{2}\right\rangle \left\langle \bar{\mu}\bar{1}\right\rangle =\left\langle \bar{1}\bar{2}\right\rangle $
and $\left\langle \mu3\right\rangle \left\langle \bar{\mu}\bar{3}\right\rangle =-k_{3u}$
(from \ref{componentsp}).

Taking the difference between the SDYM result in \eqref{sdym3ptsp} and the full YM result in \eqref{fullYM3pt} then gives 
\begin{equation}
\mathcal{\tilde{A}}_{3}^{{\rm SDYM}}(1^+,2^+,3^-)-\mathcal{A}_{3}^{{\rm YM}}(1^+,2^+,3^-)=\varepsilon_{1}^{+}\varepsilon_{2}^{+}\varepsilon_{3}^{-}\frac{k_{3u}}{k_{1u}k_{2u}}\left(\frac{X_{1,2}}{2k_{3u}}\right).\label{3ptdiscrep}
\end{equation}
This discrepancy can be removed by subtracting the following total
derivative term from the scalar action of SDYM:
\begin{equation}
\delta\mathcal{L}=\frac{1}{2}\partial_{z}{\rm Tr}\left[\left(\partial_{u}^{-1}\bar{\Phi}\right)\left\{ \Phi,\Phi\right\} \right].
\end{equation}
Indeed, using the bulk-to-boundary propagators in section \ref{sec:SDYMfeynman}, we easily see that the contribution to the 3-point correlator due to this interaction vertex is given by
\begin{equation}
\label{boundaryterm}
\delta\mathcal{A}_{3}(1^+,2^+,3^-)=\frac{1}{2}\int_{0}^{\infty}dz\partial_{z}\left(\frac{1}{k_{3u}}X_{1,2}e^{-k_{123}z}\right)=\frac{X_{1,2}}{2k_{3u}},
\end{equation}
which indeed gives \eqref{3ptdiscrep} after dressing the scalar correlator according to \eqref{3ptdressing}. Note that the right-hand-side of \eqref{boundaryterm} vanished in the flat space limit since it does not contain an energy pole.

\subsection*{4-point SDYM}
As an illustration of the spinor-helicity notation we compute the full SDYM scalar correlator at four points and show that it is void of a total energy pole. This is expected as it is well known that the amplitude for this helicity configuration is zero in the flat space limit and thus provides a useful consistency check. The full 4-point color-ordered correlator receives a contribution from two diagrams, one from the $s$-channel and the other from the $t$-channel. The expression for the $s$-channel diagram was computed in section \ref{SDYM-corr} and the final result is given as 
\begin{eqn}\label{SDYMs-1}
\mathcal{A}_4^{(s)}(1^+,2^+,3^+,4^-) = \frac{X_{1, 2} X_{3, 4}}{E E_L^{(s)} E_R^{(s)}}~,
\end{eqn}
where $E = k_{1234}$, $E_L^{(s)} = k_{12} + |\vec k_{12}|$ and $E_R^{(s)} = k_{34} + |\vec k_{12}|$. Similarly, the contribution from the $t$-channel is given as
\begin{eqn}
\mathcal{A}_4^{(t)}(1^+,2^+,3^+,4^-)  = \frac{X_{1, 4} X_{3, 2}}{E E_L^{(t)} E_R^{(t)}}~,
\end{eqn}
where $E_L^{(t)}=k_{23} + |\vec k_{23}|$ and $E_R^{(t)} = k_{14}+ |\vec k_{23}|$. By adding the two we obtain the full correlator 
\begin{eqn}\label{SDYMcorrfull1}
\mathcal{A}_4(1^+,2^+,3^+,4^-) = \frac{1}{E} \left( \frac{X_{1, 2} X_{3, 4}}{E_L^{(s)} E_R^{(s)}} + \frac{X_{1, 4} X_{3, 2}}{ E_L^{(t)} E_R^{(t)}} \right) ~.
\end{eqn}
To simplify the expression we note the following relation: 
\begin{eqn}
k_{ij}^2 - |\vec k_{ij}|^2 = 2 (k_i k_j - \vec k_i \cdot \vec k_j)= -4 \braket{ij} \braket{\bar i \bar j}~,
\end{eqn}
which results in the following simplification for the denominators of \eqref{SDYMcorrfull1},
\begin{eqn}
E_L^{(s)} E_R^{(s)} &= E_L^{(s)} E + 4 \braket{12} \braket{\bar 1 \bar 2}, \\
E_L^{(t)} E_R^{(t)} &= E_L^{(t)} E + 4 \braket{23} \braket{\bar 2 \bar 3}~.
\end{eqn}
Combined with \eqref{Xijspinor} we find that
\begin{eqn}
\mathcal{A}_4(1^+,2^+,3^+,4^-) = \frac{\braket{\mu 1} \braket{\mu 2} \braket{\mu 3} \braket{\mu 4}}{E E_L^{(t)} E_R^{(t)} E_L^{(s)} E_R^{(s)} } &\Big[  E E_L^{(t)} \braket{\bar 1 \bar 2} \braket{\bar 3 \bar 4} + E E_L^{(s)} \braket{\bar 2 \bar 3} \braket{\bar 4 \bar 1}\\
&+ 4  \braket{\bar 1\bar 2} \braket{\bar 2 \bar 3} \big( \braket{12}\braket{\bar 4 \bar 1} + \braket{23} \braket{\bar 3 \bar 4} \big) \Big]~.
\end{eqn}
The term in the second line can be simplified using the momentum conservation equation given in equation \eqref{energynon}. This results in the following simplified expression for the correlator:
\begin{eqn}
\mathcal{A}_4(1^+,2^+,3^+,4^-) =  \frac{\braket{\mu 1} \braket{\mu 2} \braket{\mu 3} \braket{\mu 4}}{E_L^{(s)} E_L^{(t)} E_R^{(s)} E_R^{(t)}} \Big( E_L^{(t)} \braket{\bar 1 \bar 2} \braket{\bar 3 \bar 4}  + E_L^{(s)} \braket{\bar 2 \bar 3} \braket{\bar 4 \bar 1} + 4 \braket{\bar 1 \bar 2} \braket{\bar 2 \bar 3} \braket{2 \bar 4} \Big)~.
\end{eqn}
By restoring the factors of $X_{i,j}$ we obtain
\begin{eqn}\label{SDYM-full}
\mathcal{A}_4(1^+,2^+,3^+,4^-) =  \frac{1}{E_L^{(s)} E_L^{(t)} E_R^{(s)} E_R^{(t)}} \Big( E_L^{(t)} X_{1,2} X_{3,4} + E_L^{(s)} X_{2,3} X_{4,1} + 4 X_{1,2} X_{3,4} \frac{\braket{\bar 4 2} \braket{\bar 2 \bar 3}}{\braket{\bar 4 \bar 3}}  \Big)~.
\end{eqn}
Hence the full SDYM scalar correlator does not contain a total energy pole and therefore vanishes in the flat space limit. It would be interesting to lift this to a spinning correlator and compare it to the 4-point correlator of full YM, which we leave for future work.  

\subsection*{4-point SDG}

Finally, let us compute the tree-level 4-point scalar correlator in SDG. The $s$-channel contribution to the 4-point function was computed in section \ref{sec:correlators} where the final result was expressed in terms of the differential operators given in \eqref{SDG-4pt1}. By explicitly evaluating the derivatives in \eqref{SDG-4pt1} we obtain the following:
\begin{align}\label{SDG-4ptexp}
\mathcal{M}_4^{(s)}(1^+,2^+,3^+,4^-) &= X_{1,2} X_{3,4} \Bigg[ X_{1,2} X_{3,4} \left\{  \frac{2}{E^2} +  \frac{1}{E E_L^{(s)}} + \frac{1}{E E_R^{(s)}}   + \frac{1}{E_L^{(s)} E_R^{(s)}} \right\} \\
&\qquad \qquad  -i X_{1,2} (k_{3w} - k_{12w}) \Big\{\frac{1}{E} +  \frac{1}{E_R^{(s)}} \Big\} - i X_{3,4} (k_{1w} - k_{2w}) \Big\{\frac{1}{E} +  \frac{1}{E_L^{(s)}} \Big\} \nno \\
&\qquad \qquad - (k_{1w} - k_{2w}) (k_{3w} - k_{12w}) \Bigg] \frac{1}{E E_L^{(s)} E_R^{(s)}}, \nno
\end{align}
where we use the notations $E,E_L^{(s)},E_R^{(s)}$ are defined below equation \eqref{SDYMs-1}. The complexity of the expression above shows the simplicity of using the differential operators in equation \eqref{SDG-4pt1}. As discussed around \eqref{SDG-4pt-flat1}, the flat space limit of this correlator is obtained by evaluating the residue of the $\frac{1}{E^3}$ pole. However, since the full amplitude is known to vanish in flat space, we expect that the $\frac{1}{E^3}$ pole in the correlator cancels out after summing over all diagrams. Apart from the $s$-channel contribution given above, the full correlator receives a contribution from $t$ and $u$-channels, which are obtained by the replacements $(1234) \to (2341)$  and $(1234) \to (1324)$ respectively in \eqref{SDG-4ptexp}. The full correlator can be expressed in terms of psinor variables by performing a similar set of manipulations as the SDYM case. Adding the $\frac{1}{E^3}$ contribution from each channel then gives
\begin{eqn}\label{SDG-full1}
&\frac{2}{E^3} \left( \frac{X_{1,2}^2 X_{3,4}^2}{E_L^{(s)} E_R^{(s)}} 
+ \frac{X_{1,4}^2 X_{2,3}^2}{E_L^{(t)} E_R^{(t)}}
+ 
\frac{X_{1,3}^2 X_{2,4}^2}{E_L^{(u)} E_R^{(u)}}\right)\\
&= 
\frac{2 X_{1,2}^2 X_{3,4}^2 \braket{\bar 1 \bar 4} \braket{\bar 1 \bar 3} }{E^2 E_L^{(s)} E_R^{(s)} E_L^{(t)} E_R^{(t)} E_L^{(u)}  \braket{\bar 1 \bar 2} \braket{\bar 3 \bar 4}}\Big[  - E - \braket{\bar 2 4} \braket{\bar 4 \bar 3} + 2 \braket{\bar 2 3} \braket{\bar 3  \bar 4}  \Big],
\end{eqn}
which shows that the $\frac{1}{E^3}$ pole drops out as expected so full correlator indeed vanishes in the flat space limit. Lifting the SDG scalar correlator to a spinning correlator is a very nontrivial task which we leave to future work.

\end{appendix}
\bibliographystyle{JHEP}
\bibliography{references}
\end{document}